\documentclass[mnsc,nonblindrev]{informs3}

\renewcommand{\theARTICLETOP}{}

\OneAndAHalfSpacedXI

\usepackage{natbib}
 \bibpunct[, ]{(}{)}{,}{a}{}{,}%
 \def\bibfont{\small}%

\usepackage{booktabs} 
\usepackage[ruled]{algorithm2e} 

\SetAlFnt{\small}
\SetAlCapFnt{\small}
\SetAlCapNameFnt{\small}
\SetAlCapHSkip{0pt}
\IncMargin{-\parindent}

\usepackage{subcaption}
\usepackage{graphicx}

\usepackage[usenames,svgnames]{xcolor}
\usepackage{enumitem}

\usepackage[]{quoting}

\usepackage{fancyhdr} 
\usepackage{url} 
\usepackage[normalem]{ulem}

\usepackage{multicol}
\usepackage{tikz} 
\usetikzlibrary{shapes,decorations}
\usetikzlibrary{arrows}
\usepackage{pgfplots} 
\pgfplotsset{compat=newest} 

\usepackage{hyperref}

\newcommand{\expec}{\ensuremath{\mathbf{E}}}
\newcommand{\ind}{\ensuremath{\mathbf{I}}}
\newcommand{\reals}{\ensuremath{\mathbb{R}}}
\newcommand{\naturals}{\ensuremath{\mathbb{N}}}
\newcommand{\defeq}{\ensuremath{\triangleq}}

\usepackage{etoolbox}
\AtBeginEnvironment{quote}{\small}

\newcommand{\Pidb}{\ensuremath{\Pi_{\mathsf{AP}}}}
\newcommand{\Piob}{\ensuremath{\Pi_{\mathsf{SM}}}}

\newcommand{\join}{\ensuremath{\mathsf{join}}}
\newcommand{\leave}{\ensuremath{\mathsf{leave}}}
\newcommand{\high}{\ensuremath{\mathsf{H}}}
\newcommand{\low}{\ensuremath{\mathsf{L}}}
\newcommand{\NN}{\mathbb{N}_0}

\newcommand{\fim}{\ensuremath{\mathsf{fi}}}
\newcommand{\nim}{\ensuremath{\mathsf{ni}}}
\newcommand{\ar}{\ensuremath{\mathsf{ap}}}
\newcommand{\sm}{\ensuremath{\mathsf{sm}}}
\newcommand{\tsm}{\ensuremath{\mathsf{tsm}}}
\newcommand{\cM}{\ensuremath{\mathsf{ar}}}

\usepackage{comment}
\usepackage[suppress]{color-edits}
\addauthor{vmr}{MediumBlue}
\addauthor{vm}{red}
\addauthor{ki}{green}

\addauthor[suppress]{ja}{magenta}

\usepackage{mathtools}

\TheoremsNumberedThrough     

\EquationsNumberedThrough    

\begin{document}



\RUNAUTHOR{Anunrojwong, Iyer, and Manshadi}
\RUNTITLE{Information Design for Congested Social Services}

\TITLE{Information Design for Congested Social Services: Optimal Need-Based Persuasion}

\ARTICLEAUTHORS{%
\AUTHOR{Jerry Anunrojwong}
\AFF{Columbia Business School, New York, NY, \EMAIL{jerryanunroj@gmail.com}}
\AUTHOR{Krishnamurthy Iyer}
\AFF{Industrial and Systems Engineering, University of Minnesota, Minneapolis, MN,  \EMAIL{kriyer@umn.edu}}
\AUTHOR{Vahideh Manshadi}
\AFF{Yale School of Management, New Haven, CT, \EMAIL{vahideh.manshadi@yale.edu}}
} 

\ABSTRACT{%
We study the effectiveness of information design in reducing
congestion in social services catering to users with varied levels of need. In the absence of price discrimination and centralized admission, the provider relies on sharing information about wait times to improve welfare.  We consider a stylized model with heterogeneous users who differ in their private outside options: {\em low-need} users have an acceptable outside option to the social service, whereas {\em high-need} users have no viable outside option.  Upon arrival, a user decides to wait for the service by joining an unobservable first-come-first-serve queue, or leave and seek her outside option. To reduce congestion and improve social outcomes, the service provider seeks to persuade more low-need users to avail their outside option, and thus better serve high-need users.  We characterize the Pareto-efficient signaling mechanisms and compare their welfare outcomes against several benchmarks. We show that if either type is the overwhelming majority of the population, information design does not provide improvement over sharing full information or no information. On the other hand, when the population is sufficiently heterogeneous, information design not only Pareto dominates full-information and no-information mechanisms, in some regimes it also achieves the same welfare as the ``first-best'', i.e., the Pareto-efficient centralized admission policy with knowledge of users' types.
}%





\maketitle
\section{Introduction}\label{sec:intro}

Social services often face the challenge of congestion due to their
limited capacity relative to their demand. The congestion partly stems from the inclusionary intent of such services: a toll-free road is available to all citizens even those who can afford alternative tolled
ones; a broad range of low- and middle-income households are eligible to apply for public housing; urgent care centers admit patients with
varied levels of condition severity. How can a social service provider
reduce congestion and thus the efficiency loss associated with service
delay?

In this context, the two controls commonly used for managing
congestion, i.e., pricing and centralized admission control, are
inapplicable due to fairness
and implementation considerations.  However, the service provider may have control over information about the status of the system to be shared with users. \kiedit{As such, the service
provider can leverage this informational advantage to influence consumer's decision in seeking the social service.}

\begin{figure}
\centering
\includegraphics[width=0.7\textwidth]{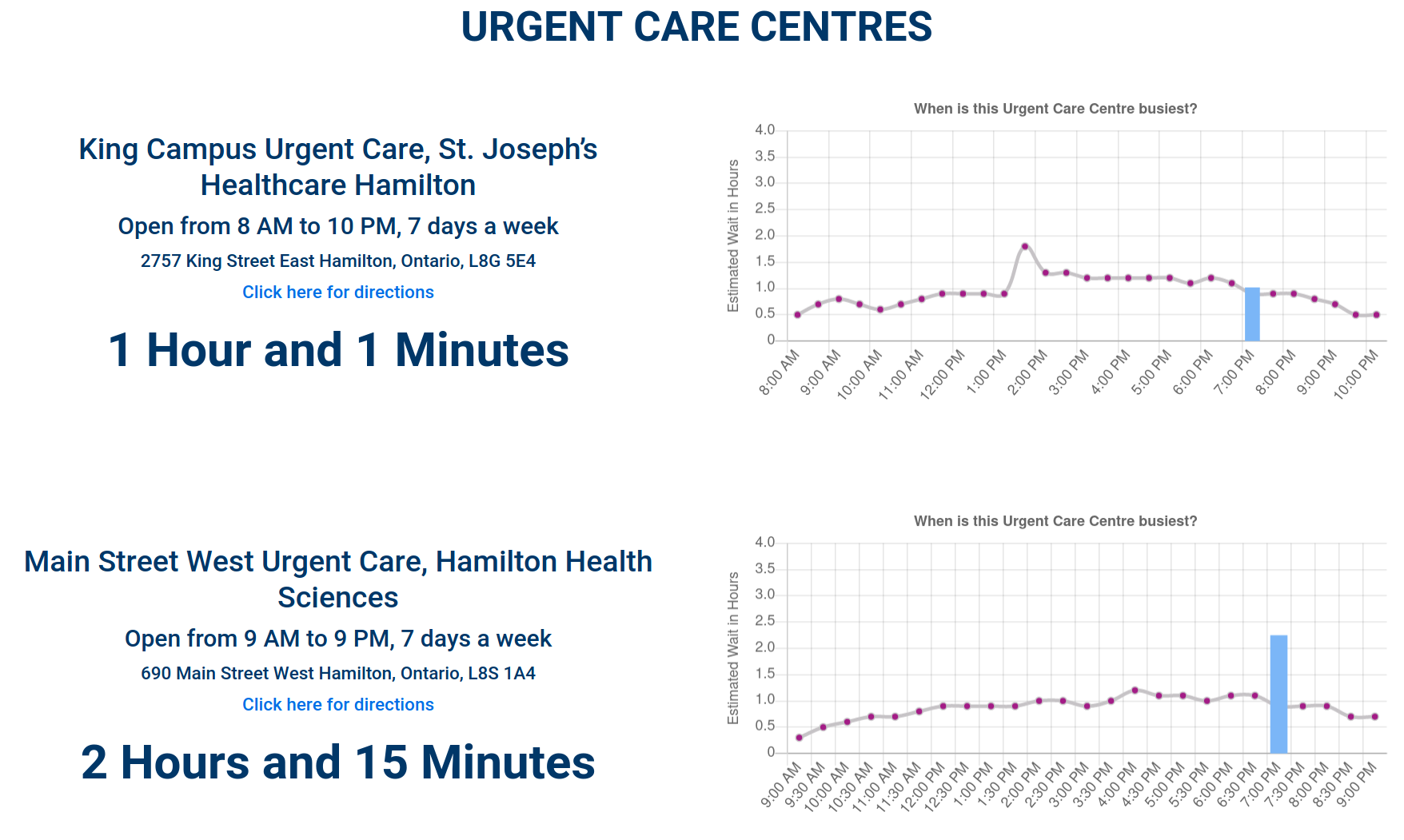}
\caption{Screenshot of the wait-time dashboard program for the Hamilton healthcare system.}\label{fig:motivation}
\end{figure}

\kiedit{\subsection{Motivating Examples}}
\label{subsec:motivation}
\kiedit{
A wide range of information provision policies are employed in practice. In the context of urgent care, some hospitals aim to provide real-time estimate of wait-time to patients. For example, see Figure \ref{fig:motivation} for a snapshot of the  \href{https://www.hamiltonemergencywaittimes.ca/}{Hamilton Healthcare System}'s wait-time dashboard which we discuss further below
(see also \href{https://jfkmc.com/covid-19/index.dot}{JFK medical center} and \href{https://www.smchealth.org/general-information/medical-emergencies}{San Mateo Medical Center} which employ similar programs). On the other hand, in the context of public housing, certain authorities provide no wait-time information (see, e.g., \href{http://www.haca.net/applicants/current-wait-list-applicants/}{Housing Authority of the County of Alameda}) while others provide average estimates (see e.g., \href{https://affordablehousingonline.com/public-housing-waiting-lists/New-York}{New York Public Housing and Project-Based Voucher Waiting Lists}). We highlight that in the above applications---which we broadly refer to as social services---managing congestion by ``pricing out'' users or by controlling admissions is impractical or undesirable}.

Through information provision, service providers aim to  not only inform users about their wait-time, but also to  ``help'' users decide whether to seek  the service. We take the Hamilton Healthcare System as our leading  example: As reported in~\citet{HamiltonExample}, upon launching their wait-time dashboard program, managers envision that 
providing wait-time information is particularly useful for patients with {\em less severe} conditions who can use this information to decide whether to currently seek care at a particular emergency center. Here, we highlight a quote from a manager\footnote{The quoted manager is Dr. Greg Rutledge who is the chief of emergency medicine in one of Hamilton's regional hospitals (St. Joseph’s Health Care).}:
\begin{quote}
    \textit{``There are still [going to]  be people who have services like nephrology, or their heart doctors, or their lung doctors, who should go regardless of the wait time,} [$\ldots$] \textit{But for those that have less serious conditions, they can decide not only where but when to go.''}
\end{quote}

\kiedit{The above insights highlight that in applications such as emergency care, there are fundamental differences in the level of need in the user population: some have no choice but to seek the service regardless of the congestion level whereas others can forgo the service if they perceive that the wait-time is too long.}


\kiedit{It is this fundamental heterogeneity of need that healthcare systems rely on when using wait-time dashboard programs, like that of the Hamilton healthcare system, to manage congestion. In this context, rather than providing full-information, one can design dashboards that provide \emph{coarsened} information about the congestion. For instance, a dashboard may announce that the wait-time is above or below a threshold $x$, or in between a sequence of thresholds $x_1, x_2, \ldots, x_k$. Such coarsened information induce a belief that the congestion level might be high even in times of moderate congestion, and thus could persuade away users with less severe need from seeking service, resulting in reduced congestion overall. In this paper, we study the  effectiveness of such information provision policies. }


\subsection{{Overview of Our Work}}

To investigate the effectiveness of information design in improving
welfare for a congested social service, we develop a stylized model
that captures the key features of such a system. We consider a single
server queueing system where users arrive according to a Poisson
process and their service times are i.i.d. and exponentially
distributed.\footnote{\vmredit{In our base model, we focus on the first-come-first-serve queuing discipline. However, motivated by applications such as urgent care systems, we also study a preemptive priority discipline in Section~\ref{subsec:priority} and demonstrate the robustness of our results under such a queuing discipline.}}  Upon arrival, each user decides to either wait for the
service by joining an unobservable queue or seek her outside
option. To capture the disparity that users face with regard to the quality of their outside
options, we categorize users into two groups: (1) {\em high-need}
users that have no feasible outside option and (2) {\em low-need}
users that have a viable alternative.
Both types incur higher waiting
costs upon joining a longer queue. Upon arrival, a high-need user always joins as
she does not have any other choice.  However, a low-need
user makes a {\em join} or {\em leave} decision to maximize her
expected utility. Even though an arriving user does not observe the
queue, her decision relies on her belief about the queue size based on
the information shared by the service provider.

We assume the service provider has complete information about the
status of the queue and he can decide how much of this information he will share with the arriving user. Sharing the information fully
may lead to bad welfare outcomes because a utility-maximizing user
does not internalize the negative externality that she imposes on
others~\citep{Naor}. Instead, the service
provider can use the lever of information sharing to influence users'
beliefs about the queue size and consequently their
decisions.  We adopt the framework of Bayesian persuasion
or information design\footnote{We use the terms Bayesian
    persuasion and information design
    interchangeably.}~\citep{kamenica} in which the service provider
commits\footnote{{Since we focus on minimizing congestion, the service provider's preferences are aligned with the {\em ex ante} preferences of the users. We believe this preference alignment makes the social service provider more likely to keep its commitment.}} to a signaling mechanism in response to which users follow an
equilibrium strategy. The welfare of each type is thus determined by
the signaling mechanism and the corresponding equilibrium response of
the users. Because high-need users always join the queue, the service
provider does not need to know user types to implement a signaling
mechanism. 


Our analysis follows the standard approach (see
e.g. \cite{lingenbrink2019optimal, bergemannM16,
  candogan-drakopoulos}) which allows us to only consider {\em
  obedient} binary signaling mechanisms where upon the arrival of a
user, the service provider makes a ``$\join$'' or ``$\leave$''
recommendation and the user finds it incentive compatible to follow
that recommendation. Further, it builds on
\cite{lingenbrink2019optimal} to establish an equivalence between the
class of obedient binary signaling mechanisms and the set of steady
state distributions that satisfy certain linear
constraints. To ensure welfare
improvement for {\em both} types, we focus on Pareto-efficient
signaling mechanisms and establish
structural results for any such mechanisms. Under mild monotonicity assumptions on
utility functions, we show
that any Pareto-efficient signaling mechanism has a threshold
structure (Theorem~\ref{thm:threshold-signaling-mechanism}).

With these structural results, we compare the optimal signaling
mechanism against the benchmarks of full information sharing and no
information sharing. Our analysis reveals several intriguing insights into effectiveness of information design. First, there exists a signaling mechanism that Pareto-dominates full-information sharing unless the latter remains Pareto-efficient
even if the service provider is allowed to disregard user
incentives (Proposition~\ref{prop:info-design-power} and Theorem~\ref{thm:full-info-comparison}). However, if the
population is mostly comprised of low-need users the welfare gain due
to information design is fairly limited (Proposition~\ref{prop:homogen}). This dichotomy stems from the intuition that in the absence of high-need users, a low-need user cannot be persuaded to leave if the queue length falls below the threshold up to which she would have joined under full information.  On the other hand, when high-need users are present it is possible to persuade more low-need users to leave.  Second, there exists a signaling mechanism that Pareto-dominates no information sharing only if the arrival rate of high-need users does not exceed a threshold.  However, if high-need users constitutes the overwhelming majority, then interestingly, no information is Pareto efficient even when the provider is allowed to disregard user incentives (Proposition~\ref{prop:info-design-power} and Theorem~\ref{thm:no-info-comparison}).  The
main intuition behind this result is that with abundance of demand
from high-need users, the system is so congested that a low-need user does not need much persuasion to choose her outside option over
the social service. Conversely, if the system is not overcrowded by
high-need users, information design proves effective over sharing no
information.  Putting these insights together, {\em we conclude that
signaling is particularly effective if the user population 
shows
sufficient heterogeneity in  need.}

To further study the power of information design, we compare its
Pareto frontier with that of a strong benchmark in which the service
provider implements a Pareto efficient {\em admission policy}
disregarding the user's incentives.  Interestingly, we show that if
the arrival rate of high-need users is higher than a threshold, the two Pareto frontiers indeed coincide. Even if the arrival rate of high-need users is below that threshold, the two Pareto frontiers
show considerable overlap ({Theorem}~\ref{thm:achieve-first-best}). 
  This further illustrates the effectiveness of
information design: any Pareto-efficient signaling mechanism that
belongs to the overlapping regions of the frontier achieves the same
welfare outcomes as those of an admission policy that can not only
observe the user types, but also enforce the join or leave decision
without regard to their incentives. Further, in such\vmdelete{overlapping}
cases, no user is indifferent between their recommended action and the
alternative, implying that the signaling mechanism primarily plays the
role of a coordination device. This is in contrast with usual
persuasion settings, where the optimal signaling mechanism extracts
all user surplus for \jadelete{at least} some signals.

To highlight our comparative insights, in Section~\ref{sec:linear-costs}, we complement our theoretical  findings with illustrative numerical examples  (see Figures~\ref{fig:welfare-comparison}-\ref{fig:heatmap} and their related discussions). Additionally, in Section~\ref{sec:extensions}, we describe how our model can be generalized to incorporate 
\kiedit{finite outside option for high-need users (in Section~\ref{subsec:finite}) and 
heterogeneity in service rates (in Section~\ref{subsec:priority}).\footnote{\kiedit{Further, in Appendices~\ref{subsec:abandon} and~\ref{subsec:greater-heterogeneity}, we study two other extensions: respectively, exogenous abandonment and more than two types of users.}}
We analytically or numerically show our qualitative insights hold in these richer models (see Proposition~\ref{prop:finite-info},  Figures~\ref{fig:finite:1}-\ref{fig:priority-queue}, and their related discussions).}

\subsection{Managerial insights}

In summary, our work investigates the effectiveness of information
design as a potential approach for reducing congestion in social
services offered to users heterogeneous in their needs. Using a
stylized model, we show that by implementing a Pareto-efficient
signaling mechanism, the service provider can achieve Pareto
improvement in the welfare by persuading more low-need users to seek
outside option, thereby reducing congestion. We also identify
conditions under which information design not only outperforms the
simple mechanisms of full or no information sharing, but also achieves
the same welfare outcomes as centralized admission policies that know
each user's need for the service. 

\kiedit{As wait-time dashboard programs have become prevalent means for congestion management in service systems, there is a natural impulse to design systems that accurately estimate and share complete information. 
\kiedit{However, contrary to the general wisdom, our results show that sharing accurate information could in fact be uniformly detrimental to all the users. Instead, revealing partial information, say in the form of thresholds and/or intervals, can improve the welfare outcomes across all users. 
Dashboard programs based on such coarsened information would also be practically appealing as they alleviate the need to accurately estimate wait-times in real-time, a task that has been documented to be significantly challenging in  practice~\citep{ang2016accurate, StanfordExample}.}
Thus, our results imply that information design not only alleviates the need for accurate wait-time estimation, this benefit comes at no welfare cost.}

\vmredit{Lastly, we discuss two practical concerns one may have about disclosing partial information for social services: repugnance and information leakage. With regard to the former, given that most information provision policies commonly used in practice do not follow full information disclosure, we do not envision that implementing our proposed policies  
would be perceived differently.\footnote{\vmredit{As discussed in Section~\ref{subsec:motivation}, some public housing authorities choose to disclose no information. Also, only some urgent care systems aim to provide real-time information that relies on (often) inaccurate estimations.}} With regard to the latter, we emphasize that we only focus on public signaling mechanisms which removes the possibility of information leakage across agents. However, information leakage over time can happen: if an agent strategically waits upon arrival and observes more than one signals before deciding to join or leave, she may be able to infer the state of the system more accurately. Nevertheless, our prescribed policy of only disclosing whether the congestion level is above/below a threshold would still perform well: observing a few signals only reveals extra information if the congestion level is close to the threshold, and thus the signal changes. Consequently, our policy still persuades away those who arrive when the congestion level is sufficiently above the threshold.\footnote{\vmredit{Developing a model that captures information leakage while being grounded in practice as well as characterizing the optimal information design are interesting directions for future work.}}}

\subsection{Related Work}

Our work relates to and contributes to several streams of literature.

\textit{Information Design: } Like ours, in many other settings
service providers and platforms have access to more information than
their customers. As such, informational aspects of service and
platform operations have been studied in many applications. Adopting
the framework of Bayesian persuasion pioneered by \citet{kamenica},
\citet{lingenbrink2018signaling} and \citet{drakopoulos2018persuading}
study effectiveness of information design for influencing the
customers' time of purchase in order to maximize the platform's
revenue. 
In a similar context, \citet{kuccukgul2019engineering} study information design
for time-locked sales campaigns on online platforms.\vmdelete{In their setting, the customers are uncertain about 
a product's value and the platform seeks to influence their purchase decision by 
strategically sharing information about previous customers' actions, with the goal of maximizing its own revenue.}
Focusing on two-sided platforms, \citet{bimpikis2020information} examine the impact of information design on supply-side decisions towards the goal of  increasing platform's revenue.
\citet{kremer} and \citet{crowdsourcing} focus on information
design in a sequential learning setting with the goal of maximizing
social welfare. In the context of misinformation on social platforms,
\citet{candogan-drakopoulos} study how the platform can optimally
signal the content accuracy while incentivizing desirable levels of
user engagement in the presence of positive network
externalities. 
Outside the framework of Bayesian persuasion,  for dynamic contests, \citet{bimpikis2019designing} show that the information disclosure policy used to inform participants about the status of competition substantially impacts the outcome. 
\citet{kanoria-saban-facilitating} show that a two-sided matching platform
can significantly improve welfare by hiding information about the
quality of a user's potential partners. In another interesting
direction, \citet{nahum} show that in two-sided matching, the presence
of experts who can reveal information can lead to an inferior outcome
for everyone in two-sided matching even if the use of such experts is
optional.

Closest to our setting is the \vmdelete{recent} work of
\citet{lingenbrink2019optimal} that study optimal signaling for
services with unobservable queues. Even though our work builds on the
machinery developed in \citet{lingenbrink2019optimal}, there are also
key differences which we discuss next. \citet{lingenbrink2019optimal}
are concerned with maximizing the service provider's revenue using
information sharing as well as static pricing. As such, the goal of an
optimal signaling mechanism in that setting is to persuade more
customers to join the queue. However, in our setting the service
provider uses information sharing mechanism to improve welfare
outcomes by persuading more low-need users to leave. Further,
\citet{lingenbrink2019optimal} mainly focus on a setting with
homogeneous users whereas we study a setting with different user
types. Relatedly, \citet{anunrojwongIL2019} study persuasion of
non-expected-utility maximizing agents, and apply it to study
throughput maximization in queues where customers' disutility depends
on the variance of their waiting-times.

Finally, \citet{das2017congestion} also study how optimal information
sharing mechanisms can reduce congestion in a traffic network when a
user chooses a path among the set of paths some of which have
uncertain states. In particular, the authors consider a static setting
where a continuum of users simultaneously decide on the path they wish
to take to minimize their own cost, and show that all public signaling
mechanisms yield the same outcomes as full information (or no
information). Our paper complements this work by considering a dynamic
setting in which users of different types sequentially arrive over
time. Upon arrival of each user, the service provider sends a
state-dependent signal. We show that public signaling can be effective
in improving welfare outcomes when compared to special mechanisms of
full information and no information.

\textit{Strategic Behavior in Queueing Systems:} Following the seminal
work of \citet{Naor}, a stream of literature has focused on analyzing
queuing systems where users are strategic. (See the surveys by
\citet{rationalqueueing} and \citet{ibrahim2018sharing}, and the references therein.) In particular,
\citet{hassin17} study mechanisms for profit maximization in an
$M/M/1$ queue with homogeneous customers, and establish the optimality
of an information sharing mechanism that, along with appropriate
prices, makes ``high-low'' announcements where arriving customers
receive a ``low'' announcement if and only if the queue-length is
below the Naor's socially optimal threshold. Our work differs in two
main respects: first, as our application context is social services,
our model ignores pricing as a lever (effectively taking prices as
exogenous) but optimizes over all information sharing mechanisms; and
second, our objective is Pareto-improvement of (customer) welfare
rather than profit maximization, the two objectives usually being
opposed.  Focusing on the information sharing aspect (for an
unobservable queue), \citet{Allon} consider a cheap talk setting where
the service provider does not have commitment power. Additionally, as
discussed above, \cite{lingenbrink2019optimal} consider information
design in conjunction with pricing in order to maximize the service
provider's revenue. 
\kiedit{Finally, recently~\citet{che-tercieux-queue-design} study the optimal design of a queuing system where the planner decides on several aspects, including the queue discipline, entry, abandonment, and information sharing. Interestingly, they show that the optimal design is to follow FCFS, recommend users to join up to a threshold (in queue size), and never recommend abandonment for a user in the queue.}

\textit{Dynamic Allocation of Social Goods:} Our paper is also related to
the literature on dynamic allocation of social goods such as public
housing~\citep{kaplan1984managing} and donated
organs~\citep{ashlagi2013kidney, ashlagi2019matching}.  Recently,
~\citet{dynamic-matching-waiting-lists-leshno}
and~\citet{arnosti2017design} consider settings where the user has a
heterogeneous preference over arriving goods and thus she faces a
trade-off between waiting longer and accepting a less preferred
good.\footnote{While our contribution is theoretical, there is also an extensive
literature on practical aspects of provision and prioritization of
social services. For example, \cite{vi-spdat} examine the validity
and reliability of a widely-used homelessness vulnerability
assessment. \cite{criteria-renal-elderly} outline criteria for kidney
transplantation in elderly patients.} (Similar trade-off exists in dynamic matching as studied in
\citet{doval2018efficiency} and \citet{baccara2018optimal}.)  These
papers focus on designing efficient allocation mechanisms such as
waitlist mechanisms. We complement this literature by studying the
role that information sharing can play in improving welfare for social
services. 
\kiedit{Finally, the recent work of~\citet{ashlagi2020optimal} studies dynamic allocation of heterogeneous items  where an agent's value for an item is pair specific, i.e., jointly depends on the agent type and the quality of the item. The information design aspect of~\citet{ashlagi2020optimal} differs from ours in that it is concerned with information disclosure about 
the (unobservable) quality of an arriving item. In contrast, in this paper, we assume that the service  rate is known to the user and we focus on the information disclosure with regard to the (unobservable) congestion level. }


\textit{Mechanism Design without Money:} Our work investigates the power of information design to
reduce congestion in social services, where the usage of monetary
payments to shape agents' incentives is either impractical or
unpalatable. As such, it is broadly related to the growing literature
on mechanism design without money. Motivated by wide-ranging
applications, this stream of literature studies resource allocation
without relying on monetary payments.  For examples of static
settings, see \citet{procaccia2009approximate, prendergast2017food}
; dynamic settings are studied in
\citet{Peng-Santiago, From-Monetary-to-Non-Monetary}

\section{Model}
\label{sec:model}

In the following, we describe a model of information design for
improving welfare outcomes in a queueing setting with heterogeneous
users. Our model builds upon that of \citet{lingenbrink2019optimal},
who study revenue maximization in a related queueing setting.

Consider a service provider who provides a social service to a stream
of users arriving over time. Due to capacity constraints, the arriving
users possibly wait in an unobservable queue for service, where they
are served on a first-come-first-serve (FCFS) basis by a single
server. Each user's service time is independently and identically
distributed as an exponential distribution with rate
one.\footnote{This normalization of the service rate to one is without loss of
  generality.}

Arriving users must decide whether to join the queue and wait for the
service or to leave for an outside option. Upon joining, we assume
there is no abandonment: if a user joins the queue she will stay until
service completion. To describe users' utility, we start with
discussing their outside options. We model the users as belonging to
one of two groups which differ in the quality of their outside
options. Specifically, we assume that each user is either a (1) {\em
  high-need} user, who has no viable outside option, which we model by
letting their utility for taking the outside option be $-\infty$; or a
(2) {\em low-need} user who has a viable outside option whose utility
we normalize to $0$. We denote a user's type as $\high$ if they are
high-need, and by $\low$ if they are low-need. We assume that users of
type $i \in \{\high, \low\}$ arrive according to an independent
Poisson process with rate $\lambda_i$, with
$\lambda = \lambda_{\low} + \lambda_{\high}$ denoting the total
arrival rate. To avoid trivialities, we assume $\lambda_{\low} > 0$.
In our analysis, we also assume that $\lambda \leq 1$, to capture the
setting where the social service is not under-capacitated.

On joining the queue to obtain service, each user receives a net
utility composed of the benefit from the social service and a cost of
waiting until service completion. Formally, the utility function of a
type $i \in \{\low, \high\}$ user is given by
$u_i : \naturals_0 \to \reals$, where $u_i(n)$ denotes her utility on
joining a queue with $n$ users already in the system, either in queue or
being served.\footnote{Here, $\naturals_0 = \naturals \cup \{0\}$
  denotes the set of non-negative integers.}  We make the natural
assumptions that $u_i(0) >0$, and
$\lim_{k \rightarrow \infty} u_i(k) <0$. Further we make the following
assumption:
\begin{assumption}[Positive and diminishing incremental waiting costs]
\label{as:utility_assumptions} 
The utility functions satisfy the following monotonicity assumptions:
\begin{enumerate}
    \item For each type $i \in \{\high, \low\}$, the utility function $u_i(n)$ is  strictly decreasing in $n$.  
    \item The difference $u_\low(n) - u_\low(n+1)$ is non-increasing in $n$. 
\end{enumerate}
\end{assumption}
We remark that the monotonicity assumption on the utility of both types is natural and it reflects the fact that waiting for service completion imposes a waiting cost on the users. The second condition requires that while each additional user ahead in queue imposes greater waiting costs on a $\low$-type user, the incremental cost decreases with more users ahead in queue. We note that the linear utility function, i.e., $u_\low(n) = 1 - c (n+1)$ for some $c > 0$, satisfies both the conditions.

We assume that the users are strategic and Bayesian in their joining decisions. Because high-need users have no viable outside option, any such arriving user always joins the queue for service. On the other hand, the low-need users may decide to leave for the outside option, based on their beliefs about the queue state. Since the queue is unobservable to the users, the service provider seeks to leverage his informational advantage to influence the low-need users' decision, with the goal towards improving welfare outcomes. To that end, the service provider commits to a {\em signaling mechanism} as follows: the service provider selects a set of possible signals $\mathbb{S}$, and a mapping $\sigma : \naturals_0 \times \mathbb{S} \to [0,1]$, such that, if there are $n$ users already in queue upon the arrival of a user, he sends a signal $s \in \mathbb{S}$ to the user with probability $\sigma(n,s) \in [0,1]$. (We require $\sum_{s \in \mathbb{S}} \sigma(n,s) = 1$ for all $n$.) Note that since high-need users in our model have no viable outside option and hence always join the queue, the service provider can implement a signaling mechanism without the knowledge of user types. 

Given the signaling mechanism, we require the low-need users' choices to constitute an equilibrium. Informally, the equilibrium requires that in the steady state that arises from the users' actions, each low-need user is acting optimally. To elaborate further, given the steady state distribution $\pi$, we require that a low-need user joins the queue upon receiving a signal $s \in \mathbb{S}$ if and only if her expected utility from joining $\expec_{\pi}[ u_\low(n) | s]$ is greater than zero, the utility of her outside option. (We assume that ties are broken in favor of joining; we note that due to negative externalities users in the queue impose on each other, the welfare under other tie-breaking rules can only be better.) Note that the steady state distribution $\pi$ itself is determined endogenously in equilibrium from the users' actions. To avoid unnecessary notational burden, we refrain from formally defining the equilibrium for general signaling mechanisms, and point the reader to \cite{lingenbrink2019optimal}. Instead, using standard arguments based
on the revelation principle (see e.g. \cite{lingenbrink2019optimal, bergemannM16, candogan-drakopoulos}), 
one can show that it suffices to consider {\em obedient} binary signaling mechanisms. These are the mechanisms where the signals are limited to ``$\join$'' and ``$\leave$'' --- which we represent as $1$ and $0$ respectively --- and for which in the resulting user equilibrium, a high-need user always joins, and a low-need user joins upon receiving signal $1$ and leaves otherwise. We describe such mechanisms more formally next. 

First note that a binary signaling mechanism can be described by $\{ p_n : n \geq 0\}$, where $p_n$ denotes the probability that a $\low$-type user receives the signal $s=1$ (``$\join$''), when the queue length is $n$ upon her arrival. Assuming that all users follow their recommendation, let $\pi = \{ \pi_n : n \geq 0\}$ denote the resulting steady-state distribution. By elementary queueing theory, the steady state distribution satisfies the following detailed-balance conditions~\citep{queueing-theory-textbook}:
\begin{align}
\label{eq:detailed-balance}
    \pi_{n+1} = (\lambda_\low p_n  + \lambda_\high )\pi_n, \quad \text{for all  $n  \geq 0$.}
\end{align}
Given the steady-state distribution and using Bayes' rule, an arriving
$\low$-type user receiving the signal $s=1$ (``$\join$'') believes the
queue-length is $n \geq 0$ with probability
$\pi_n p_n / \sum_{k \in \NN} \pi_k p_k$. Similarly, an arriving
$\low$-type user receiving the signal $s=0$ (``$\leave$'') believes the
queue-length is $n \geq 0$ with probability
$\pi_n (1 - p_n) / \sum_k \pi_{k \in \NN} (1 - p_k)$.

For a $\low$-type user, let $U_\low(s, a)$ denote her expected utility upon receiving a signal $s \in \{0, 1\}$ and choosing an action $a \in \{\join, \leave\}$. Note that we have $U_\low(s, \leave) = 0$. (Recall that $\low$-type's outside option is normalized to zero.) On the other hand, we have
\begin{align*}
    U_\low(1, \join) &=  \sum_{n \in \NN}  \frac{\pi_n p_n}{\sum_{k \in \NN} \pi_{k}p_k } u_\low(n) = \frac{\sum_{n \in \NN} (\pi_{n+1} - \lambda_\high \pi_n) u_\low(n)}{\sum_{n \in \NN}  (\pi_{n+1} - \lambda_\high \pi_n) },  \\
    U_\low(0, \join) &=  \sum_{n \in \NN}  \frac{\pi_n (1-p_n)}{\sum_{k \in \NN} \pi_{k} (1 - p_k) } u_\low(n) = \frac{\sum_{n \in \NN} (\lambda \pi_n -  \pi_{n+1}) u_\low(n)}{\sum_{n \in \NN}  (\lambda \pi_n - \pi_{n+1}) }. 
\end{align*}
Here, the second equality in each line follows from the fact that $\lambda_{\low} \pi_n p_n = \pi_{n+1} - \lambda_{\high} \pi_n$ and $\lambda_{\low} \pi_n(1-p_n) = \lambda \pi_n - \pi_{n+1}$, which follow from the detailed-balance condition~\eqref{eq:detailed-balance}. 

In an obedient binary signaling mechanism, a $\low$-type user must find it incentive compatible to follow the service provider's recommendations. Thus, in such a mechanism, we must have the following {\em obedience} constraints: $U_\low(1, \join) \geq U_\low(1, \leave) = 0$ and $U_\low(0, \join) \leq U_\low(0, \leave) = 0$. This in turn yields the following constraints on the steady-state distribution $\pi$: 
\begin{align}
    J(\pi) & \defeq \sum_{n=0}^\infty (\pi_{n+1} - \lambda_\high \pi_n) u_\low(n) \geq 0, \tag{\textsf{JOIN}} \label{eq:obedience1}\\
    L(\pi) & \defeq \sum_{n=0}^\infty (\lambda \pi_n -  \pi_{n+1}) u_\low(n) \leq 0 \tag{\textsf{LEAVE}} \label{eq:obedience2}
\end{align}
Using the preceding constraints, the following result, from \cite{lingenbrink2019optimal}, establishes a correspondence between obedient binary signaling mechanisms and a set of all distributions satisfying obedience constraints. We omit the proof for brevity.
\begin{lemma}[\citet{lingenbrink2019optimal}]\label{lem:equivalence} For any obedient binary signaling mechanism, the steady-state distribution $\pi$ satisfies the following conditions:
\begin{enumerate}
    \item Distributional constraints: $\sum_{n \in \NN} \pi_n = 1$ and $\pi_n \geq 0$ for all $n \geq 0$;
    \item Detailed-balance constraints: $\lambda_{\high} \pi_n \leq \pi_{n+1} \leq (\lambda_{\high} + \lambda_{\low}) \pi_n$ for all $n \in \NN$; and 
    \item Obedience constraints \eqref{eq:obedience1} and \eqref{eq:obedience2} as defined above.
\end{enumerate}
Conversely, for any distribution $\pi$ satisfying the preceding sets of constraints, there exists an obedient binary signaling mechanism $\{ p_n : n \geq 0\}$, with $p_n = \frac{\pi_{n+1} - \lambda_\high \pi_n}{\lambda_{\low} \pi_n}$ whenever $\pi_n > 0$ (and arbitrary otherwise). 
\end{lemma}
We let $\Piob$ denote  the set of all distributions that satisfy the
three sets of constraints mentioned above. (Here, $\mathsf{SM}$ stands for signaling mechanism.) Here, the second constraints arise from the detailed-balance conditions \eqref{eq:detailed-balance} and the fact that $p_n \in [0,1]$ for all $n$.


In addition to simplifying notation, the preceding result enables us to describe the user welfare in an obedient binary signaling mechanism purely in terms of the resulting distribution $\pi \in \Piob$. In particular, for any $\pi \in \Piob$, the welfare of type $i$ users, denoted by $W_i(\pi)$, is given by 
\begin{align}
W_\low(\pi) &= \lambda_{\low}  \sum_{n=0}^\infty \pi_n \left( \frac{\pi_{n+1} - \lambda_\high \pi_n}{\lambda_{\low} \pi_n} \right) u_\low(n) = \sum_{n=0}^\infty (\pi_{n+1} - \lambda_{\high} \pi_n) u_{\low}(n) =  J(\pi) \label{eq:Wel1}\\
W_\high(\pi) &=  \lambda_\high \sum_{n=0}^\infty  \pi_n u_\high(n). \label{eq:Wel2}
\end{align}
Here, the first line follows from the fact that the arrival rate of $\low$-type users is $\lambda_{\low}$ and that if the queue-length is $n$, which occurs with probability $\pi_n$ in steady state, an arriving $\low$-type user joins the queue with probability $(\pi_{n+1} - \lambda_{\high} \pi_n)/ \lambda_{\low} \pi_n$ and receives utility $u_{\low}(n)$. Similarly, the second line follows from the fact that a $\high$-type user always joins upon arrival. 

Since we focus on a social service setting, we seek to understand the effectiveness of information design in improving the welfare outcomes for {\em both} types. In this context, we use the following definition of Pareto efficiency:
\begin{definition}[Pareto Efficiency]
\label{def:pareto}
For any two $\pi, \hat{\pi} \in  \Piob$, we say $\hat{\pi} $ Pareto-dominates $\pi$, if $W_i(\hat{\pi}) \geq W_i(\pi)$ for $i \in \{\low, \high\}$ with a strict inequality for at least one $i$. Further, we say a distribution $\pi \in \Piob$ is Pareto-efficient within the class $\Piob$ if and only if there exists no $\hat{\pi} \in \Piob$ that Pareto-dominates $\pi$.
\end{definition}
Hereafter, we frequently abuse the terminology to say an obedient binary signaling mechanism is Pareto-efficient (within the class of such mechanisms), if the corresponding steady state distribution (as per Lemma~\ref{lem:equivalence}) is Pareto-efficient\footnote{The criterion of Pareto-dominance evaluates signaling mechanisms at the {\em ex ante} stage, i.e., upon a user arrival and before any information sharing or joining decisions. 
Such a criterion is natural when the focus is on  assessing welfare outcomes in the aggregate, and in settings where the system state emerges endogenously from user behavior.} within the class $\Piob$. 


For our comparative analysis, we look at two specific signaling
mechanisms that capture the two extremes of information sharing: 
\begin{enumerate}
\item[(a)] Full-information mechanism ($\fim$): Here, the service
provider always reveals the queue-length to an arriving user. Consequently, $\low$-type users join the queue at all queue-lengths $k$ with $u_\low(k) \geq 0$. Letting $m_\fim$ denote the smallest integer $k$ for which $u_\low(k) < 0$, the corresponding steady state distribution $\pi^\fim$ satisfies $\pi^\fim_{n+1} = \lambda \pi^\fim_n$ for $n < m_\fim$, and $\pi^\fim_{n+1} = \lambda_{\high} \pi^\fim_n$ otherwise.
\item[(b)] No-information mechanism ($\nim$): Here, the service provider reveals no information to the users. Consequently, the strategy of an arriving $\low$-type user is independent of $n$. Letting $p^\nim$ denote the probability with which a user joins the queue in a symmetric equilibrium, the corresponding steady state distribution $\pi^\nim$ satisfies $\pi^\nim_n = (p^\nim \lambda_\low  + \lambda_\high )^n \pi^\nim_0$ for all $n \geq 0$. In Lemma~\ref{lem:ni} (in Appendix~\ref{ap:structure}), we characterize the joining probability $p^\nim$ in an equilibrium.
\end{enumerate}

In the following, we also consider {\em admission policies}, where the
service provider can enforce the joining or leaving of any user,
regardless of the user's type or her incentives. While such enforcement
is clearly practically infeasible, it serves as a benchmark against
which the welfare outcomes of signaling mechanisms can be
compared. Formally, an admission policy can be described by the class
of distributions $\Pidb$ that satisfy the distributional and the
detailed-balance constraints from Lemma~\ref{lem:equivalence}, but
need not satisfy the obedience constraints~\eqref{eq:obedience1} and
\eqref{eq:obedience2}. (Here, $\mathsf{AP}$ stands for admission policy.) Analogous to Definition~\ref{def:pareto}, we define Pareto-efficiency within the class $\Pidb$. Observe that $\Piob \subseteq \Pidb$, i.e., any signaling mechanism is also an admission policy (one that also respects user incentives), and hence any signaling mechanism $\pi \in \Piob$ that is Pareto-efficient within \vmdelete{the class of admission policies} $\Pidb$ is also Pareto-efficient within \vmdelete{the class of signaling mechanisms} $\Piob$, but the converse may not hold. This observation motivates our choice of $\Pidb$ as a welfare benchmark. 

Before we end this section, we note that both $\Pidb$ and $\Piob$ are
closed and convex, and the welfare functions as defined in
\eqref{eq:Wel1} and \eqref{eq:Wel2} are linear in $\pi$. Thus, the
sets $\{ (W_\low(\pi), W_\high(\pi)), \pi \in \Piob\}$ and
$\{ (W_\low(\pi), W_\high(\pi)), \pi \in \Pidb\}$ are also convex. As
a consequence, it follows that any $\hat{\pi}$ that is Pareto-efficient
within the class of signaling mechanisms $\Piob$ (or admission
policies $\Pidb$), is a solution to the (linear) optimization
problem that maximizes the convex combination of the two user types'
welfare over $\Piob$ (respectively, $\Pidb$). \kiedit{In particular, let
$W(\pi, \theta) \defeq \theta W_\low(\pi) + (1-\theta) W_\high(\pi)$ for
all $\pi \in \Pidb$ and for $\theta \in [0,1]$. Then, each Pareto-efficient signaling
mechanism maximizes $W(\pi, \theta)$ over $\Piob$ for
some $\theta \in [0,1]$, and each maximizer of $W(\pi, \theta)$ over $\Piob$ for $\theta \in (0,1)$ is a Pareto-efficient signaling mechanism~\citep[Proposition 16.E.2]{mas1995microeconomic}. Similarly, any Pareto-efficient admission policy is a solution to $\max_{\pi \in \Pidb} W(\pi, \theta)$ for some $\theta \in [0,1]$ (and each maximizer of $W(\pi, \theta)$ over $\Pidb$ for $\theta \in (0,1)$ is a Pareto-efficient admission policy).}\footnote{\kiedit{We remark that for the extreme cases of $\theta \in \{0,1\}$, 
not all maximizers of $W(\pi,\theta)$ over $\Pidb$ (and $\Piob$) may be Pareto-efficient. We discuss $\theta =0$ here (the other extreme is similar). Since the optimization problem in this case puts no weight on the welfare of $\low$-type users,  there could be two maximizers $\pi$ and $\pi'$ such that 
$W_\high(\pi) = W_\high(\pi')$ but $W_\low(\pi) > W_\low(\pi')$. In that case, $\pi'$ is Pareto-dominated by $\pi$ and hence not Pareto-efficient. The recent work of~\citet{che2020characterizing} provides a refinement to address this issue: first optimize  $\high$-type's welfare and then among the optima, find the one that maximizes $\low$-type's welfare. We follow the same refinement.}} Furthermore, for any Pareto-efficient $\pi$,
the specific $\theta \in [0,1]$ for which $\pi$ maximizes
$W(\cdot,\theta)$ captures the relative importance the service
provider ascribes to improving the welfare of the two types. In this
context, for a given $\theta$, we refer to the admission policy that
achieves the maximum as the ``first-best''. 

\section{Structural Characterization}\label{sec:structural-characterization}

In this section, we provide structural characterizations of the
Pareto-efficient signaling mechanisms and admission policies. We use
these structural characterization in Sections~\ref{sec:comparison}
and~\ref{sec:first-best} to evaluate the effectiveness of signaling
mechanisms in improving welfare outcomes, and compare its performance
against admission policies and simple signaling mechanisms.

Before we begin, for the sake of completeness, we state the following
technical result that establishes the existence of Pareto-efficient
signaling mechanisms and admission policies. The proof follows from
the observation that the sets $\Pidb$ and $\Piob$ (or closures of some relevant
subsets) are weakly compact, and hence the maximizers of $W(\pi, \theta)$ over
these sets exist for all $\theta \in [0,1]$. It is straightforward to show that these maximizers are Pareto-efficient within their respective class. The formal proof is
provided in Appendix~\ref{ap:structural-characterization}.
\begin{lemma}[Existence]
\label{lem:pareto-efficient-exists} 
For $\lambda < 1$ and for each $\theta \in [0,1]$, there exists a signaling mechanism  (admission policy) that maximizes $W(\pi, \theta)$ over all $\pi \in \Piob$ (resp., $ \pi \in \Pidb$). If further $\lim_{n\to\infty} u_i(k) = -\infty$ for each $i \in \{\low, \high\}$, then the result also holds for $\lambda =1$. 
\end{lemma}
Next, we define the following threshold structure among distributions $\pi \in \Pidb$. 
\begin{definition}[Threshold Structure]
\label{def:threshold-structure} We say a given $\pi \in \Pidb$ has a {\em threshold} structure, if there exists an $m \in \NN \cup \{\infty\}$, such that $\pi_{k+1} = \lambda \pi_k$ for all $k < m$, and $\pi_{k+1} = \lambda_{\high} \pi_k$ for all $k > m$. In such a setting, we say the distribution $\pi$ has a threshold $x = m + a \in \reals_+$, where $a = (\pi_{m+1} - \lambda_{\high}\pi_m)/ \lambda_{\low} \pi_m \in [0,1]$.
\end{definition}
Informally, a distribution $\pi \in \Pidb$ has a threshold structure with threshold equal to $x = m +a \in [ m, m+1]$, if an arriving $\low$-type user is asked to join the queue with probability $1$ for all queue-length strictly less than $m$, asked to leave with probability $1$ for all queue-lengths strictly greater than $m$, and asked to join the queue with probability $a \in [0,1]$ if the queue length is exactly $m$. (Note that a threshold $\infty$ corresponds to the case where an arriving $\low$-type user is asked to join regardless of the queue-length.)

Our first result states that any Pareto-efficient signaling mechanism has a threshold structure. The proof, which is deferred to Appendix~\ref{ap:structural-characterization}, follows from a perturbation analysis similar to that in \citet{lingenbrink2019optimal}: we show that given any $\pi \in \Piob$ that does not have a threshold structure, one can perturb it to obtain a $\hat{\pi} \in \Piob$ that Pareto-dominates it. 
\begin{theorem}[Threshold structure of Pareto-efficient signaling mechanisms] \label{thm:threshold-signaling-mechanism}
Any signaling mechanism $\pi \in \Piob$ that is Pareto-efficient within the class $\Piob$ has a threshold structure, with the threshold less than or equal to the full-information threshold $m_\fim$. 
\end{theorem}
Furthermore, using the same argument as above, we obtain that the
Pareto-efficient admission policies also have a threshold structure. We
omit the proof for brevity. 
\begin{theorem}[Threshold structure of Pareto-efficient admission policies]
\label{thm:threshold-admission-rule}
Any admission policy $\pi \in \Pidb$ that is Pareto-efficient within the class $\Pidb$ has a threshold structure, with the threshold less than or equal to the full-information threshold $m_\fim$. 
\end{theorem}
Having established the threshold structure of any Pareto-efficient
distribution (either within $\Pidb$ or $\Piob$), we next state another
key structural property of Pareto-efficient distributions within
$\Piob$. The result implies that in any Pareto-efficient
$\pi \in \Piob$ that is not Pareto-efficient within $\Pidb$, the
obedience constraint that binds is the constraint
\eqref{eq:obedience2}. Put differently, it is the  \eqref{eq:obedience2} condition that acts as a hurdle for a Pareto-efficient admission policy to be implementable as a signaling mechanism. The intuition behind this result lies in the
observation, common in many congested service systems, that
$\low$-type users do not internalize the negative externalities they
impose on other users (both $\low$-type and $\high$-type) by joining
the queue. Hence, the $\low$-type users are naturally more inclined to
join the queue than leave, and the challenge in information sharing is
in ensuring that when the $\low$-type users are asked to leave, they
find it incentive compatible to do so. {The proof of this theorem is also deferred to Appendix~\ref{ap:structural-characterization}.}

\begin{theorem}[Significance of the \eqref{eq:obedience2} hurdle]
  \label{thm:pareto-or-leave}
  Suppose for a signaling mechanism $\pi \in \Piob$, the obedience
  constraint~\eqref{eq:obedience2} does not bind, i.e., $L(\pi) <
  0$. Then, $\pi$ is Pareto efficient within the class $\Piob$ of
  signaling mechanisms if and only if it is Pareto efficient within the
  class $\Pidb$ of admission policies.
\end{theorem}

In concluding this section, we note that preceding result raises the
intriguing possibility of the existence of a signaling mechanism that
is Pareto-efficient not only within the class $\Piob$ of signaling
mechanisms, but also within the broader class $\Pidb$ of admission
policies. For any such mechanism, it follows that under a practically
infeasible setting where the service provider observes the types of
users and is allowed to enforce the joining and leaving of users,
\vmdelete{without regard to their incentives,} he cannot jointly improve both
types' welfare. Put differently, the existence of such mechanisms also
implies the existence of admission policies where the $\low$-type
users' incentive constraints are satisfied for ``free''.

A trivial instance of such a scenario can arise, e.g., in cases where
$\lambda_{\high}$ is large enough, and the admission policy always
bars $\low$-type users from joining the queue. First, such an
admission policy must be Pareto-efficient, as any other policy that
lets some $\low$-type users in would necessarily reduce the welfare of
the $\high$-type users. Furthermore, such an admission policy can be
implemented as a no-information mechanism, which satisfies obedience
constraints as congestion in the queue with just the $\high$-type
users makes joining undesirable for the $\low$-type users. Excluding
such trivial scenarios, a natural question is whether there exist
signaling mechanisms that do not exclude any types, but are still
Pareto-efficient within the class of admission policies $\Pidb$. In Section~\ref{sec:linear-costs}, we present numerical examples where such mechanisms indeed exist (see Figure~\ref{fig:heatmap} and its related discussion).

\section{Mechanism Comparisons}
\label{sec:comparison}
Having characterized the structure of Pareto-efficient signaling
mechanisms, \vmdelete{in this section,} we compare such mechanisms
against various benchmarks in two different settings. First, in
Section~\ref{sec:homogeneous}, we consider the homogeneous setting
where all users have $\low$-type, i.e.,
$\lambda_\high = 0$.\footnote{The homogeneous setting where all users
  are $\high$-type is uninteresting from the point of design, as all
  users join the queue regardless of their information due to no
  viable outside option.} Then, in Section~\ref{sec:heterogeneous}, we
consider the heterogeneous setting with both types of users
present. As we discuss below, the two settings exhibit striking
contrast in the effectiveness of signaling mechanisms for welfare
improvement.\vmcomment{to do: shorten by a word or two to save a line. }

In light of
Theorems~\ref{thm:threshold-signaling-mechanism}~and~\ref{thm:threshold-admission-rule},
for ease of presentation, we use the following simplified notations
for a threshold policy: For $x \in \reals_{+}$, the threshold policy
$x$ is an admission policy that gives rise to a steady state
distribution $\pi_x \in \Pidb$ that has a threshold structure, as
defined in Definition~\ref{def:threshold-structure}, with threshold
$x$. For such a policy, with a slight abuse of notation, for
$i \in \{\low, \high\}$ we denote $W_i(\pi_x)$ simply by $W_i(x)$.  We
do the same for $J(\pi_x)$ and $L(\pi_x)$.

\subsection{Homogeneous Users}
\label{sec:homogeneous}

We start our comparative studies by analyzing the special case of
homogeneous users ($\lambda_{\high} =0$). Observe that in this
single-type setting, Pareto-dominance is equivalent to optimality, in
terms of maximizing the welfare of $\low$-type users. Consequently, we
let $\sm$ denote the {\em optimal} signaling mechanism, the one that
maximizes the welfare of $\low$-type users.

In the following proposition we compare the optimal signaling
mechanism $\sm$ with \jaedit{full-information} \jadelete{the full-information extreme} ($\fim$).  With a
slight abuse of notation, for $\mu \in \{ \fim, \sm \}$, we denote the
$W_\low(\mu)$ the $\low$-type welfare under mechanism $\mu$. We have
the following result, whose proof is provided in
Appendix~\ref{ap:comparison}.\vmcomment{to do: shorten by a word or two to save a line. } 
\jacomment{I think we should replace ``the full-information extreme'' in the first line with ``full-information''}
\begin{proposition}[Limits of information design]
\label{prop:homogen}
In the homogeneous setting, we have
\begin{align*}
\kiedit{W_\low(\fim) \geq \beta_\fim \cdot W_\low(\sm)},
\end{align*}
where $\beta_\fim \defeq \big(\sum_{n=0}^{m_\fim-1} \lambda_{\low}^n\big)\big/\big(\sum_{n=0}^{m_\fim} \lambda_{\low}^n \big) \geq 1 - \frac{1}{m_\fim+1}$, and $m_\fim$ is the full-information threshold. 
\kiedit{Further, the equality holds, i.e., $W_\low(\fim) = W_\low(\sm)$ if and only if $W_\low(m_\fim) \geq 
W_\low(m_\fim -1)$.}
\end{proposition}


The preceding proposition states that in the homogeneous setting, 
signaling mechanisms are not very effective in improving the welfare
beyond that already achieved by the full-information mechanism. To gain some
intuition, observe that in general Bayesian persuasion settings, the
performance gains are typically achieved by pooling, in the persuaded
agents' beliefs, the ``good'' and the ``bad'' states of the
system. However, in a queueing setting, the linear nature of the
underlying Markovian system precludes any such simple pooling of
states in the agents' belief: the only way for the system to reach a
bad state (one with long queue-length) is by progressing through all
intermediate queue-lengths. Because of this, agents are not easily
persuaded. Formally, the proof proceeds by showing that a threshold
mechanism with threshold $x < m_\fim-1$ will not be incentive
compatible. 
\kiedit{In particular, we will show that if $x < m_\fim-1$ the second obedience constraint, \eqref{eq:obedience2} will be violated. This can also be intuitively explained: if the threshold were below $m_\fim-1$,} then the queue will never be longer than $m_\fim$, and any user receiving a ``\leave'' signal will realize this and will want to join the queue, thus violating the \eqref{eq:obedience2} condition.  Since we have already shown (Theorem~\ref{thm:threshold-signaling-mechanism}) that the threshold of the signaling mechanism is at most $m_\fim$, we conclude that the threshold of the signaling mechanism is between $m_\fim - 1$ and $m_\fim$. Thus, any small improvement in welfare of $\sm$ over $\fim$ stems from the difference in user behavior when queue length is $m_\fim - 1$, where users always join under $\fim$ but may sometimes leave under $\sm$. \kiedit{Building on this observation, in the proof we bound the relative welfare gain by a factor $1/\beta_\fim$.}

\kiedit{In contrast, in a ``sufficiently'' heterogeneous population, the presence of $\high$-type users makes persuading the $\low$-type ones possible: the queue now consists of two types of users. Therefore, a user's belief about the queue length will now depend on the behavior of both types. Leveraging this, 
a signaling mechanism can set a threshold much lower that that of the full-information mechanism without violating the~\eqref{eq:obedience2} condition. This in turn, can result in substantial welfare gain. (For numerical examples, see Figure \ref{fig:welfare-comparison} and its related discussion in Section \ref{sec:linear-costs}.)}

Finally, in the following proposition (proven in Appendix \ref{ap:comparison}), we show that even though information design results in limited or no improvement over the full-information mechanism, it always outperforms the no-information mechanism.
\begin{proposition}[Suboptimality of the no-information mechanism]
\label{prop:homogen-noinfo} In the
    homogeneous setting, the no-information mechanism is never optimal. In
    particular, the welfare under the no-information
    mechanism $\nim$ satisfies
    $W_\low(\nim) < (1 - \lambda_\low^{m_\fim +1}) W_\low(\fim) \leq (1 - \lambda_\low^{m_\fim +1}) W_\low(\sm)$. 
  \end{proposition}


\subsection{Heterogeneous Users}
\label{sec:heterogeneous}
Next\jaedit{,} we proceed to a setting where the population is a mixture of the
two types. \jadelete{Note that in this case}\jaedit{Here}, we have two objectives, namely, the
welfare of both types.  As discussed before, to examine the
effectiveness of information design in improving the welfare of both
types, we focus on the notion of Pareto efficiency.  In the following,
we draw comparisons between Pareto-efficient optimal signaling
mechanisms with the extreme forms of sharing information, i.e., full
information and no information.\vmcomment{to do: shorten by a word or two to save a line. } \jacomment{done!}

Our main result in this section is the following proposition which provides the necessary and sufficient conditions on arrival rates under which the full-information and no-information mechanisms are Pareto-dominated.

\begin{proposition}[Power of information design]
\label{prop:info-design-power}
The following statements hold.
\begin{enumerate}
  \item {For any $\lambda_\high > 0$, there exists a signaling mechanism that Pareto dominates the full-information mechanism if and only if $\lambda_\low \in
    [\bar{\Lambda}_\low(\lambda_\high),1)$, where 
    $\bar{\Lambda}_\low(\lambda_\high) \in (0, 1- \lambda_\high]$ and  we have: 
    \begin{align*}    
    \bar{\Lambda}_\low(\lambda_\high) \geq (1-\lambda_\high)
    \tfrac{u_\low(m_\fim-1)}{ u_\low(0) - \sum_{k=1}^{m_\fim-1}
      \lambda_\high^k (u_\low(k-1) - u_\low(k))} > 0.
\end{align*}}
\item \vmredit{For $\lambda < 1$ and $\lambda_\low \in
    [\bar{\Lambda}_\low(\lambda_\high),1)$, 
    if the utility functions are such that 
     $u_\low(m_\fim-1) \leq W_\low(\fim)$, $u_\high(m_\fim) \leq W_\high(\fim)$,  $L(m_\fim-1) \leq 0$, and $W_\high(\fim) > 0$, then we have:
\begin{align*}
W_\low(\sm) \geq \beta_{\low, \sm} \cdot W_\low(\fim) \quad \quad \text{and} \quad \quad 
W_\high(\sm) \geq \beta_{\high, \sm} \cdot W_\high(\fim),  
\end{align*}
where $\beta_{\low, \sm} \defeq \left( 1 + \frac{ \lambda_\high(1-\lambda) \lambda_\low  \lambda^{m_\fim-1}}{ 1 - \lambda_\high - \lambda_\low\lambda^{m_\fim-1} }\right) > 1$ and $\beta_{\high, \sm} \defeq \left(1 + \frac{ (1-\lambda_\high)(1-\lambda) \lambda_\low  \lambda^{m_\fim-1}}{ 1 - \lambda_\high - \lambda_\low\lambda^{m_\fim-1} }\right) >1$.}
\kiedit{\item The no-information mechanism $\nim$ is Pareto-dominated by a signaling mechanism if and only if $\lambda_\high \in [0 , \bar{\Lambda}_\high)$, where
    $\bar{\Lambda}_\high$ is the unique root in $(0,1)$ of the function $g(x) \defeq \sum_{k \in \NN} (1- x) x^k u_\low(k)$. Here, $\bar{\Lambda}_\high$ is the smallest arrival rate of the $\high$-type users under which no $\low$-type user joins the queue in the equilibrium of the no-information mechanism.}
\end{enumerate}
\end{proposition}


The preceding result states that as long as the arrival rate $\lambda_\low$ of $\low$-type users is sufficiently high, the welfare of both types can be improved using information design, as compared to full-information sharing. As we discuss below (after stating Theorem~\ref{thm:full-info-comparison}), under sufficiently high $\lambda_\low$, the negative externality that a $\low$-type user imposes on other $\low$-type users exceeds the utility she receives from the service; in such settings not revealing all the information about the state helps the $\low$-type users to internalize this negative externality. 
\vmredit{To illustrate the benefit of information design in  such a regime, the second part of the proposition 
places a lower bound on the welfare gain of each type \vmdelete{if 
the threshold of the optimal signaling mechanism is even as high as ${m_\fim-1}$} under certain conditions on the utility functions of the two types.}
On the other hand, as long as the arrival rate $\lambda_\high$ of $\high$-type is not too high, information design can improve the welfare over no-information sharing. In this case, information design helps by providing enough state information to correlate $\low$-type users' actions with the queue-state. Taken together, the result implies that information design has an unambiguously positive role for welfare improvement in settings where the type composition of user population is fairly balanced. 


We defer the complete proof of Proposition~\ref{prop:info-design-power} to Appendix~\ref{ap:comparison}. The proof relies on two intermediate results that have a similar dichotomous structure, characterizing when each of the two benchmarks is Pareto efficient. We devote the rest of this section to discussing (and proving) the two results, and their relation to our main proposition. The proofs of both results are given in Appendix~\ref{ap:comparison}.

The first result presents the following dichotomy for the 
full-information benchmark.
\begin{theorem}[Information design vs. full-information]
  \label{thm:full-info-comparison} Exactly one of the following two
  statements holds:
\begin{enumerate}
\item the full-information mechanism $\fim$ is Pareto-efficient within
  the class of admission policies;
\item there exists a \emph{signaling mechanism} that Pareto-dominates
  the full-information mechanism.
\end{enumerate}
Further, the first case occurs if and only if $W_\low(m_\fim) >
W_\low(m_\fim -1)$.
\end{theorem}
To understand the implications of the preceding result, consider an
admission policy with threshold $x < m_\fim$. As $x$ increases, more $\low$-type users are served by the service provider, increasing
their utility. At the same time, the negative externality each
$\low$-type user imposes on other $\low$-type users increases as $x$
increases. (This is in addition to the negative externalities imposed
on $\high$-type users.)  The preceding result states that for the
full-information mechanism to be Pareto-efficient, the gains from
serving more $\low$-type users must dominate the negative
externalities they impose on other $\low$-type users (which is succinctly captured by
the condition $W_\low(m_\fim) > W_\low(m_\fim -1)$). Conversely, if serving more $\low$-type users imposes greater negative externality on other $\low$-type users, our result states
that information design can leverage this to improve the welfare of
{\em both} types over the full-information
mechanism. Finally, tying back to Proposition~\ref{prop:info-design-power}, for the effect of negative externality to dominate, the arrival rate $\lambda_\low$ of $\low$-type users must be large enough, as captured by the condition $\lambda_\low \geq \Bar{\Lambda}_\low(\lambda_\high)$.

Next, we obtain the following dichotomy for the no-information benchmark. 
\begin{theorem}[Information design vs. no-information]
\label{thm:no-info-comparison}
Exactly one of the following two statements holds:
\begin{enumerate}
\item the no-information mechanism $\nim$ is Pareto-efficient within
  the class of admission policies;
\item there exists a signaling mechanism that Pareto-dominates the
  no-information mechanism.
\end{enumerate}
Further, the first case occurs if and only if all $\low$-type users
choose their outside option under the no-information mechanism.
\end{theorem}
The preceding result neatly breaks the analysis into two cases. In the first case, even if no other $\low$-type users join the queue, the outside option is more desirable for a $\low$-type user. In other words, a $\low$-type user does not need much persuasion to forego the social service and avail the outside option. In such instances, any information shared by the service provider would only induce some $\low$-type user to join the queue and hence reduce $\high$-type users' welfare. Barring this exception, information design proves effective in improving welfare of both types over the no-information mechanism. Tying back to Proposition~\ref{prop:info-design-power}, we obtain that for all $\low$-type users to choose their outside option under the no-information mechanism, the system must be already overwhelmed by $\high$-type users, as captured by the condition $\lambda_\high \geq \bar{\Lambda}_\high$ on the arrival rate of $\high$-type
users.

\section{Achieving First-best}\label{sec:first-best}
Having compared the effectiveness of information design against those of the two extreme signaling mechanisms, we now investigate its limitations. Specifically, in this section, we compare signaling mechanisms against Pareto-efficient admission policies, and ask how limiting the requirement of ensuring obedience constraints is in terms of welfare improvement.

To better study this question, we consider the problem of maximizing
the weighted welfare
$W(\pi, \theta) = \theta W_\low(\pi) + (1-\theta) W_\high(\pi)$, both
over the class of admission policies and the class of signaling
mechanisms. \kiedit{As described in Section~\ref{sec:model}, each 
Pareto-efficient admission policy and signaling mechanism can be
obtained as a maximizer of $W(\pi, \theta)$ for some $\theta \in [0,1]$.} Furthermore, the specific $\theta$ for which a Pareto-efficient mechanism (or an admission policy) maximizes
$W(\pi, \theta)$ captures the relative weight placed by the service
provider in improving either type's welfare. For notational
convenience, for any $\theta \in [0,1]$, we let $\sm(\theta)$ denote
the signaling mechanism that maximizes $W(\pi, \theta)$ over
$\pi \in \Piob$, and $\ar(\theta)$ denote the admission policy\footnote{We note that for certain values of $\theta$, the maximizers of $W(\pi,\theta)$ over $\Pidb$ (and $\Piob$) may not be unique. To avoid burdensome notation, we let $\ar(\theta)$ (and $\sm(\theta)$) denote the set of all maximizers in such instances.} that
does the same over $\Pidb$.

As we discussed in the conclusion of
Section~\ref{sec:structural-characterization},
Theorem~\ref{thm:pareto-or-leave} raises the possibility that there
exists $\theta \in [0,1]$ such that $\sm(\theta) = \ar(\theta)$.  The
main point we make in this section is that, remarkably, for a wide
range of $\theta$, $\sm(\theta) = \ar(\theta)$, i.e., the signaling
mechanism $\sm(\theta)$ is Pareto-efficient within the class of
admission policies. We have the following theorem whose proof is given in Appendix~\ref{app:sec:first-best}.
\begin{theorem}[Achieving first-best]
\label{thm:achieve-first-best} 
With $\bar{\Lambda}_\high$ as defined in Proposition~\ref{prop:info-design-power}, the following holds.
\begin{enumerate}
    \item For $\lambda_\high \in  [\bar{\Lambda}_\high,1]$, we have $\sm(\theta) = \ar(\theta)$ for all $\theta \in [0,1]$.  
    \item For $\lambda_\high < \bar{\Lambda}_\high$, there exists a $\theta(\lambda_\low, \lambda_\high) \in (0, 1]$ such that for all $\theta > \theta(\lambda_\low, \lambda_\high)$ we have $\sm(\theta) = \ar(\theta)$, and for all $\theta < \theta(\lambda_\low, \lambda_\high)$ the signaling mechanism $\sm(\theta)$ is independent of $\theta$. 
\end{enumerate}
\end{theorem}
The preceding result has an interesting implication about the role of information design when signaling mechanisms achieve Pareto-dominance over $\Pidb$. \vmdelete{Note that i}In such cases, neither obedience constraint binds, since $\sm(\theta) = \ar(\theta)$. Thus, the obedience constraints impose no limitations on the service provider. In such cases, information design plays solely the role of a coordination device, directing some $\low$-type users away from the queue and others to join the queue. In neither instance the user is indifferent between the recommended action and the alternative. This is unlike what happens in typical persuasion settings, where optimality requires indifference for at least some signals.\vmcomment{to do: shorten by a word or two to save a line. }

We also note that the two cases of the proposition are exactly the same as that of Theorem~\ref{thm:no-info-comparison}. In particular, when no-information mechanism is Pareto-efficient, the set of Pareto-efficient signaling mechanisms is same as the set of Pareto-efficient admission policies. Put differently, in instances where signaling mechanisms lack the power of admission policies, no-information is Pareto-dominated by some signaling mechanisms.

Finally, the equivalence of  $\sm(\theta)$ and $\ar(\theta)$ is also appealing from an implementation point of view: the service provider can implement a signaling mechanism without the knowledge of user types. However, under an admission policy, the service provider  observes the type of each arriving user and  makes $\join$ and $\leave$ decisions on her behalf.


\section{Numerical Analysis under Linear Waiting Costs}
\label{sec:linear-costs}

To gain further comparative insights, in this section we augment our analytical results with a numerical analysis. \jadelete{To do this, we} \jaedit{We} focus on the setting of linear utilities: $u_i(k) = V_i - c_i (k+1)$ for $i \in \{\low, \high\}$, where $V_i > 0$ denotes type-$i$ users' value for the service, and $c_i > 0$ denotes the type-$i$ users' waiting costs per unit time. (Note that the utility function $u_i(k)$ includes the waiting costs incurred due to time spent in the queue, as well as due to time spent receiving the service.) Since we focus on the notion of Pareto-dominance, each users' utility can be scaled by an arbitrary positive number without any effect on our analysis. Thus, we normalize the utility functions by choosing the value $V_i =1$ for each $i \in \{\low, \high\}$. With this normalization, we further assume that $c_\low = c_\high = c \in (0,1)$; this restricts our analysis to the setting where the two user types place the same relative weights on the value of service and the cost of waiting. Making this assumption of the homogeneity of the ``inside option'' enables us to neatly isolate the effects of the heterogeneity of the users' outside option.

Before we proceed with the analysis, we note that in this setting the quantity $\bar{\Lambda}_\high$ in Proposition~\ref{prop:info-design-power} is given by $\bar{\Lambda}_\high = 1- c \in (0,1)$. Thus, Proposition~\ref{prop:info-design-power} implies that the no-information mechanism is Pareto-dominated for all $\lambda_\high \in [0, 1-c)$.

\begin{figure}
\centering
\includegraphics[width=0.7\textwidth]{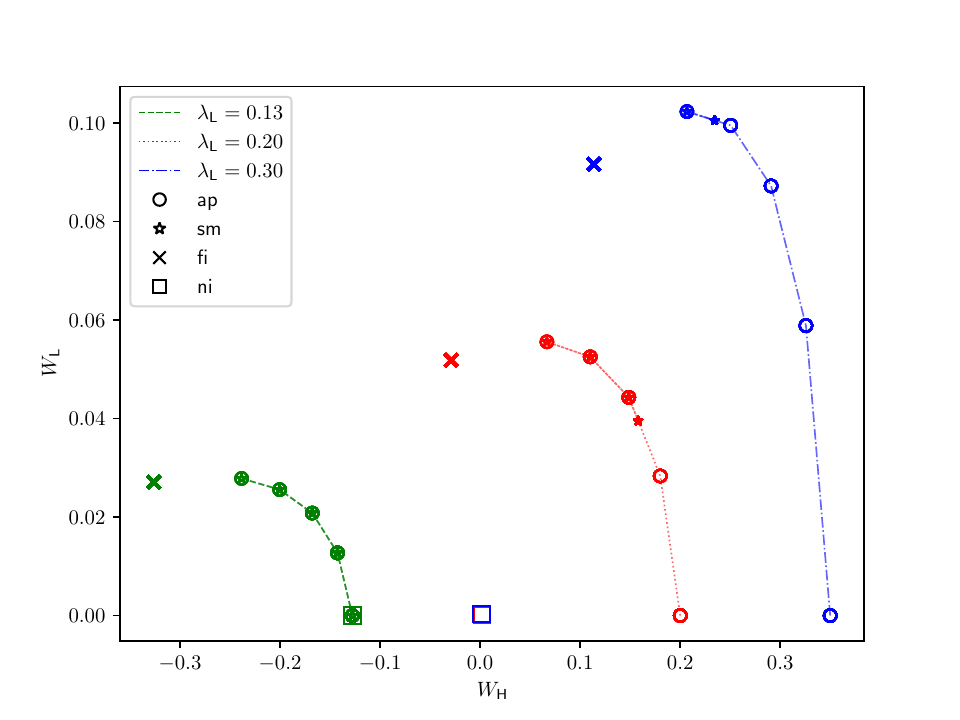}
\caption{Welfare of Pareto-efficient signaling mechanisms and admission policies for $\lambda_\low \in \{0.13, 0.20, 0.30\}$, $\lambda_\high = 1 - \lambda_\low$ and $c= 0.15$. Here, green (dashes) represents $\lambda_\low = 0.13$, red (dots) represents $\lambda_\low = 0.20$, and blue (dashdots) represents $\lambda_\low = 0.30$. Further, circles ($\circ$) represent efficient admission policies ($\ar$), stars ($\star$) represent efficient signaling mechanisms ($\sm$), cross ($\times$) represents the full-information mechanism ($\fim$), and square ($\square$) represents the no-information mechanism ($\nim$). (The no-information points for $\lambda_\low \in \{0.2, 0.3\}$ overlap, and so do those corresponding to signaling mechanisms and admission policies for each fixed $\lambda_\low$.)}\label{fig:welfare-comparison}
\end{figure}

Next, we illustrate the qualitative insights of Proposition~\ref{prop:info-design-power} and Theorems~\ref{thm:full-info-comparison},~\ref{thm:no-info-comparison},~and~\ref{thm:achieve-first-best} via numerical examples. First, in Figure~\ref{fig:welfare-comparison},  we plot the welfare of
Pareto-efficient signaling mechanisms and admission policies for
different values of $\lambda_\low \in \{0.13, 0.20, 0.30\}$, and
$c = 0.15$. We fix $\lambda_\high = 1 - \lambda_\low$ in each case, to study the extreme
setting where the service capacity exactly matches the total arrival
rate $\lambda = \lambda_\low + \lambda_\high=1$. For each value of $\lambda_\low$, we also plot the
full-information mechanism ($\fim$) and the no-information mechanism
($\nim$). First, observe that for $\lambda_\low = 0.13$, we have
$\lambda_\high > \bar{\Lambda}_\high = 1-c$, and hence, from Proposition~\ref{prop:info-design-power}, the no-information mechanism ($\nim$, green square) is
Pareto-efficient. On the other hand, we \jaedit{note} that the
full-information mechanism ($\fim$, green cross) is Pareto-dominated by
a signaling mechanism (green star).  Further, note that as
established in the first case of Theorem~\ref{thm:achieve-first-best}, $\ar(\theta) = \sm(\theta)$
for $\theta \in [0,1]$.  On the other hand, for the other two values
of $\lambda_\low$, we see that the no-information mechanism achieves
zero welfare for both types, and is Pareto-dominated in the class of
signaling mechanisms. Additionally, even though the two Pareto frontiers do not coincide, they overlap considerably,  particularly for $\lambda_\low = 0.20$. Finally, we observe that as the proportion $\lambda_\low$ of users with viable outside option increases, the
welfare of both user types increases.\vmcomment{to do: shorten by a word or two to save a line. } \jacomment{done!}

\begin{figure}
\centering
	\begin{subfigure}{0.32\linewidth}
	\includegraphics[width = \textwidth]{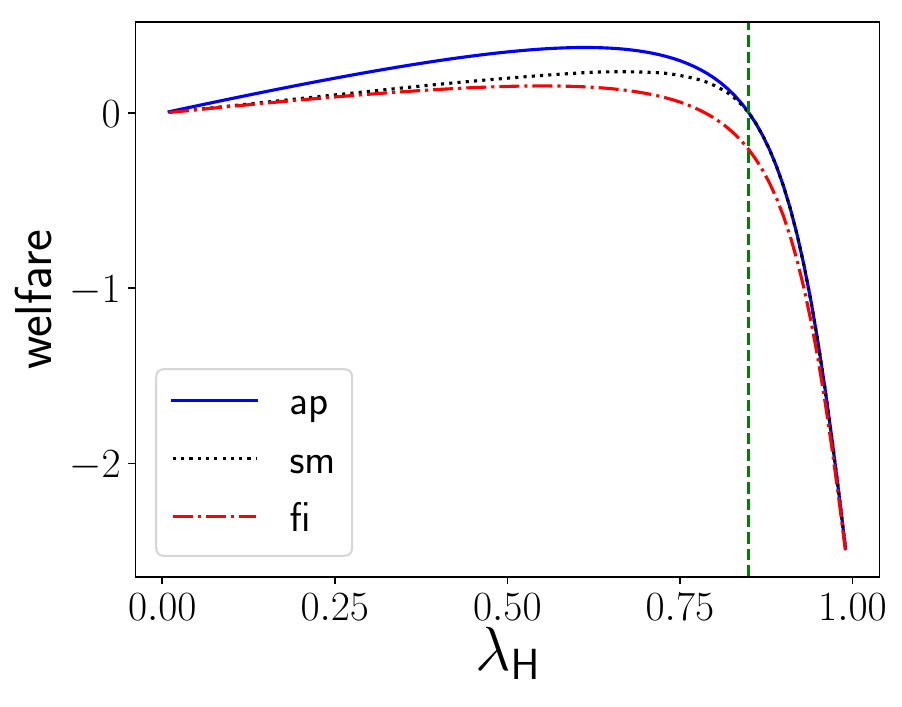}
	\caption{$\theta=0$}
	\label{fig:welfare-theta-comparison-1}
	\end{subfigure}%
	\begin{subfigure}{0.32\linewidth}
	\includegraphics[width = \textwidth]{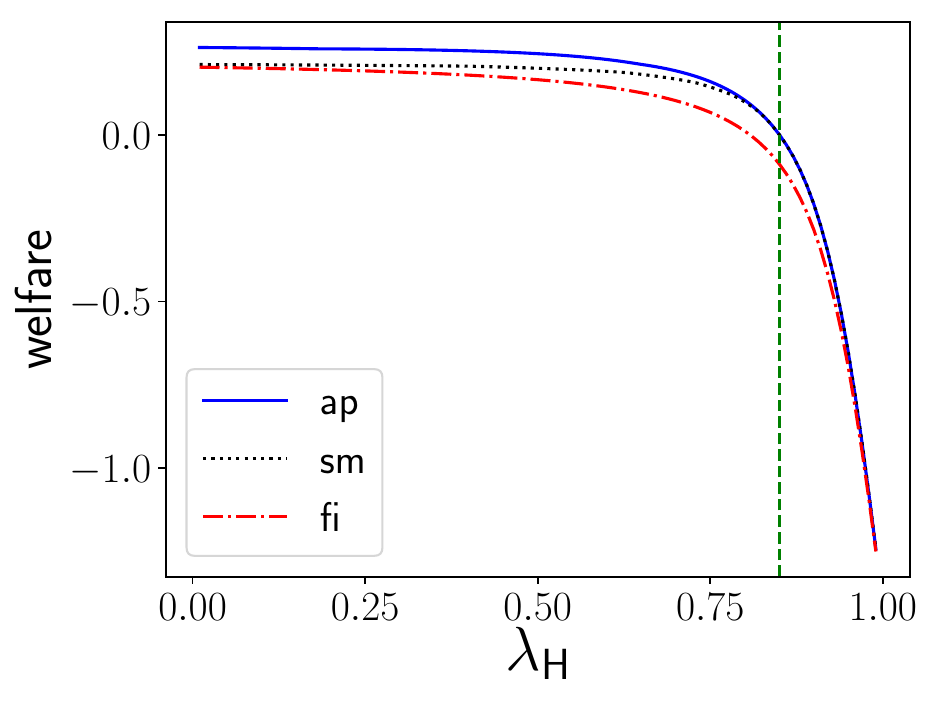}
	\caption{$\theta=0.5$}
	\label{fig:welfare-theta-comparison-2}
    \end{subfigure}%
    \begin{subfigure}{0.32\linewidth}
	\includegraphics[width = \textwidth]{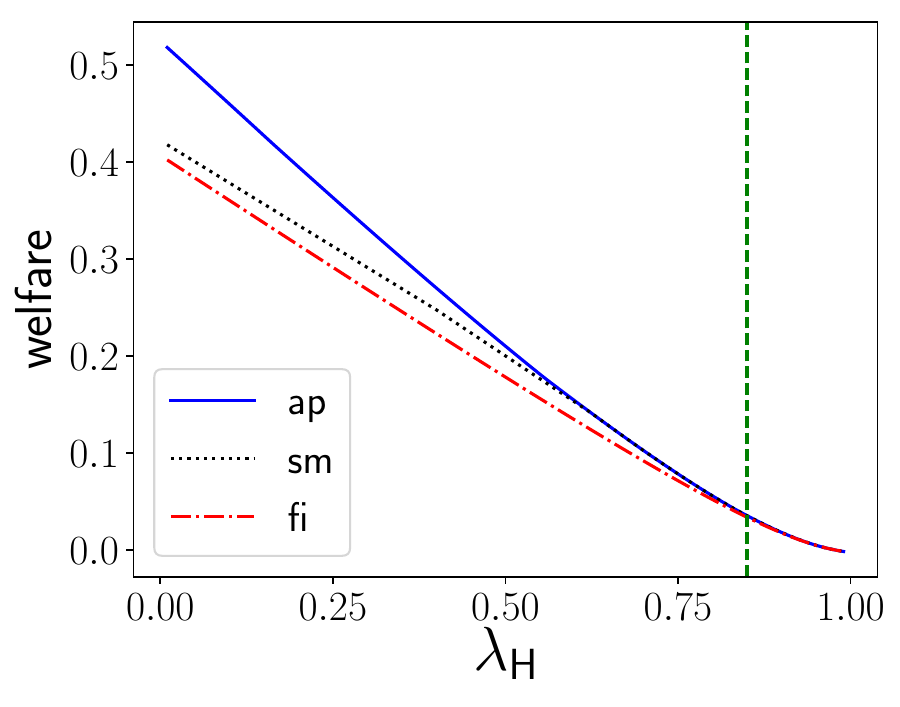}
	\caption{$\theta=1$}
	\label{fig:welfare-theta-comparison-3}
\end{subfigure}
\caption{Welfare of the Pareto-efficient signaling mechanism $\sm(\theta)$, the Pareto-efficient admission policy $\ar(\theta)$, and the full-information mechanism $\fim$ for $\theta \in \{0, 0.5, 1\}$. Here $\lambda_\low = 1- \lambda_\high$, and $c = 0.15$.}\label{fig:welfare-theta-comparison}
\end{figure}

We further complement the findings of Theorem~\ref{thm:achieve-first-best} with numerical computations
presented in Figure~\ref{fig:welfare-theta-comparison}. Here, we plot
the welfare of the signaling mechanism $\sm(\theta)$, the admission
policy $\ar(\theta)$ and the full-information mechanism $\fim$ for
$\theta \in \{0, 0.5, 1\}$, $c = 0.15$, and $\lambda =1$. Note that
$\theta = 0$ and $\theta=1$ correspond to the extreme cases where the
service provider seeks to maximize the welfare of  one type,
perhaps at the expense of the other. The case $\theta = 0.5$
corresponds to the case where the service provider values the two
types equally. Together, these three cases provide a representative
account of the service provider's potential objectives for welfare
improvement.  In these figures, in the region right of the green line,
we have $\lambda_\high > \bar{\Lambda}_\high = 1-c$. Thus, as shown in Theorem~\ref{thm:achieve-first-best}, we have $\sm(\theta) = \ar(\theta)$. For $\theta \in \{0.5, 1\}$, we see that
even for some values of $\lambda_\high < 1-c$, the two are
equal. Finally, we note that as $\lambda_\high \to 0$, we approach the
homogeneous setting, and as shown in Proposition~\ref{prop:homogen},
we observe the performance of the signaling mechanism $\sm(\theta)$
approaching that of the full-information mechanism in each case. 

\begin{figure}[htbp]
	\begin{subfigure}{0.5\linewidth}
		\centering
		\includegraphics[width = \textwidth]{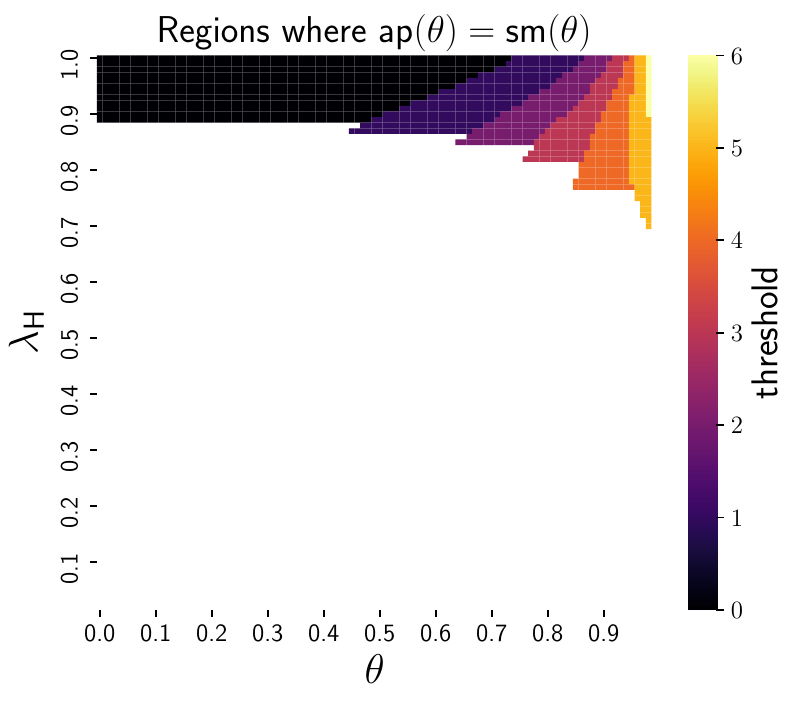}
		\caption{$c = 0.12$}
		\label{fig:heatmap-1}
	\end{subfigure}%
\begin{subfigure}{0.5\linewidth}
	\centering
	\includegraphics[width = \textwidth]{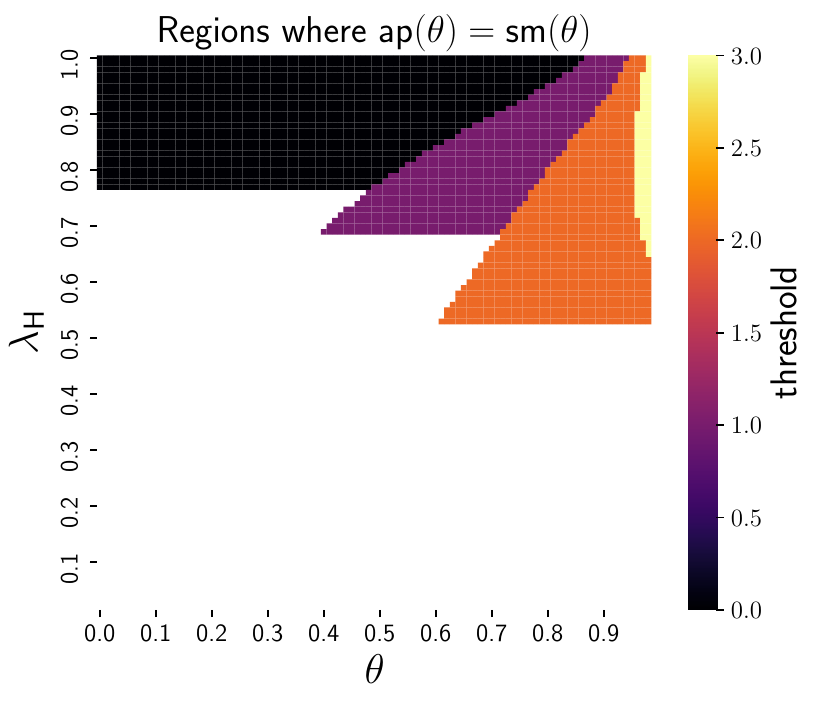}
	\caption{$c=0.24$}
	\label{fig:heatmap-3}
\end{subfigure}
\caption{Regions of the $(\theta, \lambda_{\high})$ plane for which $\sm(\theta) = \ar(\theta)$, i.e., the signaling mechanism $\sm(\theta)$ is Pareto-efficient within $\Pidb$. The colors represent the value of the threshold in $\ar(\theta)$.}\label{fig:heatmap}
\end{figure}

Finally, in Figure~\ref{fig:heatmap}, we plot for each  $c \in \{0.12, 0.24\}$, the values of $(\theta, \lambda_\high)$ for which the Pareto-efficient admission policy $\ar(\theta)$ is the same as the Pareto-efficient signaling mechanism $\sm(\theta)$. (Here, again $\lambda_\low = 1 - \lambda_\high$.) In other words, for these values, information design plays mainly the role of a coordination device, inducing users to coordinate towards a better welfare outcome. In particular, neither  obedience constraints bind for such values of $(\theta, \lambda_\high)$. Observe that, as shown in Theorem~\ref{thm:achieve-first-best}, for any fixed $\lambda_\high$, the values of $\theta$ for which this holds is an interval of the form $(\theta(\lambda_\low, \lambda_\high), 1]$. In particular, for $\lambda_\high > 1-c$, this is the entire interval $[0,1]$. Conversely, for small enough values of $\lambda_\high$, i.e., as we approach the homogeneous setting, we observe that this set is empty.

In Figure~\ref{fig:heatmap}, the threshold corresponding to each point $(\theta, \lambda_\high)$ is proportional to the color intensity used at the point. (Higher thresholds have lighter colors.) We observe that 
the threshold is non-zero for intermediate values of $\lambda_\high$ or sufficiently large $\theta$. This highlights that it is possible to have $\sm(\theta) = \ar(\theta)$ while letting some $\low$-type users join the queue. Further, note that for any fixed value of $\lambda_\high$, as $\theta$ increases, the threshold value in the Pareto-efficient mechanism $\sm(\theta) = \ar(\theta)$ increases; as more weight is placed on $\low$-type users' welfare, the Pareto-optimal signaling mechanism asks $\low$-type users to join the queue for a larger range of queue-length values. (Also, as stated in Theorem~\ref{thm:achieve-first-best}, in the complement  interval $[0 , \theta({\lambda_\low, }\lambda_\high))$ the threshold for $\sm(\theta)$ is independent of $\theta$.)  We also note that for any fixed $\theta$, the values of $\lambda_\high$ for which $\sm(\theta) = \ar(\theta)$ is fairly complex, with it being a union of two intervals for some values of $\theta$.

\section{Extensions}
\label{sec:extensions}

In this section, we explain how our model can be 
\kiedit{generalized in two dimensions: (i) having  $\high$-type users with finite outside option (Section~\ref{subsec:finite}), 
and (ii) incorporating heterogeneity in service rates, along with a priority service discipline (Section~\ref{subsec:priority}).
Additionally, in Appendices~\ref{subsec:abandon} and~\ref{subsec:greater-heterogeneity}, we generalize our framework to allow for user abandonment and more than two types, respectively.
Through analytical and numerical results, we illustrate that our key qualitative insights about the effectiveness of information design hold under all of the aforementioned generalizations. }

\subsection{\textcolor{black}{Fully Persuadable Population}}
\label{subsec:finite}

\kiedit{In our baseline model, we capture the level of need of $\high$-type users  by making the extreme assumption that $\high$-type users have an outside option of $-\infty$. Consequently, such users cannot be not persuaded to avail the outside option in the model. While such an assumption seems reasonable in certain contexts, such as the urgent care application discussed in Section~\ref{subsec:motivation}, it is natural to examine the power of information design in settings where the entire user population is persuadable.}

\kiedit{Toward that goal, we extend our model to incorporate finite outside options for both user types, with the $\high$-type users having a worse outside option compared to the $\low$-type users. In particular, normalizing the $\low$-type users' utility for the outside option to be zero, we denote the $\high$-type users' utility for the outside option as $\ell_\high < 0$. Furthermore, for the ease of notation, we denote an $\high$-type user's {\em incremental} utility for obtaining the service (over the outside option) as $u_\high(k) \defeq u_\high^{O}(k) - \ell_\high$, where $u_\high^{O}(k)$ denotes the utility from availing the service when $k$ users are already ahead in queue. In addition to Assumption~\ref{as:utility_assumptions} and paralleling the second condition therein, we assume that the difference $u_{\high}(n) - u_{\high}(n+1)$ is non-increasing in $n$. Further, to gain structural insights, we make a {\em utility dominance} assumption, which requires that the utility of the $\high$-type users dominates that of the $\low$-type users at all queue lengths, i.e., $u_\high(k) > u_\low(k)$ for all $k \geq 0$. (We observe that this dominance assumption is automatically satisfied in our baseline model with $\ell_\high = -\infty$.)}


\kiedit{
In practice, a social service provider may not always be able to observe the type of a user. Moreover, ethical concerns may limit a service provider from making information provision depend on the users' outside options.  Such limitations may make private signaling infeasible. \jaedit{Due} to such \jadelete{practical} considerations, we focus on {\em public signaling mechanisms}. Note that in our baseline model,  public and private signaling are the same because high-need users always join irrespective of the belief.}\vmcomment{to do: shorten by a word or two to save a line. } \jacomment{done!}

\kiedit{
Note that the utility dominance assumption implies that no (obedient) public signaling mechanism can provide a signal under which the $\low$-type joins but the $\high$-type does not join. Thus, we can restrict ourselves to signaling mechanisms that use three signals, denoted by $s \in \{0, 1, 2\}$, where signal $s=0$ recommends no user type join, $s=1$ recommends only the $\high$-type joins, while $s=2$ recommends both user types join. For any signaling mechanism, let $\pi_{k,j}$ denote the steady state probability that the queue length is $k$ and the signal sent is $j$. By analyzing the underlying birth-death chain of the queue, one can show that the steady-state distribution $\pi \defeq \{ \pi_{k,j} : k \geq 0, j=0,1,2\}$ satisfies the following balance conditions:
\begin{align}
\lambda \pi_{k,2} + \lambda_\high \pi_{k,1} &= \sum_{j \in \{0,1,2\}} \pi_{k+1, j}, \quad \text{for $k\geq 0$.} \tag{\textsf{BALANCE}}\label{eq:balance-fully-persuade}
\end{align}}
\kiedit{Furthermore, for $i \in \{\high, \low\}$ and $j \in \{0, 1, 2\}$, define $S_{i, j}(\pi) \defeq \sum_{k=0}^\infty \pi_{k,j} u_i(k)$. Note that $S_{i,j}(\pi)$ is proportional to the expected utility obtained by a type-$i$ user for joining the queue upon receiving the signal $j$. Due to the utility dominance assumption, we have $S_{\high, j}(\pi) \geq S_{\low, j}(\pi)$. Thus, for $\pi$ to be obedient, it must satisfy 
\begin{align}
    S_{H,0}(\pi) \leq 0, \quad \quad  S_{\low, 1}(\pi) \leq 0 \leq S_{\high, 1}(\pi), \quad \quad \text{and} \quad \quad S_{\high, 2}(\pi) \geq 0.
    \tag{\textsf{OBEDIENCE}}\label{eq:obd}
\end{align}
With the above definitions, we have the welfare functions as 
\begin{align*}
    W_\low(\pi) = \lambda_\low S_{\low, 2}(\pi), \quad \quad \text{and} \quad \quad W_\high(\pi) = \lambda_\high (S_{\high, 1}(\pi) + S_{\high,2}(\pi)).
\end{align*}}
\kiedit{Following the same argument as described in Section~\ref{sec:model}, we can 
obtain the Pareto frontier of signaling mechanisms by solving the following linear program for each $\theta \in [0,1]$: 
\begin{align*}
    \max_{\pi} & \qquad \theta W_\low(\pi) + (1-\theta) W_\high(\pi) \\  
    \text{subject to,} &  \qquad \text{\eqref{eq:obd}},\qquad \kiedit{\text{\eqref{eq:balance-fully-persuade}}} \\
     &\sum_{k=0}^{\infty} \sum_{j=0}^2 \pi_{k,j} = 1,\quad  \pi_{k,j} \geq 0 \quad \text{for all $j=0,1,2$ and $k\geq 0$.}
\end{align*}}

\kiedit{Finally, we define the class of threshold signaling mechanisms as follows: For
$0 \leq x \leq y$  with $x = m + a$ and $y= n + b$ where $m, n \in \naturals_0$ and $a,b \in [0, 1)$, a threshold mechanism with thresholds $x,y$ is
given by}
\jadelete{
\begin{align*}
  \pi_{k,2}
  &= \begin{cases}
    Z \lambda^k  & \text{for $0 \leq k < m$;}\\
    Z \lambda^{m} a & \text{for $k=m$;}\\
    0 & \text{otherwise,}
  \end{cases}\\
  \pi_{k,1}
  &= \begin{cases}
    Z \lambda^{m} (1-a) & \text{for $k=m$;}\\
    Z \lambda^{m} \lambda_a \lambda_\high^{k- m-1} & \text{for $m+1
      \leq k < n$;}\\
    Z \lambda^{m} \lambda_a \lambda_\high^{n- m-1}b & \text{for $k=n$;}\\
    0 & \text{otherwise,}
  \end{cases}\\
  \pi_{k,0}
  &= \begin{cases}
    Z \lambda^{m} \lambda_a \lambda_\high^{n- m-1} (1-b) & \text{for $k=n$;}\\
    Z \lambda^{m} \lambda_a \lambda_\high^{n- m} b & \text{for $k=n+1$;}\\
    0 & \text{otherwise,}
  \end{cases}
\end{align*}
}
\vmcomment{to do: express the above with indicator function to save $1/3$ of a page. }
\jaedit{
\begin{align*}
    \pi_{k,2} &= Z\lambda^{k} \ind\{0 \leq k < m\} + Z\lambda^m a \ind\{k=m\} \\
    \pi_{k,1} &= Z \lambda^{m} (1-a) \ind\{k = m\} + Z \lambda^{m} \lambda_a \lambda_\high^{k- m-1} \ind\{m+1
      \leq k < n\} + Z \lambda^{m} \lambda_a \lambda_\high^{n- m-1}b \ind\{k=n\} \\
     \pi_{k,0} &= Z \lambda^{m} \lambda_a \lambda_\high^{n- m-1} (1-b) \ind\{k=n\} + Z \lambda^{m} \lambda_a \lambda_\high^{n- m} b \ind\{k=n+1\}
\end{align*}
}
\kiedit{where 
$\lambda_a\defeq \lambda_\high + \lambda_\low a$ and
$Z = Z(x,y)$ is a normalizing constant.\jaedit{\footnote{$Z(x,y)$ is given by $1/Z(x,y) =  \sum_{k\leq m} \lambda^k  + \lambda^m \lambda_a \sum_{k=m+1}^{n} \lambda_\high^{k-m-1} + \lambda^m \lambda_a \lambda_\high^{n-m} b.$}} We denote the preceding mechanism as $\mathsf{Th}(x,y)$ and the corresponding welfare functions as $W_\high(x,y)$ and $W_\low(x,y)$.\jaedit{\footnote{The welfare functions are given by $W_\low(x, y) = \lambda_\low Z(x,y) \left( \sum_{k<m} \lambda^k u_\low(k)  + \lambda^m a u_\low(m)\right)$ and \\ $W_\high(x, y) = \lambda_\high Z(x,y) \left( \sum_{k=0}^m \lambda^k u_\high(k) + \lambda^m \lambda_a \sum_{k=m+1}^{n-1} \lambda_\high^{k-m-1} u_\high(k) +  \lambda^m \lambda_a \lambda_\high^{n-m-1} b\cdot  u_\high(n) \right)$. }}
}


\kiedit{In the rest of this section, we first analytically establish the effectiveness of signaling mechanisms compared
 to the benchmarks of  full information and no
information (in Proposition~\ref{prop:finite-info}). Then, we conduct numerical analysis to illustrate the robustness of our insights (with respect to $\high$-type's outside option) and also discuss the structure of Pareto-efficient signaling mechanisms. }

\subsubsection{\textcolor{black}{Mechanism Comparisons.}} \kiedit{The following proposition, which is in nature similar to Proposition~\ref{prop:info-design-power},  compares signaling mechanisms to the extreme cases of full and no-information mechanisms. \kiedit{Characterization of these two mechanisms in this modified model naturally follows the ones specified in Section~\ref{sec:model}, and for the sake of brevity we do not repeat these definitions. For $i \in \{\high, \low\}$, let $m_i$ denote the smallest value of $k$ for which $u_{i}(k) < 0$.}}
\kiedit{\begin{proposition}
\label{prop:finite-info}
 The following statements hold.
 \begin{enumerate}
  \item The full-information mechanism is Pareto-efficient within the class
  of signaling mechanisms if and only if it is Pareto-efficient within
  the class of admission policies. Furthermore, if
  $W_{\high}(m_\low, m_\high-1) \geq W_{\high}(m_\low, m_\high)$ or
  $W_{\low}(m_\low-1, m_\high) \geq W_{\low}(m_{\low},
  m_\high)$, then the full-information mechanism is Pareto-dominated by a
  threshold signaling mechanism.
  \item The no-information mechanism is never Pareto-efficient in the class
  of admission policies. Furthermore, suppose under the no-information
  mechanism, $\low$-type users join with positive
  probability. Then, the no-information mechanism is Pareto-dominated
  in the class of signaling mechanisms.
  \end{enumerate}
\end{proposition}}
\kiedit{The first part of the above proposition implies that if at least one \jaedit{of} the types benefits from lowering its threshold below the full-information threshold, then there exists a signaling mechanism more effective than sharing full information. The second part implies that if without any information, some $\low$-type users still join (e.g., the system is not too overcrowded), then information design is more effective that sharing no information.  Together, these two parts confirm that the power of information design persists in a population where all users are strategic and may decide not to join. The proof of the above proposition builds on the ideas used in the proof of Theorems~\ref{thm:full-info-comparison} and~\ref{thm:no-info-comparison} but also substantially departs from those proofs due to the difference in obedience constraints. 
The proof is presented in Appendix~\ref{app:sec:finite}.}

\subsubsection{\textcolor{black}{Numerical Analysis.}} 
To further examine the effectiveness of signaling, we turn our attention to numerical analysis. We consider a setting analogous to that in Section~\ref{sec:linear-costs} with arrival rates $\lambda_\low = 0.2$ and $\lambda_\high = 1 - \lambda_\low$. In particular, we assume that $u_{\low}(k) = 1-c(k+1)$ and $u_{\high}(k) = 1-c(k+1)-\ell_{\high}$ with $c = 0.13$ and $\ell_{\high} \in \{-1,-5,-10\}$.

\begin{figure}
	\begin{subfigure}{0.33\linewidth}
    	\centering
    	\includegraphics[height=0.8\linewidth]{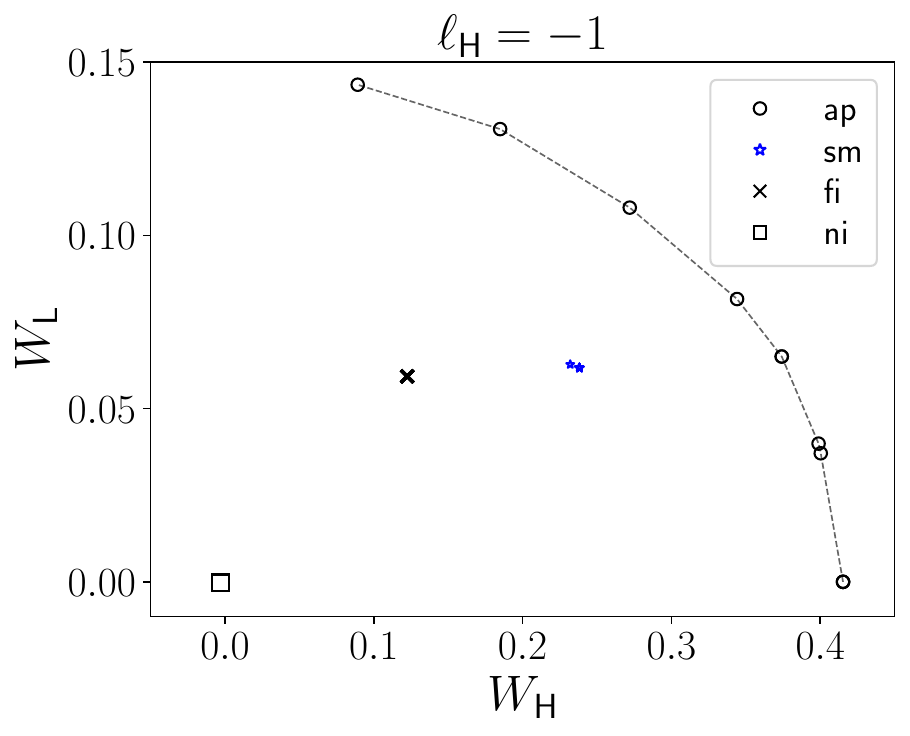}
    \end{subfigure}%
	\begin{subfigure}{0.33\linewidth}
    	\centering
    	\includegraphics[height=0.8\linewidth]{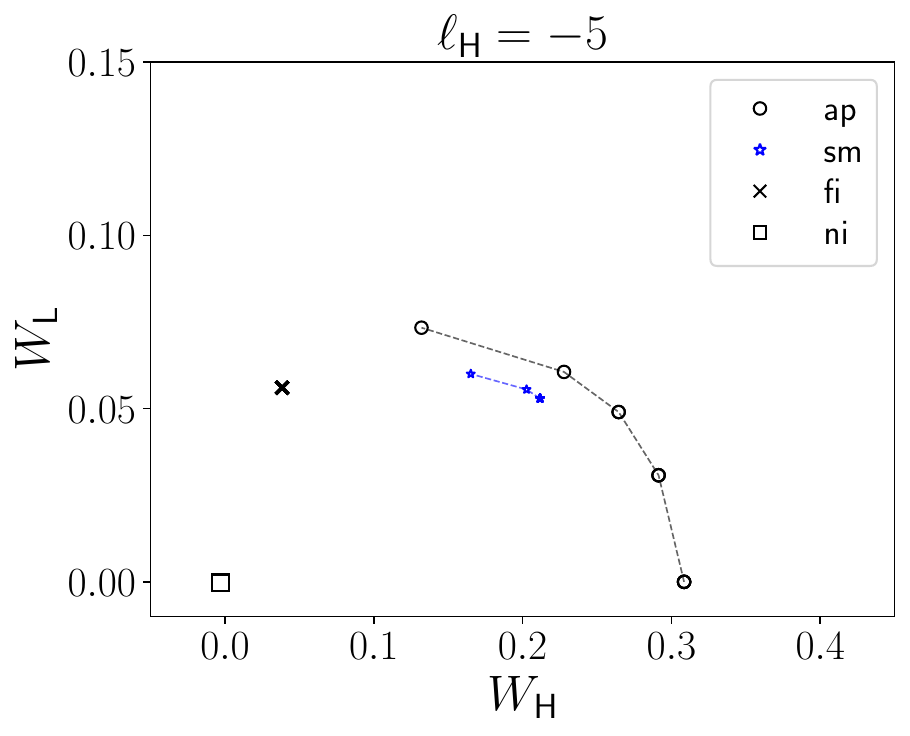}
    \end{subfigure}%
	\begin{subfigure}{0.33\linewidth}
    	\centering
    	\includegraphics[height=0.8\linewidth]{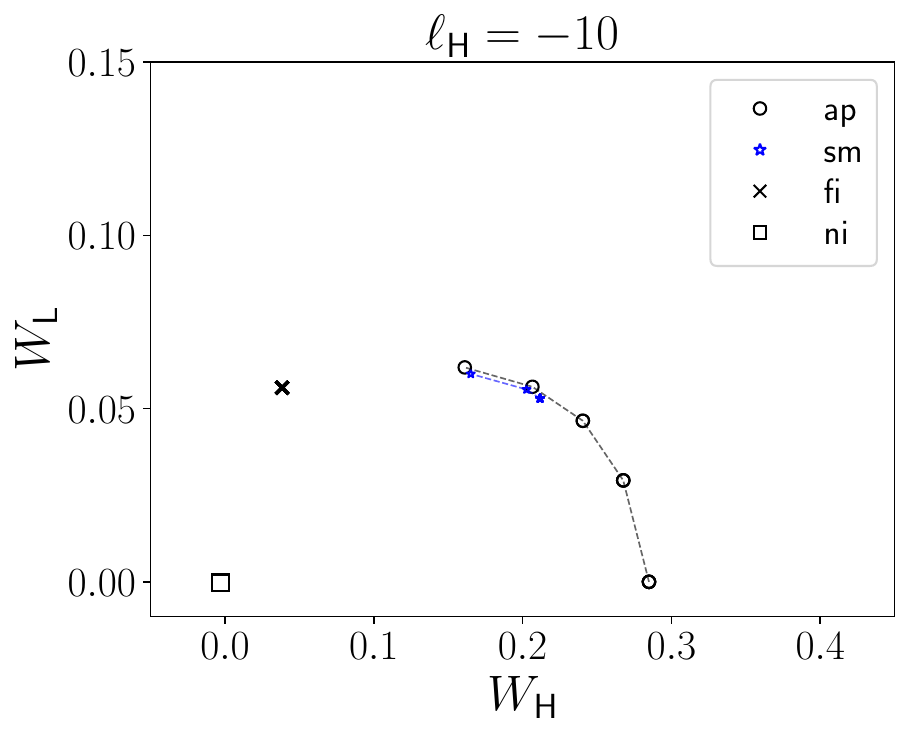}
    \end{subfigure}%
	\caption{Welfare of Pareto-efficient signaling mechanisms and admission policies for $\lambda_\low = 0.2, \lambda_\high = 1-\lambda_\low$, $c= 0.13$, $\ell_\high = -1$ (left panel) $\ell_\high = -5$ (middle panel), and $\ell_\high = -10$ (right panel). In all three panels, circles ($\circ$) represent efficient admission policies ($\ar$), stars ($\star$) represent efficient signaling mechanisms ($\sm$), cross ($\times$) represents the full-information mechanism ($\fim$), and square ($\square$) represents the no-information mechanism ($\nim$).}
	\label{fig:finite:1}
\end{figure}

\kiedit{In Figure~\ref{fig:finite:1}, we plot the welfare of Pareto-efficient signaling mechanisms and admission policies, along with those of the full-information and the no-information mechanisms, for different values of the outside option $\ell_\high$. First, we observe that in each setting, there exists a signaling mechanism that Pareto dominates the full-information and the no-information mechanism, thus demonstrating the effectiveness of information design in a fully persuadable user population. More importantly, we observe that as the $\high$-type users' outside option worsens, the Pareto frontier of admission policies approaches the Pareto frontier of the signaling mechanisms. Since the value of $\ell_\high$ reflects the $\high$-type users' need for the service, this numerical observation supports our broader conclusion that signaling is more effective when the user population is more heterogeneous in their need for service.}

\kiedit{We conclude this discussion by noting that in this generalized setting, the  optimal signaling mechanism may not be a threshold mechanism. In Appendix~\ref{sec:finite:structure}, we present two counter examples which show that the optimal signaling can be ``slightly'' different from $\mathsf{Th}(x,y)$. However, our numerical analysis suggests that (1) such non-threshold mechanisms only arise when $|\ell_{\high}|$ is relatively small, and (2) even in those cases, there exists a threshold mechanism that achieves a nearly identical welfare to that of the optimal signaling mechanism (See Figure~\ref{fig:tsm:gap} in Appendix~\ref{sec:finite:structure}). Finally, we remark that in the first part of Proposition~\ref{prop:finite-info}, we provide sufficient conditions under which there exists a threshold mechanism that Pareto dominates full-information mechanism, implying the effectiveness of threshold mechanisms even if they are not optimal.   }

\subsection{\textcolor{black}{Heterogeneity in Service Rates}}\label{subsec:priority}

\kiedit{Our baseline model assumes that the user types differ only in their utility (for both inside and outside options), but the service times across user types is homogeneous. However, in certain contexts, the heterogeneity in the need also translates to a heterogeneity in the service times. For instance, a critical patient arriving to an emergency room (ER) not only has higher need for service, but also requires longer service times, as compared to a non-critical patient. Given this practical concern, we consider a model where the user types not only differ in their need for service (through heterogeneity in their utility for service and outside option), but also in the service times -- the service times of a type-$i$ user, for $i \in \{\high, \low\}$ is assumed to be exponentially distributed with rate $\mu_i > 0$. To ensure stability, we assume that $\mu_i > \lambda_i$ for $i \in \{ \high, \low\}$.}

\kiedit{In such a setting, and especially when the user types are observable upon joining, it is reasonable to consider a preemptive priority service discipline, where any $\high$-type user has priority over a $\low$-type user. (For instance, such a discipline is natural in the ER setting mentioned above.) Under a preemptive priority service discipline, the state of the queue can be succinctly described as $(m,n)$, where $m\geq 0$ is the number of $\high$-type users and $n\geq 0$ is the number of $\low$-type users in the queue. We let $u_i(m, n)$ denote the utility of a user of type $i\in \{\high, \low\}$ for joining the queue, and assume that the outside option for the $\low$-type user is zero, while that of the $\high$-user is $-\infty$.\footnote{\vmredit{In Appendix~\ref{sec:priority:app}, we discuss the above setting without the priority scheme.}}}

\begin{figure}
	\begin{subfigure}{0.48\linewidth}
    	\centering
    	\includegraphics[height=0.75\linewidth]{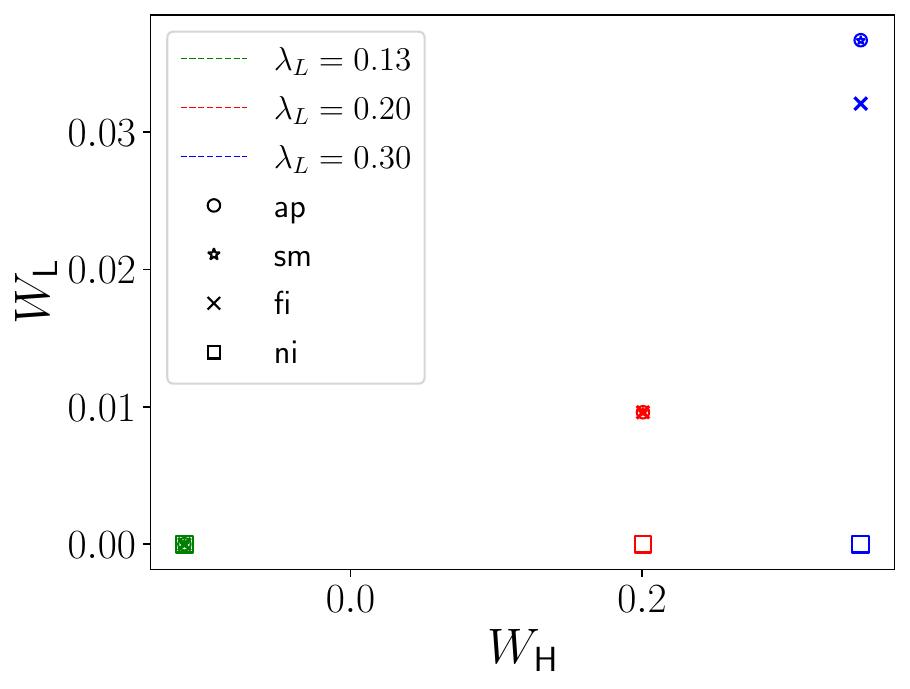}
    \end{subfigure}%
	\begin{subfigure}{0.48\linewidth}
		\centering
		\includegraphics[height=0.75\linewidth]{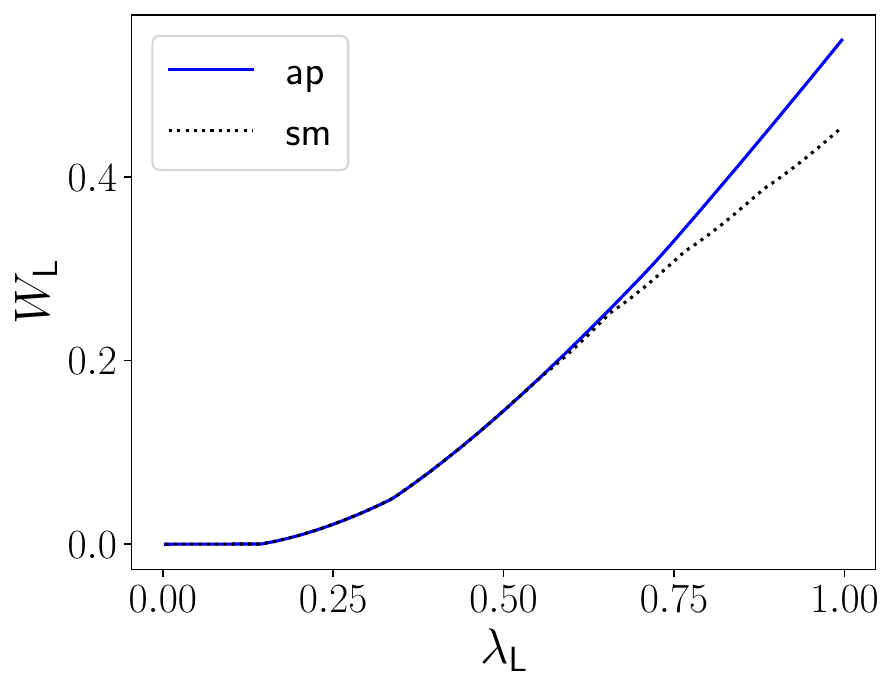}
	\end{subfigure}
	\caption{
	Left: Welfare of Pareto-efficient signaling mechanisms and admission policies for a preemptive priority queue 
	with $\mu_\low = 1.05, \mu_\high = 1$ and $c_\low = c_\high = 0.15$. The colors and shapes have the same definition as in Figure~\ref{fig:welfare-comparison}. Right: 
	The welfare of $\low$-type users under $\sm$ and $\ar$ for $\lambda_{\low} \in [0,1]$ and $\lambda_{\high} = 1 - \lambda_{\low}$.}
\label{fig:priority-queue}
\end{figure}

\kiedit{Similar to our baseline model, to find the optimal signaling mechanisms in this setting we formulate an infinite linear program, consisting of the obedience constraints and the balance conditions. Let $\pi_{m,n}^s$ denote the steady state probability that the queue state is $(m,n)$ and the signal sent to an arrival is $s \in \{0,1\}$, where the signal $s = 1$ recommends that the $\low$-type user join the queue, and $s=0$ recommends that she choose the outside option. (Note that in this model, as in our baseline model, the $\high$-type user always joins the queue.) Then, the balance conditions for the priority queue (with preemption) can be written as follows: for all $n\geq 0$,
\begin{align}\label{eq:pr-balance}
 (\lambda_\high + \mu_\high) \sum_{s=0,1} \pi_{m,n}^s   + \lambda_\low \pi_{m,n}^1 &= \mu_\high \sum_{s=0,1} \pi_{m+1, n}^s + \lambda_\high \sum_{s=0,1} \pi_{m-1,n}^s + \lambda_\low \pi_{m, n-1}^1\ind\{n> 0\} , \quad \text{for $m\geq 1$,} \notag\\
 (\lambda_\high + \mu_\low\ind\{n>0\}) \sum_{s=0,1} \pi_{0,n}^s + \lambda_\low \pi_{0,n}^1 &= \mu_\high \sum_{s=0,1} \pi_{1,n}^s + \mu_\low \sum_{s=0,1}\pi_{0,n+1}^s + \lambda_\low \pi_{0,n-1}\ind\{n>0\}, \quad \text{for $m=0$.} \tag{\textsf{Pr-BAL}}
\end{align}
Given the steady state distribution $\pi = \{\pi_{m,n}^s : m,n\geq0, s=0,1\}$, the obedience constraints can be written as
\begin{align}\label{eq:pr-obey}
    \sum_{m,n\geq 0} \pi_{m,n}^1 u_\low(m,n) \geq 0, \qquad \sum_{m,n\geq 0} \pi_{m,n}^0 u_\low(m,n) \leq 0. \tag{\textsf{Pr-OBD}}
\end{align}
Here, the first constraint requires that the $\low$-type user finds is optimal to join the queue upon receiving the signal $s=1$, while the second constraints requires it is optimal to choose the outside option upon receiving the signal $s=0$. Finally, the welfare of the two types can then be written as 
\begin{align*}
    W_\high(\pi) = \lambda_\high \sum_{m,n\geq 0} \sum_{s=0,1} \pi_{m,n}^s u_\high(m,n),  & &
    W_\low(\pi) = \lambda_\low \sum_{m,n\geq 0} \pi_{m,n}^1 u_\low(m,n).
\end{align*}
\kiedit{Note that since $\high$-type users have (preemptive) priority over the $\low$-type users, from their perspective, the queue only consists of $\high$-type users. Consequently, the $\low$-type users do not impose any externality on $\high$-type users. 
Combined with the fact that $\high$-type users always join, this implies that the welfare of $\high$-type users is unaffected by  the signaling mechanism. On the other hand, information design can still impact the welfare of $\low$-type users, and hence a Pareto-efficient mechanism maximizes the welfare of $\low$-type users \vmdelete{. Such a mechanism can be found} by solving the following linear program:}\vmcomment{to do: shorten by a word or two to save a line.}\vmcomment{done!} 
\begin{align*}
    & \max_\pi \qquad  W_\low(\pi)\\
    \text{subject to, } &\eqref{eq:pr-obey}, \quad \eqref{eq:pr-balance}\\
    \sum_{m,n\geq 0} & \sum_{s=0,1} \pi_{m,n}^s = 1, \quad  \pi_{m,n}^s\geq 0, \quad \text{for all $m,n\geq 0$ and $s=0,1$.}
\end{align*}
For our numerical analysis of this model, we consider the case of linear waiting costs. From basic queueing analysis, we obtain $u_\high(m,n) = 1 - c_\high\left(\frac{m+1}{\mu_\high}\right)$ and 
\begin{align*}
    u_\low(m,n) = 1 - c_\low \left( \frac{m}{\mu_\high - \lambda_\high} + \frac{(n+1)\mu_\high}{\mu_\low(\mu_\high - \lambda_\high)}\right).
\end{align*}
In Figure~\ref{fig:priority-queue}, we plot the welfare of the Pareto-efficient signaling mechanisms (stars) and admission policies (circles) for $\mu_\high=1,\mu_\low = 1.05$, $c_\high = c_\low = 0.15$, $\lambda_\high = 1 - \lambda_\low$ and for different values of $\lambda_\low \in \{0.13, 0.20, 0.30\}$. In each case, we also plot the full-information mechanism (\fim, cross) and the no-information mechanism (\nim, square). }
\kiedit{\vmdelete{From the figure, w}We observe that our qualitative insights continue to hold: when the population is sufficiently heterogeneous (e.g., when $\lambda_\low = 0.30$), the full-information and no-information mechanisms are Pareto dominated by a signaling mechanism. This illustrates the power of information design over these benchmarks even with heterogeneity in service times.}\vmcomment{to do: shorten by a word or two to save a line.}\vmcomment{done!}

\kiedit{Furthermore, when $\lambda_\low = 0.30$ or $0.20$, the  Pareto-efficient  signaling mechanism coincides with the Pareto-efficient admission policy. To highlight the power of signaling in achieving the first-best (i.e., the Pareto-efficient admission policy), in the right panel of Figure~\ref{fig:priority-queue}, 
\kiedit{we plot the welfare of $\low$-type users under 
the Pareto-efficient signaling mechanism ($\sm$) and the 
Pareto-efficient admission policy ($\ar$) as $\lambda_{\low}$
varies from $0$ to $1$ (with $\lambda_{\high} = 1 - \lambda_{\low}$). We observe that when $\lambda_{\low}$ is low (i.e., the system is over-crowded by $\high$-type users) then neither $\sm$ or $\ar$ let any $\low$-type user in, both achieving welfare of $0$. However, for moderate values of $\lambda_{\low}$, $\sm$ and $\ar$ still coincide but now achieve positive welfare by letting some $\low$-type users join. \vmredit{In Appendix~\ref{sec:priority:app}, we build on this numerical exercise to confirm that our qualitative findings remain the same under a wide range of gap between the service rates.}}}

\section{Conclusion}
\label{sec:conclude}
Social services often share two common features: they have limited
capacity relative to their demand, and they aim to serve users with
varied levels of needs.  Reducing congestion for such services using
price discrimination or admission control is often not feasible in this
setting. However, the service provider can use its informational
advantage, about service availability and wait times, to influence
users decisions in seeking the service by choosing what information to
reveal.  How effective will such a lever be? Our work seeks to answer
this question.  Adopting the framework of Bayesian persuasion, we
study information design in a queueing system that serves users who are
heterogeneous in their need for the service.  We show that, by and
large, information design provides a Pareto-improvement in welfare of
all user types when compared to simple mechanisms of sharing full
information or no information.  Further, we show that information
design can go beyond and achieve the ``first-best'': it can achieve
the same welfare outcomes as those of centralized admission policies
that observe each user's type, and disregard user incentives.  \kiedit{Finally, we show that our qualitative findings -- on the benefits of well-designed information disclosure policies in the presence of need heterogeneity --  continue to hold under various extensions to our model motivated from practical concerns. In sum, our results comprehensively exhibit a promising role for information design in improving welfare outcomes in congested social services.}

\bibliographystyle{informs2014}
\bibliography{references}

%
%
%

\newpage

\begin{APPENDIX}{}



\section{Proofs from Section~\ref{sec:structural-characterization}}
\label{ap:structural-characterization}

In this section, we present the missing proofs from Section~\ref{sec:structural-characterization}.

\proof{{Proof of Lemma~\ref{lem:pareto-efficient-exists}.}}
The proof is immediate for $\lambda < 1$, since the sets $\Pidb$ and $\Piob$ are compact (under the topology of weak convergence) and $W(\pi, \theta)$ is continuous in $\pi$ for each $\theta \in [0,1]$. To see this, note that for any $\pi \in \Pidb$, by Lemma~\ref{lem:positive-pi} (stated at the end of this section), we have $\pi_k \leq \lambda^k \pi_0$ for all $k$. Hence, for $\lambda < 1$, Prohorov's theorem~\citep{aliprantisB2006} directly implies compactness of $\Pidb$ and $\Piob$. 

Thus, for the rest of the proof, suppose $\lambda =1$. Let $\Pidb^\fim$ and $\Piob^\fim$ denote the set of admission policies and signaling mechanisms that are not Pareto-dominated by the full-information mechanism. We again use Prohorov's theorem to show that these sets are relatively compact, implying the existence of a maximizer of $W(\cdot, \theta)$ over the closures of these sets. The result then follows since any maximizer of $W(\pi, \theta)$ over the closure of $\Pidb^\fim$ ($\Piob^\fim$) is also a maximizer over $\Pidb$ (resp., $\Piob$).

Thus, we first show that the set of distributions $\Pidb^\fim$ is tight. Fix an $\epsilon > 0$. Let $W_\low(\fim)$ and $W_\high(\fim)$ denote the welfare of each type under the full information mechanism. Next, fix some large enough $N$ to be chosen later. Consider a steady-state distribution $\pi \in \Pidb$. We have the following expressions:
\begin{align*}
    W_\high(\pi) &= \lambda_\high \sum_{n=0}^\infty \pi_n u_\high(n)\\
    &\leq  \lambda_\high u_\high(0)  \left( \sum_{n < N} \pi_n \right) +  \lambda_\high u_\high(N) \left( \sum_{n \geq N} \pi_n \right)\\
    &\leq \lambda_\high u_\high(0) + \lambda_\high u_\high(N) \left( \sum_{n \geq N} \pi_n \right).
\end{align*}
where the first inequality follows from Assumption~\ref{as:utility_assumptions} and the second follows because $u_\high(0) > 0$. Similarly, we have for large enough $N > m_\fim$, 
\begin{align*}
    W_\low(\pi) &= \sum_{n=0}^\infty (\pi_{n+1} - \lambda_\high \pi_n) u_\low(n)\\
    &\leq u_\low(0) \left(\sum_{n< N} (\pi_{n+1} - \lambda_\high \pi_n)\right)  + u_\low(N)  \left(\sum_{n\geq N} (\pi_{n+1} - \lambda_\high \pi_n) \right)\\
    &= u_\low(0)\left(  \pi_{N} -  \pi_0 + (1- \lambda_\high) \sum_{n=0}^{N-1} \pi_{n} \right)    + u_\low(N)  \left( - \lambda_\high \pi_{N} + (1- \lambda_\high) \sum_{n>N} \pi_{n}  \right)\\
    &\leq (1-\lambda_\high) u_\low(0)   + u_\low(N)  \left( -  \pi_{N} + (1- \lambda_\high) \sum_{n\geq N} \pi_{n}  \right)\\
    &\leq \lambda_\low u_\low(0)   + u_\low(N)  \left( -  \frac{1}{N+1} + \lambda_\low \sum_{n\geq N} \pi_{n}  \right),
\end{align*}
where the final inequality follows from (1) $\lambda_\low = 1- \lambda_\high$, (2) $u_\low(N) < 0$, and (3) $(N+1) \pi_N \leq \sum_{n=0}^N \pi_n \leq 1$ because of the detailed-balance conditions $\pi_n \leq \pi_{n-1}$. Thus, for $N \geq N^\epsilon_0 = \frac{2}{\epsilon \lambda_\low }$, we obtain
\begin{align*}
    W_\low(\pi) &\leq \lambda_\low u_\low(0)   + \lambda_\low  u_\low(N) \left( -  \frac{\epsilon }{2}  +  \sum_{n\geq N} \pi_{n}  \right).
\end{align*}
Since $\lim_{n\to\infty} u_i(n) = -\infty$, let $N_1^\epsilon$ be large enough so that $\max\{ u_\low(k), u_\high(k) \} < -\frac{2}{\epsilon^2} $ for $k \geq N_1^\epsilon$. Then, for all $N \geq N^\epsilon = \max\{N_0^\epsilon, N_1^\epsilon\}$, we have
\begin{align*}
    W_\high(\pi) &\leq \lambda_\high \left(u_\high(0) - \frac{2}{\epsilon^2} \sum_{n \geq N} \pi_n\right)\\
    W_\low(\pi) &\leq  \lambda_\low\left( u_\low(0) - \frac{2}{\epsilon^2} \left(-\frac{\epsilon}{2} + \sum_{n \geq N} \pi_n\right)\right).
\end{align*}
Now, for any $\pi \in \Pidb$, if $\sum_{n\geq N} \pi_n \geq \epsilon$, then we have $W_i(\pi) \leq \lambda_i u_i(0) - \frac{\lambda_i}{\epsilon}$ for $i \in \{\low, \high\}$.  For small enough $\epsilon >0$, we obtain that $W_i(\pi) < W_i(\fim)$ and hence $\pi$ is Pareto dominated by the full-information mechanism, implying $\pi \not\in \Pidb^\fim$. Thus, we conclude that for all small enough $\epsilon>0$, there exists an $N^\epsilon$ such that for all $\pi \in \Pidb^\fim$, we have $\sum_{n\geq N^\epsilon} \pi_n < \epsilon$. Thus, the set of distributions $\Pidb^\fim$ is tight. 

Using Prohorov's theorem, we then conclude that $\Pidb^\fim$ is relatively compact (under weak topology). The set $\Piob^\fim$, being a subset of $\Pidb^\fim$, is also relatively compact. Since $W(\pi, \theta)$ is continuous in $\pi \in \Pidb^\fim$ for any $\theta \in [0,1]$, we obtain that the maximizer of $W(\pi, \theta)$ over the closure of $\Pidb^\fim$ (and, separately, $\Piob^\fim$) exists and is Pareto-efficient within $\Pidb$ (resp., $\Piob$).~\Halmos\endproof

\proof{{Proof of Theorem~\ref{thm:threshold-signaling-mechanism}.}} Let $\pi \in \Pidb$ be such that there exists an $m\geq 0$ with $\pi_{m+1} < \lambda \pi_m$ and $\lambda_\high \pi_{m+1} < \pi_{m+2}$. In words, this implies that under $\pi$, an arriving $\low$-type user is asked to leave with positive probability if the queue length is $m$, and asked to join with positive probability if the queue length is $m+1$. We now show that such a $\pi$ cannot be Pareto-efficient within $\Piob$. We do this by constructing an $\hat{\pi} \in \Piob$ that Pareto-dominates $\pi$.

Towards that end, consider the following perturbation of $\pi$ for small enough $\delta > 0$:
\begin{align*}
    \hat{\pi}_k = \begin{cases}  \pi_k & \text{if $k < m+1$;}\\
    \pi_{m+1} + \delta \sum_{n > m+1} \pi_n & \text{if $k = m+1$;}\\
    \pi_k (1- \delta) & \text{if $k  > m+1$.}
    \end{cases}
\end{align*}
First, it is straightforward to verify that $\hat{\pi}$ satisfies the detailed balance constraints in Lemma~\ref{lem:equivalence} for all small $\delta > 0$. 
In addition, we have
\begin{align*}
    J(\hat{\pi}) &= \sum_{k=0}^\infty ( \pi_{k+1} - \lambda_\high \pi_k) u_\low(k) + \delta \left(\sum_{k > m+1} \pi_k \right) (u_\low(m)  - \lambda_\high  u_\low(m+1))\\  
    &\quad - \delta \pi_{m+2} u_\low(m+1) - \delta \sum_{k > m+1}  ( \pi_{k+1} - \lambda_\high \pi_k) u_\low(k) \\
    &= J(\pi) + \delta \cdot \sum_{k > m+1} \pi_k  \cdot \left( ~ u_\low(m)   - u_\low(k-1) - \lambda_\high ( u_\low(m+1) -   u_\low(k) ) ~ \right) .
\end{align*}

Now, as $\lambda_\high <  1$, for any $k > m+1$, we have
\begin{align*}
  u_\low(m) - u_\low(k-1) - \lambda_\high (u_\low(m+1) - u_\low(k))
  &> u_\low(m) - u_\low(k-1) - u_\low(m+1) - u_\low(k)\\  
  &= (u_\low(m)  - u_\low(m+1))  - (u_\low(k-1) - u_\low(k)) \\
  &\geq 0,
\end{align*}
where we have used Assumption~\ref{as:utility_assumptions} in both
inequalities.  Specifically, the first inequality follows from the
fact that $u_\low(k)$ is strictly decreasing in $k$ and hence
$u_\low(m+1) - u_\low(k) > 0$, and the second inequality follows from
the fact that $u_\low(n) - u_\low(n+1)$ is
non-increasing in $n$. Using this and the fact that
$\pi_{m+2} > \lambda_{\high} \pi_{m+1} \geq 0$, we obtain that
$J(\hat{\pi}) > J(\pi) \geq 0$. Hence the obedience constraint
\eqref{eq:obedience1} holds for $\hat{\pi}$.

By similar algebraic steps, we have
\begin{align*}
    L(\hat{\pi}) &= L(\pi) - \delta \cdot \sum_{k > m+1} \pi_k \cdot (~u_\low(m)   - u_\low(k-1) - \lambda ( u_\low(m+1) -   u_\low(k) )  ~) .
\end{align*}
Using the fact that $\lambda \leq 1$, by a similar argument as before, we obtain that the parenthetical term is non-negative, and hence $L(\hat{\pi}) \leq L(\pi) \leq 0$. Thus, the obedience constraint \eqref{eq:obedience2} also holds for $\hat{\pi}$. Taken together, this implies we have $\hat{\pi} \in \Piob$. 

Next, note that 
\begin{align*}
    W_\high(\hat{\pi}) &= \lambda_\high \sum_{n=0}^\infty \hat{\pi}_n u_\high(n) \\
    &= W_\high(\pi) + \lambda_\high \delta \cdot \left( \sum_{k > m+1} \pi_k \cdot ( u_\high(m+1) - u_\high(k)) \right).
\end{align*}
Since $u_\high(k)$ is strictly decreasing in $k$, we obtain $W_\high(\hat{\pi}) \geq W_\high(\pi)$. Finally, we have $W_\low(\hat{\pi}) = J(\hat{\pi}) > J(\pi) = W_\low(\pi)$. Thus, we obtain that $\hat{\pi}$ Pareto-dominates $\pi$. 

From the above, we conclude that for any Pareto-efficient signaling mechanism $\pi \in \Piob$, it must be the case that 
whenever there exists an $m \geq 0$ with $\pi_{m+1} < \lambda \pi_m$, we have $\pi_{m+2} = \lambda_\high \pi_{m+1}$. This implies that $\pi$ must have one of the following two structures:
\begin{enumerate}
    \item for all $m \geq 0$, we have $\pi_{m+1} = \lambda \pi_m$; OR
    \item there exists an $m \geq 0$ such that $\pi_{k+1} = \lambda \pi_k$ for $k < m$, $\pi_{m+1} < \lambda \pi_m$ and $\pi_{k+1} = \lambda_\high \pi_k$ for $k > m$. 
\end{enumerate}
In the first case, we have $\low$-type users being asked to join the queue for all queue length, implying that $\pi$ trivially has a threshold structure (with threshold equal to $\infty$). In the second case, the $\low$-type users are asked to join with probability $1$ for queue-lengths strictly less than $m$ and asked to leave with probability $1$ for queue-lengths strictly greater than $m$. Again, this implies a threshold structure for $\pi$, with threshold in the interval $[m, m+1]$. 

Having shown the threshold structure of Pareto efficient signaling mechanisms, next we show  that the corresponding threshold is  less than or equal to the full-information threshold $m_\fim$. Let $\pi \in \Piob$ have a threshold structure, with a threshold $x > m_\fim$, where $x = m+a$ with $m \in \NN$ and $a \in [0,1]$. Thus, we have $\pi_{k+1} = \lambda \pi_{k}$ for all $k < m$, and $\pi_{k+1} = \lambda_\high \pi_k$ for all $k > m$.  Note, we allow $m = \infty$, which captures the case where $\pi_{k+1} = \lambda \pi_{k}$ for all $k \in \NN$. Observe that $x > m_\fim$ implies that $m \geq m_\fim$, and hence the threshold structure of $\pi$ implies $\pi_{m_\fim} > 0$.

We prove that such a distribution $\pi$ cannot be Pareto efficient by constructing a $\hat{\pi} \in \Piob$ which Pareto dominates $\pi$.
Consider $\hat{\pi}$ defined as follows:
\begin{align*}
    \hat{\pi}_k &= \begin{cases} \frac{1}{Z} \pi_k & \text{if $k \leq m_\fim$;}\\
    \frac{1}{Z} \lambda_\high^{k-m_\fim} \pi_{m_\fim} & \text{if $k > m_\fim$,}
    \end{cases}
\end{align*}
where $Z = \sum_{k \leq m_\fim} \pi_k  + \pi_{m_\fim} \sum_{k > m_\fim} \lambda_\high^{k-m_\fim}$. Using the detailed balance constraints in Lemma~\ref{lem:equivalence}, it follows that $\pi_k \geq \lambda_\high^{k - m_\fim} \pi_{m_\fim}$ for all $k > m_\fim$. Thus, as $\sum_k \pi_k = 1$, we have $Z \leq 1$. 

Next, consider
\begin{align*}
J(\hat{\pi}) = \sum_{k=1}^\infty \left(\hat{\pi}_{k +1} - \lambda_\high \hat{\pi}_k\right) u_\low(k) &= \frac{1}{Z} \sum_{k<m_\fim } \left(\pi_{k +1} - \lambda_\high \pi_k\right) u_\low(k) \\
&>  \frac{1}{Z} \sum_{k=1}^\infty \left(\pi_{k +1} - \lambda_\high \pi_k\right) u_\low(k) = \frac{1}{Z} \cdot J(\pi).
\end{align*}
Here, the inequality follows from the fact that $u_\low(k)  < 0$ for $k \geq m_\fim$ and that  $\pi_{m_\fim +1 } - \lambda_\high \pi_{m_\fim} = \lambda_\low \min\{x - m_\fim, 1\} \pi_{m_\fim } >0$. Since $J(\pi) \geq 0$ and $Z \leq 1$, we obtain that $J(\hat{\pi}) > J(\pi) \geq 0$. Hence, the obedience constraint \eqref{eq:obedience1} holds for $\hat{\pi}$. Moreover, the threshold structure of $\pi$ implies 
\begin{align*}
    L(\hat{\pi}) &= \sum_{k=1}^\infty \left(\lambda \hat{\pi}_k - \hat{\pi}_{k +1} \right) u_\low(k)   = \sum_{k= m_\fim}^\infty \left(\lambda \hat{\pi}_k - \hat{\pi}_{k +1} \right) u_\low(k) \leq 0.
\end{align*}
Thus, the obedience constraint \eqref{eq:obedience2} also holds for $\hat{\pi}$. Hence, we obtain that $\hat{\pi} \in \Piob$. 

Furthermore, for $\ell \leq m_\fim$, we have $ \sum_{k \leq \ell} \hat{\pi}_k  = \frac{1}{Z} \cdot \sum_{k \leq \ell} \pi_k$. Since, $Z \leq 1$, this implies $ \sum_{k \leq \ell} \hat{\pi}_k  \geq  \sum_{k \leq \ell} \pi_k $ for all $\ell  \leq m_\fim$. For $\ell > m_\fim$, after some algebra, we obtain
\begin{align*}
    \sum_{k \leq \ell} \hat{\pi}_k - \sum_{k \leq \ell} \pi_k  =    \frac{1}{Z}\left( \sum_{q \leq m_\fim} \sum_{k> \ell} \pi_q \left( \pi_k - \pi_{m_\fim}\lambda_\high^{k - m_\fim}\right) + \sum_{q= m_\fim +1}^\ell \sum_{k>\ell} \pi_{m_\fim} \lambda_{\high}^{q- m_\fim} \left(\pi_k - \lambda_{\high}^{k-q} \pi_q\right)  \right). 
\end{align*}
In Lemma~\ref{lem:positive-pi} (stated at the end of this section), we show that $\pi_k \geq \pi_{q} \lambda_{\high}^{k-q}$ for all $k > q$. Thus, the right-hand side is non-negative, and hence, $ \sum_{k \leq \ell} \hat{\pi}_k  \geq  \sum_{k \leq \ell} \pi_k $ for $\ell  > m_\fim$ as well. Together, this implies that $\hat{\pi}$ is stochastically dominated by $\pi$. Since $u_\high(k)$ is strictly decreasing in $k$, we have 
\begin{align*}
    W_\high(\hat{\pi}) &= \lambda_\high \sum_{k=0}^\infty \hat{\pi}_k u_\high(k) \geq \lambda_\high \sum_{k=0}^\infty \pi_k u_\high(k) = W_\high(\pi).
\end{align*}
Finally, since $W_\low(\hat{\pi}) = J(\hat{\pi}) > J(\pi) = W_\low(\pi)$, we conclude that $\hat{\pi} \in \Piob$ Pareto-dominates $\pi$, and hence $\pi$ cannot be Pareto-efficient within $\Piob$. 
\Halmos\endproof

\begin{lemma}\label{lem:positive-pi} For any $\pi \in \Pidb$, and for any $k  >q \in \NN$, we have $\pi_k \geq \lambda_{\high}^{k-q} \pi_q$ and $\pi_k \leq \lambda^{k-q} \pi_q$.  In particular, when $\lambda_\high > 0$, for any $\pi \in \Pidb$, we have $\pi_k \in (0,1)$ for all $k \in \NN$.
\end{lemma}
\proof{{Proof.}}
The proof follows immediately from the detailed balance constraints in Lemma~\ref{lem:equivalence}. 
\Halmos\endproof

\proof{{Proof of Theorem~\ref{thm:pareto-or-leave}.}} 
Since
$\Piob \subset \Pidb$, any signaling mechanism $\pi \in \Piob$ that is
Pareto efficient within the class of admission policies must be so
within the class of signaling mechanisms. Thus, it remains to show
that for any signaling mechanism $\pi \in \Piob$ with $L(\pi)<0$, if
$\pi$ is Pareto dominated by an admission policy, then it is Pareto
dominated by a signaling mechanism.

By Theorem~\ref{thm:threshold-signaling-mechanism}, we obtain that if $\pi$ does not have a threshold structure, or if it has a threshold structure with threshold greater than the full-information threshold $m_\fim$, then $\pi$ is Pareto dominated within the class $\Piob$ of signaling mechanisms, and there is nothing to prove. Hence, suppose that $\pi$ has a threshold structure with threshold smaller or equal to $m_\fim$.  This in turn implies that $J(\pi) > 0$, as a $\low$-type user always receives non-negative utility upon joining the queue, and receives a positive utility if the queue is empty (which occurs with positive probability).

Since $\pi$ is not Pareto-efficient within the class $\Pidb$, there
exists an admission policy $\hat{\pi} \in \Pidb$ that Pareto-dominates
$\pi$. In particular, we have $W_i(\hat{\pi}) \geq W_i(\pi)$ for
$i \in \{\high, \low\}$, with at least one inequality strict.

Next, let $\tilde{\pi} = (1- \epsilon) \pi + \epsilon \hat{\pi}$ for
some $\epsilon \in (0,1]$ to be chosen later. By convexity of $\Pidb$,
we have $\tilde{\pi} \in \Pidb$. Furthermore, by linearity, we have
$J(\tilde{\pi}) = (1 -\epsilon) J(\pi) + \epsilon J(\hat{\pi})$ and
$L(\tilde{\pi}) = (1-\epsilon) L(\pi) + \epsilon L(\hat{\pi})$. Since
$J(\pi) > 0$ and $L(\pi) < 0$, for all small enough $\epsilon > 0$ we
have $J(\tilde{\pi}) \geq 0$ and $L(\tilde{\pi}) \leq 0$. Thus, the
obedience constraints~\eqref{eq:obedience1} and \eqref{eq:obedience2}
hold for $\tilde{\pi}$, and hence $\tilde{\pi} \in \Piob$. Finally,
again by linearity, we have
\begin{align*}
  W_\low(\tilde{\pi}) &= (1-\epsilon) W_\low(\pi) + \epsilon W_\low(\hat{\pi}) \geq W_\low(\pi)\\
  W_\high(\tilde{\pi}) &= (1-\epsilon) W_\high(\pi) + \epsilon W_\high(\hat{\pi}) \geq W_\high(\pi),
\end{align*}
with at least one inequality strict. Thus, we obtain that the
signaling mechanism $\tilde{\pi}$ Pareto-dominates $\pi$ and hence
$\pi$ is not Pareto-efficient within the class $\Piob$.
\Halmos\endproof

\section{Structural Results} 
\label{ap:structure}
Before proceeding to present the missing proofs of Sections~\ref{sec:comparison} and \ref{sec:first-best}, we present three structural results. First, in Lemma~\ref{lem:ni}, we characterize the equilibrium structure under the no-information mechanism. Then, in Lemma~\ref{lem:unimodal}, we study the shape of welfare functions $W_{\low}(\cdot)$ and $W_{\high}(\cdot)$ for threshold mechanisms. Finally, in Lemma~\ref{lem:leave-worsens}, we study the  function $L(\cdot)$ defined in \eqref{eq:obedience2}. We remark that the last two lemmas are used in the proofs of results in Sections \ref{sec:comparison} and \ref{sec:first-best}.

\begin{lemma}[Equilibrium structure under no-information mechanism]
\label{lem:ni}
For $p \in [0,1]$, let $\pi(p) \in \Pidb$ be given by $\pi_n(p) = (1- \lambda_\low p - \lambda_\high) (\lambda_\low p + \lambda_\high)^n$. Then, the steady state distribution under the no-information mechanism $\nim$ is given by $\pi(p^\nim) \in \Piob$, for $p^\nim \in [0,1]$ that satisfies the following conditions:
\begin{enumerate}
    \item if $\sum_{k=0}^\infty \lambda^k u_\low(k)   \geq 0$ then $p^\nim = 1$;
    \item if $\sum_{k=0}^\infty \lambda_\high^k u_\low(k) \leq 0$ then $p^\nim = 0$;  
    \item otherwise, $p^\nim \in (0,1)$ satisfies $\sum_{k=0}^\infty (\lambda_\low p^\nim + \lambda_\high)^k u_\low(k)= 0$.
\end{enumerate}
Here, $p^\nim \in [0,1]$ denotes the probability under the no-information mechanism that a $\low$-type user joins the queue upon arrival.
\end{lemma}
\proof{{Proof of Lemma~\ref{lem:ni}.}}
First note that an arriving $\low$-type user has no information about the queue length. Therefore, a symmetric equilibrium strategy consists of a probability $p$ with which she joins the queue.  Let $\pi_n(p)$ be  the steady-state distribution corresponding to such a strategy. By detailed balance constraint, we have: 
\begin{align*}
    \pi_{n+1}(p) = (\lambda_{\high} + p \lambda_{\low})  \pi_{n}(p), ~ n \in \NN
\end{align*}
This implies $\pi_n(p) = (1- \lambda_\low p - \lambda_\high) (\lambda_\low p + \lambda_\high)^n$, $n \in \NN$. A $\low$-type users chooses $p$ that maximizes her utility. This gives rise to the three cases listed in the statement of the lemma. 
\Halmos\endproof

\begin{lemma}[Properties of welfare functions]
\label{lem:unimodal}
\begin{enumerate}
    \item The welfare function $W_\high(x)$ is strictly decreasing in $x \in \reals_+$. 
    \item 
    The welfare function $W_\low(x)$ is unimodal over $x \in \reals_{+}$. Furthermore, $W_\low(x)$ is monotone between consecutive integers, initially increasing up to a maximum, and then decreasing.
    \item The function $W(x, \theta) = \theta W_\low(x) + (1 - \theta)  W_\high(x)$ attains its maximum at an integer $m \leq m_\fim$.
\end{enumerate}
\end{lemma}

\proof{{Proof of Lemma~\ref{lem:unimodal}.}}  The proof of the first statement follows from the fact
the steady-state distribution under the threshold policy $x$ is
stochastically dominated by that under the threshold policy
$\hat{x} > x$. Since $u_\high(k)$ is strictly decreasing in $k$, we
thus obtain that $W_\high(x) > W_\high(\hat{x})$.

For the second statement, we show that (i) $W_\low(x)$ is monotone between consecutive integers, and (ii) $W_\low(x)$ is  unimodal over integers, initially increasing up to a maximum, and then decreasing. Together, theses two properties imply the unimodality of $W_\low(x)$ for $x \in \reals_{+}$. Note that for $x = m + a$, where $m \in \NN$
and $a \in [0,1)$, we have
\begin{align*}
  W_\low(x) = \lambda_\low \cdot \frac{1}{\sum_{k=0}^{m} \lambda^k + \frac{\lambda^{m} (\lambda_\high + a \lambda_\low )}{1 - \lambda_\high}} \cdot \left( \sum_{k=0}^{m-1} \lambda^k u_\low(k)  + a \lambda^m  u_\low(m)\right).
\end{align*}
Since this is of the form
$\frac{\alpha + \beta a}{\gamma + \delta a}$, where
$\alpha, \beta, \gamma, \delta$ are independent of $a$, we obtain that
$W_\low(m+a)$ is monotone in $a \in [0,1)$. Thus, we conclude that
$W_\low(x)$ is monotone between consecutive integers. It is
straightforward to verify that $W_\low(x)$ is continuous, and hence
the maximum of $W_\low(x)$ is attained at an integer.

For $m \in \NN$, we have
\begin{align*}
  W_\low(m) = \lambda_\low \cdot \frac{1}{\sum_{k=0}^{m} \lambda^k + \frac{\lambda^{m} \lambda_\high }{1 - \lambda_\high}} \cdot  \sum_{k=0}^{m-1} \lambda^k u_\low(k) = \lambda_\low \cdot \Gamma(m) \cdot \Lambda(m) = \lambda_\low \Phi(m), 
\end{align*}
where $\Gamma(m) = \frac{1}{\sum_{k=0}^{m} \lambda^k + \frac{\lambda^{m}
    \lambda_\high }{1 - \lambda_\high}}$, $\Lambda(m) =
\sum_{k=0}^{m-1} \lambda^k u_\low(k)$,  and $\Phi(m) = \Gamma(m) \Lambda(m)$.

In the following, we show $\Phi$ is unimodal by establishing that if
$\Phi$ decreases at some integer $m \in \NN$, then it decreases at all
integers $k \geq m$. Towards that goal, for any function
$f : \NN \to \reals$, let $\Delta f(m) \defeq f(m) - f(m-1)$ denote the
finite difference at $m$. Then, we have
\begin{align*}
  \Delta \Gamma(m) &= \frac{1}{\sum_{k=0}^{m} \lambda^k + \frac{\lambda^{m}
    \lambda_\high }{1 - \lambda_\high}} - \frac{1}{\sum_{k=0}^{m-1} \lambda^k + \frac{\lambda^{m-1}
                     \lambda_\high }{1 - \lambda_\high}} = - \lambda^{m-1} \left( \frac{\lambda - \lambda_\high }{1 - \lambda_\high}  \right)\Gamma(m) \Gamma(m-1), \\
  \Delta \Lambda(m) &= \lambda^{m-1} u_\low(m-1), 
\end{align*}
and hence,
\begin{align}\label{eq:phi-difference}
  \Delta \Phi(m)
  &= \Lambda(m) \Delta \Gamma(m) + \Gamma(m-1) \Delta \Lambda(m)\notag\\
  &= -  \lambda^{m-1} \left( \frac{\lambda - \lambda_\high }{1 - \lambda_\high}  \right)\Gamma(m) \Gamma(m-1) \Lambda(m)  + \lambda^{m-1} u_\low(m-1) \Gamma(m-1)\notag\\
  &=  \lambda^{m-1}\Gamma(m-1) \left( u_\low(m-1) -  \left( \frac{\lambda - \lambda_\high }{1 - \lambda_\high}  \right)\Phi(m)\right).
\end{align}
Substituting $\Phi(m) = \Phi(m-1) + \Delta \Phi(m)$ into
\eqref{eq:phi-difference} and after some algebra, we obtain for all
$k \in \NN$,
\begin{align}\label{eq:phi-difference-2}
   \left( 1 + \left( \frac{\lambda - \lambda_\high }{1 - \lambda_\high}  \right) \lambda^{k-1}\Gamma(k-1)\right)\Delta \Phi(k) &= \lambda^{k-1}\Gamma(k-1) \left( u_\low(k-1) -  \left( \frac{\lambda - \lambda_\high }{1 - \lambda_\high}  \right)\Phi(k-1)\right).
\end{align}

Now, suppose $\Delta \Phi(m) = \Phi(m) - \Phi(m-1) \leq 0$ for some
$m \geq 1$. Since $\Gamma(m-1)$ is positive,
from~\eqref{eq:phi-difference} we obtain
$u_\low(m-1) \leq \left( \frac{\lambda - \lambda_\high }{1 -
    \lambda_\high} \right)\Phi(m)$.  Using the expression
\eqref{eq:phi-difference-2} with $k =m+1$, we get
\begin{align*}
  \left( 1 + \left( \frac{\lambda - \lambda_\high }{1 - \lambda_\high}  \right) \lambda^{m}\Gamma(m)\right) \Delta \Phi(m+1) &= \lambda^{m}\Gamma(m) \left( u_\low(m) -  \left( \frac{\lambda - \lambda_\high }{1 - \lambda_\high}  \right)\Phi(m)\right)\\
                                                                                                                             &\leq \lambda^{m}\Gamma(m) \left( u_\low(m) -  u_\low(m-1)\right)\\
  &\leq 0,
\end{align*}
where we have used
$u_\low(m-1) \leq \left( \frac{\lambda - \lambda_\high }{1 -
    \lambda_\high} \right)\Phi(m)$ in the first inequality, and the
fact that $u_\low(\cdot)$ is decreasing in the second inequality. This
in turn implies that $\Delta \Phi(m+1) \leq 0$.

Thus, by induction, we obtain that if $\Delta \Phi(m) \leq 0$ then
$\Delta \Phi(m+k) \leq 0$ for all $k \geq 0$. This proves the
unimodality of $\Phi$ and hence that of $W_\low = \lambda_\low
\Phi$. Finally, note that
$W_\low(1) - W_\low(0) = \lambda_\low (1 -
\lambda_\high)\frac{u_\low(0)}{1 - \lambda_\high + \lambda } > 0$.
Thus, $W_\low(m)$ initially increases up to a maximum, and then
decreases subsequently.

For the third statement, note that for $x = m + a$, where $m \in \NN$ and $a \in [0,1)$, we have
\begin{align*}
    W(x, \theta) 
    = \theta W_\low(x) + (1 -\theta) W_\high(x) & \\
    = \frac{1}{\sum_{k=0}^m \lambda^k + \frac{\lambda^m(\lambda_\high+a \lambda_\low)}{1 - \lambda_\high}} & \left(  \theta \lambda_\low \left(\sum_{k=0}^{m-1} \lambda^k u_\low(k) + a \lambda^m u_\low(m)\right)  \right.\\
    & \quad \left. + (1 -\theta) \left(\lambda_\high \sum_{k =0}^m \lambda^k u_\high(k) + (a \lambda_\low + \lambda_\high) \sum_{k>m} \lambda^m \lambda_\high^{k-1-m}u_\high(k)\right)\right).
\end{align*}
Again, this is of the form $\frac{\alpha + \beta a}{\gamma + \delta a}$, where
$\alpha, \beta, \gamma, \delta$ are independent of $a \in [0,1)$. Thus, we obtain that
$W(m+a, \theta)$ is monotone in $a \in [0,1)$, and hence
$W(x, \theta)$ is monotone between consecutive integers. Since $W( x, \theta)$ is continuous in $x$, the maximum of $W(x, \theta)$ is attained at an integer.~\Halmos\endproof

\begin{lemma}[Properties of the~\eqref{eq:obedience2} function]
\label{lem:leave-worsens}
For $x \in \reals_+$, the function $L(x)$ is strictly decreasing as long as it is non-negative, subsequent to which it stays negative. Formally, we have $L(x) \leq \max \{ \inf_{0 \leq u \leq x} L(u) , 0\}$. 
\end{lemma}
\proof{{Proof of Lemma~\ref{lem:leave-worsens}.}}
Consider a threshold policy $x = m+a$, where $m \in \NN$ and $a \in [0,1)$. We have
\begin{align*}
    L(x) &= \sum_{k=0}^\infty (\lambda \pi_k - \pi_{k+1}) u_\low(k)\\
    &= \lambda_\low \pi_m (1-a) u_\low(m) + \lambda_\low \sum_{k=1}^\infty \pi_{m+k} u_\low(m+k)\\ 
    &=  \lambda_\low  \cdot  \frac{   \lambda^m (1-a) u_\low(m) +   \lambda^m (\lambda_\high + \lambda_\low a)  \sum_{k=1}^\infty\lambda_\high^{k-1} u_\low(m+k)    }{\sum_{k=0}^{m} \lambda^k + \frac{\lambda^{m} (\lambda_\high + a \lambda_\low )}{1 - \lambda_\high}}.
\end{align*}
Since this is a ratio of two linear functions of $a$, we obtain that
it is monotone in $a$, and hence, it suffices to analyze $L(x)$ as a
function over integers. After some algebra, we have
\begin{align*}
  \frac{1}{\lambda_\low} L(m)
  &=   \frac{  \lambda^m      \sum_{k=0}^\infty\lambda_\high^{k} u_\low(m+k)} {\sum_{k=0}^{m} \lambda^k + \frac{\lambda^{m} \lambda_\high }{1 - \lambda_\high} }\\
  &= \frac{\lambda^m \sum_{k=0}^\infty \lambda_\high^k }{\sum_{k=0}^{m} \lambda^k + \frac{\lambda^{m} \lambda_\high }{1 - \lambda_\high}} \cdot  \frac{       \sum_{k=0}^\infty\lambda_\high^{k} u_\low(m+k)    }{\sum_{k=0}^\infty \lambda_\high^k}.
\end{align*}
Now, both factors on the right-hand side are strictly decreasing in $m$. Further, the first factor is positive. If $L(m)$ is non-negative, then the second factor is non-negative, and hence $L(m+1) - L(m) < 0$. On the other hand, if $L(m) < 0$, then the second factor is negative, and since it is decreasing, we obtain $L(m+1) < 0$ as well. Thus, we conclude that $L(x)$ is strictly decreasing as long as it is non-negative, subsequent to which it stays negative. Formally, we have $L(x) \leq \max\{ \inf_{0 \leq u \leq x} L(y) , 0\}$. 
\Halmos\endproof

\section{Proofs from Section~\ref{sec:comparison}}
\label{ap:comparison}

\proof{{Proof of Proposition~\ref{prop:homogen}.}}
First note that $0 < W_{\low}(\fim) \leq W_{\low}(\sm)$ simply follows from the observation that $\fim$ is a feasible signaling mechanism. Thus its welfare is a lower bound on that achieved by the optimal signaling mechanism $\sm$. 

Next, we prove $ W_{\low}(\fim) \geq \beta_\fim W_{\low}(\sm)$. The proof consists of two steps. In the first step, we show $x_\sm \geq m_\fim -1$, where $x_\sm \in \reals_+$ is the threshold of the $\sm$ mechanism. We prove this by showing that the second obedience constraint, \eqref{eq:obedience2}, will not be satisfied if the threshold is below $m_\fim -1$. More precisely, let $\pi^{\sm}$ denote the steady-state distribution corresponding to $\sm$ mechanism, and let $x_\sm = m + a$ where $m \in \NN$ and $a \in [0,1)$. Then 
\begin{align*}
    L(\pi^\sm) &= \begin{cases} \lambda_\low \pi_{m} (1 - a) \cdot u_\low(m) + \lambda_\low \pi_{m+1} \cdot u_\low(m+1), & \text{if $a > 0$;}\\
    \lambda_\low \pi_{m} \cdot u_\low(m), & \text{if $a = 0$.}
    \end{cases}
\end{align*}
Here, the first case follows from the fact that, under the optimal signaling mechanism $\sm$, a user is asked to leave with probability $1 - a$ if the queue length equals $m$, which occurs with probability $\pi_{m}$, and asked to leave with probability $1$ if the queue-length equals $m+1$, which occurs with probability $\pi_{m+1}$. The second case follows analogously. 

Since  $\pi^\sm \in \Piob$, we have $L(\pi^\sm) \leq 0$. This condition, along with the fact that $u_\low(\cdot)$ is strictly decreasing, forces  $u_\low(m+1) < 0$ if $a > 0$, and $u_\low(m) \leq 0$ and $u_\low(m + 1) < 0$ if $a = 0$. In both cases, we have $m+1 \geq m_\fim$, and hence $x_\sm = m+a \geq m_\fim - 1 +a \geq m_\fim -1$. 
Further, from Theorem~\ref{thm:threshold-signaling-mechanism}, we have $x_\sm \leq  m_\fim$. Putting these two together, we obtain
\begin{align*}
x_{\sm} \in [m_{\fim}-1, m_{\fim}].    
\end{align*}
Since $W_\low(x)$ is monotone between integers (as established in Lemma~\ref{lem:unimodal}), we thus obtain that \kiedit{$W_\low(\fim) = W_\low(\sm)$ if and only if $W_\low(m_\fim) \geq W_\low(m_\fim -1)$. Furthermore, we have $W_\low(\sm) \leq \max\{ W_\low(m_\fim -1) , W_\low(m_\fim)\}$}. Now, 
\begin{align*}
    W_\low(m_\fim -1) &= \frac{\sum_{n=0}^{m_\fim-2} \lambda_{\low}^n u_{\low}(n)}{\sum_{n=0}^{m_\fim-1} \lambda_{\low}^n }\\
    &\leq \frac{\sum_{n=0}^{m_\fim-1} \lambda_{\low}^n u_{\low}(n)}{\sum_{n=0}^{m_\fim-1} \lambda_{\low}^n }
    = \left(\frac{\sum_{n=0}^{m_\fim} \lambda_{\low}^n }{\sum_{n=0}^{m_\fim-1} \lambda_{\low}^n }\right) \cdot  \frac{\sum_{n=0}^{m_\fim-1} \lambda_{\low}^n u_{\low}(n)}{\sum_{n=0}^{m_\fim} \lambda_{\low}^n }
    = \frac{1}{\beta_\fim} \cdot W_\low(\fim).
\end{align*}
Here, in the inequality follows from the fact that $u_\low(m_\fim-1) \geq 0$, and the first and the second equalities follow from the definition of a threshold mechanism. In the final equality, we have used the definition of $\beta_\fim$. Thus, taken together, we obtain $W_\low(\fim) \geq  \beta_\fim W_\low(\sm)$. The statement of the proposition follows after noting that $\beta_\fim \geq 1 - \frac{1}{m_\fim+1}$ for all $\lambda_\low \leq 1$. 
\Halmos\endproof

  \proof{{Proof of Proposition~\ref{prop:homogen-noinfo}.}} Recall that $p^\nim$ denotes the probability with
  which a $\low$-type user joins under the no-information
  mechanism. Since $\lambda_\high=0$, we note that $p^\nim > 0$, as a
  $\low$-type user will find it optimal to join the queue if no other
  such user does so. Consequently, we consider the cases $p^\nim = 1$ and $p^\nim \in (0,1)$. Suppose $p^\nim = 1$. Then we have
  \begin{align*}
    W_\low(\nim)
    &= \sum_{n \in \NN} (1- \lambda_\low) \lambda_\low^n u_\low(n)\\
    &\leq \sum_{n < m_\fim} (1- \lambda_\low) \lambda_\low^n u_\low(n) + u_\low(m_\fim) \lambda_\low^{m_\fim}\\
    &= \left( \sum_{n=0}^{m_\fim} (1 - \lambda_\low) \lambda_\low^n\right) \cdot \frac{\sum_{n =0}^{m_\fim-1} \lambda_\low^n u_\low(n)}{\sum_{n=0}^{m_\fim} \lambda_\low^n } + u_\low(m_\fim) \lambda_\low^{m_\fim}\\
    &= \left(  1 - \lambda_\low^{m_\fim +1} \right) \cdot  W_\low(\fim) + u_\low(m_\fim) \lambda_\low^{m_\fim}\\
    &< \left(  1 - \lambda_\low^{m_\fim +1} \right).
  \end{align*}
  Here, we use the fact that $u_\low(k)$ is decreasing $k$ in the
  first inequality, and the final inequality follows from
  $u_\low(m_\fim) < 0$. On the other hand, if $p^\nim \in (0,1)$, we
  have
  $W_\low(\nim) = 0 < (1 - \lambda_\low^{m_\fim +1} )
  W_\low(\fim)$. \Halmos\endproof


\proof{{Proof of Proposition~\ref{prop:info-design-power}.}}  Recall from
Theorem~\ref{thm:full-info-comparison} that the full-information
mechanism is Pareto-efficient if and only if
$W_\low(m_\fim) - W_\low(m_\fim-1) > 0$. 
By a little algebra, this condition can be shown to be equivalent to
\begin{align*}
  f(\lambda_\low, \lambda_\high)\defeq  \lambda_\low  u_\low(0) -  \lambda_\low  \sum_{k=1}^{m_\fim-1} (\lambda_\high + \lambda_\low)^k (u_\low(k-1) -   u_\low(k)) -  (1 - \lambda_\high) u_\low(m_\fim-1) <0.
\end{align*}

It is straightforward to verify that $f(0, \lambda_\high) < 0$,
$f(1- \lambda_\high, \lambda_\high) = 0$,
$\partial_\low f(0, \lambda_\high) = u_\low(0) - \sum_{k=1}^{m_\fim-1}
\lambda_\high^k (u_\low(k-1) - u_\low(k)) \geq u_\low(m_\fim-1) > 0$,
and $\partial_\low^2 f < 0$ for $\lambda_\low \in [0, 1 -\lambda_\high]$, where $\partial_\low $ denotes the
partial derivative with respect to $\lambda_\low$. These facts imply that for
any fixed  \vmredit{$\lambda_\high \in (0,1)$}, the function
$f(\cdot, \lambda_\high)$ has a root
$\bar{\Lambda}_\low(\lambda_\high) \in (0, 1- \lambda_\high]$
satisfying $f(\lambda_\low, \lambda_\high) < 0$ for
$\lambda_\low < \bar{\Lambda}_\low(\lambda_\high)$ and
$f(\lambda_\low, \lambda_\high) > 0$ for
$ \bar{\Lambda}_\low(\lambda_\high) < \lambda_\low < 1 -
\lambda_\high$. Thus, we obtain that the full-information mechanism is
Pareto-efficient if and only if
$\lambda_\low < \bar{\Lambda}_\low(\lambda_\high)$. Finally, the
definition of $f$, along with some straightforward algebra, yields the
following lower-bound:
\begin{align*}
 \bar{\Lambda}_\low(\lambda_\high)   \geq  \frac{u_\low(m_\fim-1)}{ u_\low(0) -   \sum_{k=1}^{m_\fim-1} \lambda_\high^k (u_\low(k-1) -   u_\low(k))} \cdot (1 - \lambda_\high).
\end{align*}

\vmredit{In order to prove the second part, we note that the proof of the first part implies that if  $\lambda_\low < \bar{\Lambda}_\low(\lambda_\high)$, we have $J(m_\fim-1) = W_\low(m_\fim-1) \geq  W_\low(m_\fim) > 0$.
Furthermore,  the assumption $L(m_\fim-1) \leq 0$ implies that the threshold mechanism with threshold of $m_\fim - 1$ is an obedient mechanism. This further implies that the efficient signaling mechanism  has a threshold of at most  $m_\fim - 1$. To see why, suppose  $x_{\sm} > m_\fim - 1$. As established in Lemma~\ref{lem:unimodal},  $W_\high(m_\fim-1) > W_\high(x_{\sm})$ and $W_\low(m_\fim-1) \geq  W_\low(x_{\sm})$ implying that the threshold mechanism with threshold $m_\fim-1$ Pareto dominates the one with threshold $x_{\sm}$ which is a contradiction.  } 
    
\vmredit{In light of the above observations, we have $W_i(\sm) \geq W_i(m_\fim -1)$, for $i \in \{\low, \high\}$. Thus in the following, we establish a multiplicative gap between $W_i(m_\fim -1)$ and $W_i(m_\fim)$ for each $i \in \{\low, \high\}$.}

\vmredit{We start with the $\low$-type users. Observe that, for any threshold mechanism with threshold $m \leq
m_\fim$, we have
\begin{align*}
  W_\low(m) &= \frac{\lambda_\low}{Z_m} \sum_{k=0}^{m -1} \lambda^k u_\low(k),
\end{align*}
where $Z_m = \sum_{k=0}^{m-1} \lambda^k + \sum_{k=m}^\infty \lambda^m
\lambda_\high^{k-m} = \frac{1 - \lambda^m}{1 - \lambda} +
\frac{\lambda^m}{1- \lambda_\high}$. Thus, we obtain
\begin{align*}
  W_\low(m-1) - W_\low(m)
  &= \frac{\lambda_\low}{Z_{m-1}} \sum_{k=0}^{m -2} \lambda^k u_\low(k) - \frac{\lambda_\low}{Z_m} \sum_{k=0}^{m -1} \lambda^k u_\low(k)\\
  &= \frac{\lambda_\low}{Z_{m-1}} \sum_{k=0}^{m -2} \lambda^k u_\low(k) - \frac{\lambda_\low}{Z_{m-1}} \sum_{k=0}^{m -1} \lambda^k u_\low(k)  +\frac{\lambda_\low}{Z_{m-1}} \sum_{k=0}^{m -1} \lambda^k u_\low(k) - \frac{\lambda_\low}{Z_m} \sum_{k=0}^{m -1} \lambda^k u_\low(k)\\
  &=  \frac{\lambda_\low  \lambda^{m-1}    }{ (1- \lambda_\high)Z_{m-1}} \left(W_\low(m) -  (1- \lambda_\high) u_\low(m-1) \right)\\
  &= \left(\frac{ (1-\lambda) \lambda_\low  \lambda^{m-1}}{ 1 - \lambda_\high - \lambda_\low\lambda^{m-1} }\right) \left(W_\low(m) -  (1- \lambda_\high) u_\low(m-1) \right).
\end{align*}
\vmredit{Equivalently, we have}
\begin{align*}
  W_\low(m-1) = \left( 1 + \frac{ (1-\lambda) \lambda_\low  \lambda^{m-1}}{ 1 - \lambda_\high - \lambda_\low\lambda^{m-1} }\right) W_\low(m) -   \left(\frac{ (1- \lambda_\high)(1-\lambda) \lambda_\low  \lambda^{m-1}}{ 1 - \lambda_\high - \lambda_\low\lambda^{m-1} }\right)u_\low(m-1).
\end{align*}
\vmredit{Letting $m = m_\fim$ and using the assumption that  
$u_\low(m_\fim-1) \leq W_\low(\fim)$, we get}
\begin{align*}
  W_\low(m_\fim-1) \geq  \left( 1 + \frac{ \lambda_\high(1-\lambda) \lambda_\low  \lambda^{m_\fim-1}}{ 1 - \lambda_\high - \lambda_\low\lambda^{m_\fim-1} }\right) W_\low(m_\fim) = \beta_{\low, \sm} \cdot W_\low(m_\fim).
\end{align*}
Note that by definition, $1 - \lambda_\high - \lambda_\low\lambda^{m_\fim-1} > 1 - \lambda_\high - \lambda_\low > 0$ for $\lambda < 1$, implying that $\beta_{\low, \sm} > 1$.

Next, we proceed to the $\high$ type. Similar to $W_\low(m)$, we have
\begin{align*}
  W_\high(m) = \frac{\lambda_\high}{Z_m} \left(\sum_{k=0}^{m-1} \lambda^k u_\high(k) + \lambda^m \sum_{k=m}^\infty \lambda_\high^{k-m} u_\high(k)\right) \defeq \lambda_\high \frac{F_m}{Z_m}.
\end{align*}
Thus, we get
\begin{align*}
  W_\high(m-1) - W_\high(m)
  &= \frac{\lambda_\high}{Z_{m-1}} \left( F_{m-1} - F_m\right) +  \frac{\lambda_\low\lambda^{m-1}}{(1- \lambda_\high)Z_{m-1}}  W_\high(m).
\end{align*}
Furthermore,
\begin{align*}
  F_{m-1} - F_m &= -\lambda_\low \lambda^{m-1} \sum_{k=m}^\infty \lambda_\high^{k-m} u_\high(k).
\end{align*}
Thus,
\begin{align*}
  W_\high(m-1) - W_\high(m)
  &=   \frac{\lambda_\low\lambda^{m-1}}{(1- \lambda_\high)Z_{m-1}}  \left( W_\high(m) - \lambda_\high \left( \sum_{k=m}^\infty (1-\lambda_\high)\lambda_\high^{k-m} u_\high(k)    \right) \right)\\
  &= \left(\frac{ (1-\lambda) \lambda_\low  \lambda^{m-1}}{ 1 - \lambda_\high - \lambda_\low\lambda^{m-1} }\right)  \left( W_\high(m) - \lambda_\high \left( \sum_{k=m}^\infty (1-\lambda_\high)\lambda_\high^{k-m} u_\high(k)    \right) \right).
\end{align*}
This implies that
\begin{align*}
  W_\high(m-1) &= \left(1 + \frac{ (1-\lambda) \lambda_\low  \lambda^{m-1}}{ 1 - \lambda_\high - \lambda_\low\lambda^{m-1} }\right) W_\high(m) -  \left(\frac{\lambda_\high (1-\lambda) \lambda_\low  \lambda^{m-1}}{ 1 - \lambda_\high - \lambda_\low\lambda^{m-1} }\right) \left( \sum_{k=m}^\infty (1-\lambda_\high)\lambda_\high^{k-m} u_\high(k)    \right)\\
  &\geq \left(1 + \frac{ (1-\lambda) \lambda_\low  \lambda^{m-1}}{ 1 - \lambda_\high - \lambda_\low\lambda^{m-1} }\right) W_\high(m) -  \left(\frac{ \lambda_\high (1-\lambda) \lambda_\low  \lambda^{m-1}}{ 1 - \lambda_\high - \lambda_\low\lambda^{m-1} }\right) u_\high(m).
\end{align*}

\vmredit{Again, letting $m = m_\fim$ and using the assumption that  
$u_\high(m_\fim) \leq W_\high(\fim)$, we get}
\begin{align*}
  W_\high(m_\fim-1) \geq \left(1 + \frac{ (1-\lambda_\high) (1-\lambda) \lambda_\low  \lambda^{m_\fim-1}}{ 1 - \lambda_\high - \lambda_\low\lambda^{m_\fim-1} }\right) W_\high(m_\fim) = \beta_{\high, \sm} \cdot W_\high(m_\fim). 
\end{align*}}

\kiedit{Before proving the \vmredit{third} part of the proposition, we note that using Assumption~\ref{as:utility_assumptions}, along with a stochastic dominance argument, we obtain that function $g(x) \defeq \sum_{k\in \NN} (1- x) x^k u_\low(k)$ is strictly decreasing in $x \in [0,1)$. Furthermore, $g(0) = u_\low(0) > 0$ and $\lim_{x \to 1^{-}} g(x) < 0$. Thus, there exists a unique $\bar{\Lambda}_\high \in (0,1)$ such that $g(\bar{\Lambda}_\high) = 0$. }
To prove the second part of the proposition, we show that if
$\lambda_\high \geq \bar{\Lambda}_\high$, then no $\low$-type user
joins under the no-information mechanism, i.e., $p^\nim =0$. The
result then follows from Theorem~\ref{thm:no-info-comparison}.  Thus,
suppose $\lambda_\high \geq \bar{\Lambda}_\high$, and no (other)
$\low$-type user joins the queue under no-information mechanism. The
steady-state distribution $\pi$ of the queue is then that of an
$M/M/1$ queue with arrival rate $\lambda_\high$, and hence we have
$\pi_n = (1- \lambda_\high)\lambda_\high^n$ for $n\geq 0$. This
implies that the expected utility (in steady-state) of a $\low$-type
user for joining is given by
$ \sum_{k \in \NN} \pi_n u_\low(n) = \sum_{k \in \NN} (1 -
\lambda_\high) \lambda_\high^k u_\low(n) = \kiedit{g(\lambda_\high)} \leq 0$. The inequality
follows from the fact that \kiedit{$g(x)$} is decreasing in $x$
and equals zero when $x = \bar{\Lambda}_\high$. Thus, we
obtain that the optimal action for a $\low$-type user is indeed not to
join, and hence $p^\nim = 0$. This completes the
proof.~\Halmos\endproof

\proof{{Proof of Theorem~\ref{thm:full-info-comparison}.}}
  Recall that under the full-information mechanism $\fim$, the
  $\low$-type users receive the ``$\join$'' signal if and only if the
  queue-length is strictly less than the full-information threshold
  $m_\fim$. Thus, conditional on receiving the ``$\leave$'' signal,
  the queue-length is at least $m_\fim$, and the expected utility of
  the $\low$-type users for joining the queue is given by
  \[U_\low(0, \join) \leq u_\low(m_{\fim}) < 0,\] where we have used
  the definition of $m_\fim$ and the fact that $u_\low(k)$ is strictly
  decreasing in $k$. Together with the fact that the probability of
  receiving a ``$\leave$'' signal is positive under the
  full-information mechanism, we obtain that $L(m_\fim) < 0$, and hence
  the~\eqref{eq:obedience2} condition does not bind. Hence, from
  Theorem~\ref{thm:pareto-or-leave}, we conclude that the
  full-information mechanism is Pareto-efficient within the class
  $\Pidb$ if and only if it is so within the class $\Piob$. Thus, we
  obtain the dichotomy in the theorem statement.

  To show the final part of the theorem, suppose the
  full-information mechanism is Pareto-efficient within the class of
  admission policies $\Pidb$. Consider the admission policy with
  threshold $m_\fim -1$. In Lemma~\ref{lem:unimodal} (see Appendix~\ref{ap:structure}), we show that $W_\high(x)$ is strictly decreasing in the threshold $x$. Hence, we have
  $W_\high(m_\fim-1) > W_\high(m_\fim)$. Since the full-information
  mechanism is Pareto-efficient within $\Pidb$, this implies 
  $ W_\low(m_\fim) > W_\low(m_\fim -1)$.

  Conversely, suppose $W_\low(m_\fim) > W_\low(m_\fim -1)$.  In
  Lemma~\ref{lem:unimodal}, we also show that $W_\low(x)$ is unimodal, i.e, $W_\low(x)$ is increasing for small $x \in \reals_+$ and decreasing otherwise. The unimodality then implies that for all $0 \leq \hat{x} < m_\fim$, we have $W_\low(\hat{x}) < W_\low(m_\fim)$. Thus, no admission policy
  with threshold $\hat{x} < m_\fim$ Pareto dominates the
  full-information mechanism. Since any admission policy that is not Pareto-efficient is dominated by some threshold policy with threshold less than or equal to $m_\fim$,  we obtain
  that the full-information mechanism, with threshold $m_\fim$, is
  Pareto-efficient within the class $\Pidb$ of admission
  policies.
    \Halmos\endproof

\proof{{Proof of Theorem~\ref{thm:no-info-comparison}.}}  Recall that
under the no-information mechanism, a $\low$-type users joins the
queue with a fixed probability $p^\nim \in [0,1]$ irrespective of the
queue-length upon arrival. 

For $p^\nim \in (0,1)$ it is straightforward to verify that the
resulting steady-state distribution does not have a threshold
structure, and hence by
Theorem~\ref{thm:threshold-signaling-mechanism}, the no-information
mechanism is not Pareto-efficient. For $p^\nim = 1$, the resulting
steady-state distribution has a threshold structure with threshold
equal to infinity. In this case, Theorem~\ref{thm:threshold-signaling-mechanism}
implies that the no-information mechanism is not
Pareto-efficient. Thus, if $p^\nim \in (0,1]$, then the no-information
mechanism is Pareto-dominated within the class $\Piob$ of signaling
mechanisms.

Finally, suppose $p^\nim = 0$. Then, the steady-state distribution
$\pi^\nim$ is given by $\pi^\nim_n = (1-\lambda_\high)\lambda_\high^n$
for $n \geq 0$. Now, consider any other admission policy
$\hat{\pi} \in \Pidb$, where at least some fraction of $\low$-type
users are admitted into the queue. Using a coupling argument, it is
straightforward to show that $\hat{\pi}$ stochastically dominates
$\pi^\nim$. Since $u_\high(n)$ is strictly decreasing in $n$, this
further implies that $W_\high(\hat{\pi}) < W_\high(\pi^\nim)$. Hence,
it follows that the no-information mechanism $\nim$ is Pareto-efficient
within the class $\Pidb$ of admission policies.
\Halmos\endproof

\section{Proofs from Section~\ref{sec:first-best}}\label{app:sec:first-best}

\proof{{Proof of Theorem~\ref{thm:achieve-first-best}.}}  First,
suppose $\lambda_\high \in [\bar{\Lambda}_\high,1]$, and fix a $\theta \in
[0,1]$. From Theorem~\ref{thm:no-info-comparison}, we obtain that the
no-information mechanism $\nim$ is Pareto-efficient, and furthermore,
under $\nim$, all $\low$-type users choose their outside
option. Consider the admission policy $\ar(\theta)$. If $\ar(\theta)$
makes some $\low$-type users join the queue, then the welfare of
$\high$-type users can only be lower than that in $\nim$:
$W_\high(\nim) \geq W_\high(\pi)$. Thus, for $\ar(\theta)$ to be
Pareto-efficient, we must have
$W_\low(\ar(\theta)) > W_\low(\nim) = 0$. Thus, we have
$J(\ar(\theta)) = W_\low(\ar(\theta)) > 0$, and hence the obedience
constraint~\eqref{eq:obedience1} holds. Furthermore, we have
\begin{align*}
  J(\ar(\theta)) + L(\ar(\theta)) &= \lambda_\low \sum_{n \in \NN} \pi_n(\ar(\theta)) u_\low(n) \leq 0,
\end{align*}
where $\pi(\ar(\theta))$ denotes the steady-state distribution under
$\ar(\theta)$. This is because $\pi(\ar(\theta))$ stochastically
dominates the steady-state distribution under $\nim$, and hence the
right-hand side expression is less than
$\lambda_\low \sum_{n \in \NN} (1 - \lambda_\high)\lambda_\high^n u_\low(n)$, which
is non-positive as $\lambda_\high \geq \bar{\Lambda}_\high$. Since
$J(\ar(\theta)) \geq 0$, this implies that $L(\ar(\theta)) \leq 0$,
and hence $\ar(\theta)$ also satisfies the obedience
constraint~\ref{eq:obedience2}. Taken together, we obtain that
$\ar(\theta) \in \Piob$, and hence $\ar(\theta) = \sm(\theta)$.


Next, let $\lambda_\high < \bar{\Lambda}_\high$. Fix $\theta_1, \theta_2 \in [0,1]$ with $\theta_2 > \theta_1$, and let $x_i$ denote the threshold of the Pareto-efficient admission policy $\ar(\theta_i)$. In the following, we first show that $x_1 \leq x_2$. By the definition of $W(\pi, \theta)$ and $\ar(\theta)$, we have
\begin{align*}
    \theta_1 W_\low(x_1) + (1 -\theta_1) W_\high(x_1) &\geq \theta_1 W_\low(x_2) + (1 -\theta_1) W_\high(x_2),\\
  \theta_2 W_\low(x_2) + (1 -\theta_2) W_\high(x_2) &\geq \theta_2 W_\low(x_1) + (1 -\theta_2) W_\high(x_1).
\end{align*}
After some algebra, we obtain 
\begin{align*}
    W_\low(x_2) -   W_\low(x_1) \geq  W_\high(x_2) -  W_\high(x_1).
\end{align*}
Now, if $x_1 > x_2$, then from Lemma~\ref{lem:unimodal} in Appendix~\ref{ap:structure}, we obtain $W_\high(x_1) < W_\high(x_2)$. The preceding inequality would then imply $W_\low(x_1) < W_\low(x_2)$. However, this would imply that the admission policy $\ar(\theta_1)$ is Pareto-dominated by the policy $\ar(\theta_2)$, a contradiction. Thus, we obtain that $x_1 \leq x_2$. 

Next, suppose the admission policy $\ar(\theta_1)$ satisfies the obedience constraints, and hence $L(x_1) \leq 0$. In Lemma~\ref{lem:leave-worsens} (stated and proven in Appendix~\ref{ap:structure}), we establish that if $L(x) \leq 0$ then $L(u) \leq 0$ for all $u \geq x$. Since $x_1 \leq x_2$, Lemma~\ref{lem:leave-worsens} implies that
$L(x_2) \leq 0$, and hence the \eqref{eq:obedience2} condition holds for $\ar(\theta_2)$. Further, by Theorem~\ref{thm:threshold-admission-rule} we have $x_2 \leq m_\fim$, which implies $J(x_2) \geq 0$ and hence the \eqref{eq:obedience1} condition holds for $\ar(\theta_2)$. Together, we obtain that $\ar(\theta_2)$ also satisfies the obedience constraints. 

Thus, we conclude that if for some $\theta_1 \in [0,1]$ the admission policy $\ar(\theta_1)$ satisfies the obedience constraints, then so does the admission policy $\ar(\theta_2)$ for all $\theta_2 > \theta_1$. This implies the existence of (a smallest such) $\theta(\lambda_\low, \lambda_\high) \in [0,1]$ such that for all $\theta > \theta(\lambda_\low, \lambda_\high)$ we have $\sm(\theta) = \ar(\theta)$.\footnote{\kiedit{Note that in this case, the threshold of the admission policy $\ar(\theta)$ (or equivalently the signaling mechanism $\sm(\theta)$) can be positive. For numerical examples, see Figure~\ref{fig:heatmap} and its related discussion in Section~\ref{sec:linear-costs}.}} 

(Note that we allow the possibility that $\theta(\lambda_\low, \lambda_\high) = 1$.) Further, we have $\theta(\lambda_\low, \lambda_\high) > 0$, since for
$\theta = 0$, the admission policy $\ar(0)$ makes all $\low$-type
users take the outside option. However, the obedience
condition~\eqref{eq:obedience2} does not hold for $\ar(0)$ since
$\lambda_\high < \bar{\Lambda}_\high$. 

Finally, for $\theta < \theta(\lambda_\low, \lambda_\high)$, the admission policy $\ar(\theta)$ does not satisfy the obedience constraints, and hence $\sm(\theta) \neq \ar(\theta)$. Theorem~\ref{thm:pareto-or-leave} then implies that the \eqref{eq:obedience2} condition binds for all such $\theta$, i.e., $L(\sm(\theta)) = 0$. In Lemma~\ref{lem:leave-worsens}, we also prove that $L(x)$ is strictly decreasing as long as it is non-negative, and remains negative subsequently. Thus, there exists a unique threshold $\bar{x} \leq m_\fim$ (independent of $\theta$) with $L(\bar{x})=0$. From this, we conclude that $\sm(\theta)$ is the threshold mechanism with threshold $\bar{x}$ for all $\theta < \theta(\lambda_\low, \lambda_\high)$.~\Halmos\endproof

\section{\textcolor{black}{Proofs from Section~\ref{subsec:finite}}}
\label{app:sec:finite}
\kiedit{\proof{{Proof of Proposition~\ref{prop:finite-info}.}} 
\textit{Proof of Part 1:}
Let $\pi$ denote the full-information mechanism.
Suppose $\pi$ is Pareto dominated by an admission policy
$\tilde{\pi}$, i.e., $W_\high(\tilde{\pi}) \geq W_\high(\pi)$ and
$W_\low(\tilde{\pi}) \geq W_\low(\pi)$, with at least one inequality
strict. Since under the full-information mechanism $\pi$, none of the
obedience constraints bind, we obtain that for small enough
$\delta >0$, the admission policy
$ \pi_\delta = (1- \delta) \pi + \delta \tilde{\pi}$ satisfies all the
obedience constraints, and hence can be implemented as a signaling
mechanism. Using the linearity of the welfare functions, we conclude
that $\pi_\delta$ Pareto dominates the full-information mechanism
$\pi$.}

\kiedit{Finally, using Lemma~\ref{lem:one-side-dominance} stated later in this section, we obtain
$W_{\low}(m_\low, m_\high-1) > W_{\low}(m_\low, m_\high)$ and
$W_{\high}(m_\low-1, m_\high) \geq W_{\high}(m_{\low},
m_\high)$. Thus, if
$W_{\high}(m_\low, m_\high-1) \geq W_{\high}(m_\low, m_\high)$, then
we obtain that the threshold mechanism
$\mathsf{Th}(m_\low, m_\high-1)$ Pareto-dominates the full information
mechanism. On the other hand, if
$W_{\low}(m_\low-1, m_\high) \geq W_{\low}(m_\low, m_\high)$, then the
threshold mechanism $\mathsf{Th}(m_\low-1, m_\high)$ Pareto dominates
the full information. \kiedit{In either case, there exists a threshold signaling mechanism that Pareto dominates the full-information mechanism, since for any $\delta > 0$, the signaling mechanisms $(1- \delta) \mathsf{Th}(m, n-1) +\delta \mathsf{Th}(m,n)$ and $(1- \delta) \mathsf{Th}(m-1, n) +\delta \mathsf{Th}(m,n)$ both have a threshold structure.}}





\kiedit{\textit{Proof of Part 2:}
We begin by showing that the no-information mechanism
is Pareto dominated in the class of admission policies. First, suppose
under the no-information mechanism, the $\low$-type users never join,
while the $\high$-type users join with some probability $p \in
(0,1]$. In this case, the admission policy that never admits the
$\low$-type, and implements the admission rule that maximizes the
$\high$ types' welfare Pareto dominates the no-information
mechanism. Next, if the $\low$-type users join with probability
$p \in (0,1)$ under the no-information mechanism, then due to the
assumption on the utilities, the $\high$-type user always joins. Since
$p \in (0,1)$, the welfare of the $\low$-type in this case is
zero. This implies that the admission policy that never admits
$\low$-type user and always admits the $\high$-type user Pareto
dominates the no-information mechanism. Finally, suppose both types
join with probability $1$ under the no-information mechanism. In this
case, the admission policy that never admits any type above queue
length $m_\high$ and always admits below this queue length achieves
higher utility for both types, and hence Pareto dominates the
no-information mechanism.}

\kiedit{Next, suppose under the no-information mechanism, the $\low$-type
users join with positive probability. To show that the no-information
mechanism is Pareto dominated by a signaling mechanism, we split the
argument into two cases:
\begin{enumerate}
\item Suppose in equilibrium, both types join with probability $1$.
  Consider the threshold mechanism $\mathsf{Th}(m_\high, m_\high)$,
  i.e., the mechanism sends signal $2$ up to queue length $m_\high$,
  and sends signal $0$ afterwards. From a straightforward  argument, it
  follows that this mechanism is obedient, and achieves higher welfare
  for both types than the no-information mechanism.
\item Suppose in equilibrium, the $\high$-type users join with
  probability $1$ and the $\low$-type users join with probability
  $p \in (0,1)$. Letting $\pi$ denote the no-information mechanism, we
  have $\pi_{k,0} = 0$, $\pi_{k,1} > 0$ and $\pi_{k,2} > 0$ for all
  $k \geq 0$. Furthermore, we have
  $S_{\high, 1}(\pi) > S_{\low, 1}(\pi) = 0$. Thus, by
  Lemma~\ref{lem:structural_results} stated later in this section, we obtain that no-information
  mechanism is Pareto dominated by a signaling mechanism.
\end{enumerate}
Taken together, we obtain the result. \Halmos\endproof}

\kiedit{
The following lemmas are used in the proof of Proposition~\ref{prop:finite-info}.}
\kiedit{\begin{lemma}\label{lem:one-side-dominance} Suppose $m \leq m_\low$. Then, for $n \geq m$ we have
  $W_\low(m,n-1) > W_\low(m,n)$ and $W_\high(m-1,n) > W_\high(m,n)$.
\end{lemma}}
\proof{{Proof.}}  
\kiedit{First we define two auxiliary functions: $\Psi(m) \triangleq W_\low(m,n)/Z(m,n)$ and $\Phi(m,n) \triangleq W_\high(m,n)/Z(m,n)$ where $Z(m,n)$, $W_\low(m,n)$, $W_\high(m,n)$ are defined in~\eqref{eq:finite:normalization} and~\eqref{eq:threshold:finite} in section~\ref{subsec:finite}.}

\kiedit{
Let $m \leq m_\low$ and
$n\geq m \in \naturals_0$. We have
\begin{align*}
  W_\low(m, n-1) - W_\low(m,n)
  &= (Z(m,n-1) - Z(m,n))\Psi(m)\\
  &= Z(m,n)Z(m,n-1) \left(\frac{1}{Z(m,n)}  - \frac{1}{Z(m,n-1)}\right) \Psi(m)\\
  &= Z(m,n-1) W_\low(m,n) \lambda^m \lambda_\high^{n-m}.
\end{align*}}
\kiedit{Since $m \leq m_\low$, we have $W_\low(m,n) > 0$. Thus, we
obtain $W_\low(m, n-1) > W_\low(m,n)$.
Next, we have
\begin{align*}
  &W_\high(m-1, n) - W_\high(m,n)\\
  &= Z(m-1,n) \Phi(m-1,n) - Z(m,n) \Phi(m,n)\\
  &= Z(m-1,n) \left(\Phi(m-1,n)  - \Phi(m,n)\right)  + \Phi(m,n) \left(Z(m-1,n) -  Z(m,n)\right)\\
  &= Z(m-1,n) Z(m,n) \lambda_\low \lambda^{m-1} \lambda_\high\cdot \\
  &\quad \left(\left(\sum_{k=0}^{m-1}\lambda^k u_\high(k)\right) \left( \sum_{k=m}^n \lambda_\high^{k-m}    \right)      -  \left( \sum_{k=0}^{m-1}\lambda^k\right) \left(   \sum_{k=m}^{n-1}\lambda_\high^{k-m}u_\high(k)\right)       \right)\\
  &\geq Z(m-1,n) Z(m,n) \lambda_\low \lambda^{m-1} \lambda_\high \left(\sum_{k=0}^{m-1}\lambda^k \right) \left( \lambda_\high^{n-m} u_\high(m-1) + \left( \sum_{k=m}^{n-1} \lambda_\high^{k-m}    \right)  \left(u_\high(m-1)    -  u_\high(m)\right)      \right)\\
  &> 0,
\end{align*}}
\kiedit{where the final inequality follows from the fact that $u_\high$ is
strictly decreasing, and since $m\leq m_\low$, we have
$u_\high(m-1) \geq u_\low(m-1) > 0$.}

\Halmos\endproof

\kiedit{
\begin{lemma}\label{lem:structural_results} Consider a signaling
  mechanism $ \pi = \{ \pi_{k, j} : j=0,1,2; k \geq 0\}$ such that
  $\pi_{k,0} = 0$ for all $k \geq 0$ and there exists an $m \geq 0$
  with $\pi_{m, 1} > 0$ and $\pi_{m+1,2} > 0$. If in addition
  $S_{\high, 1}(\pi) > 0$, then $\pi$ is Pareto dominated by a
  signaling mechanism.
\end{lemma}}
\kiedit{
\proof{{Proof.}}  Suppose $\pi$ is as stated in the lemma statement, and
furthermore, there exists an $m \geq 0$ such that $\pi_{m, 1} > 0$ and
$\pi_{m+1, 2} > 0$. Then, for small enough $\delta > 0$, define $\tilde{\pi}$ as follows:
  \begin{align*}
    \tilde{\pi}_{k,2}
    &= \begin{cases}
      \pi_{k,2} & \text{for $k < m$;}\\
      \pi_{m,2} + \frac{\delta}{\lambda_\low} \sum_{n > m+1} \sum_j
      \pi_{n,j} & \text{for $k = m$;}\\
      (1- \delta)\pi_{m+1, 2} - \frac{\delta
        \lambda_\high}{\lambda_\low} \sum_{n \geq m+1} \sum_j
      \pi_{n,j} & \text{for $k = m +1$;}\\
      (1- \delta) \pi_{k,2} & \text{for $k > m+1$;}
    \end{cases}\\
    \tilde{\pi}_{k,1}
    &= \begin{cases}
      \pi_{k,1} & \text{for $k < m$;}\\
      \pi_{m,1} - \frac{\delta}{\lambda_\low} \sum_{n > m+1} \sum_j
      \pi_{n,j} & \text{for $k = m$;}\\
      (1- \delta)\pi_{m+1, 1} + \frac{\delta
        \lambda}{\lambda_\low} \sum_{n \geq m+1} \sum_j
      \pi_{n,j} & \text{for $k = m +1$;}\\
      (1- \delta) \pi_{k,1} & \text{for $k > m+1$;}
    \end{cases}\\
    \tilde{\pi}_{k,0} &= \pi_{k,0} - \delta \pi_{k,0} \ind\{k \geq
                        m+1\}.
  \end{align*}
Then, it is straightforward to verify that $\tilde{\pi}$ satisfies the
balance conditions, given by $\lambda \pi_{k,2} + \lambda_{\high}
\pi_{k,1} = \sum_j \pi_{k+1, j}$ for all $k\geq 0$. Furthermore, we
have, for each $i \in \{\high,\low\}$,
\begin{align*}
  S_{i,2}(\tilde{\pi}) - S_{i,2}(\pi)
  &= \frac{\delta}{\lambda_\low} \sum_{n > m+1} \left({\textstyle\sum}_j \pi_{n,j}\right)  \left( u_i(m) - u_i(n-1) - \lambda_\high ( u_i(m+1) - u_i(n))\right) \\
  &\quad - \frac{\delta \lambda_\high}{\lambda_\low} \sum_{n \geq m+1} \pi_{n,0} u_i(n)\\
  S_{i,1}(\tilde{\pi}) - S_{i,1}(\pi)
  &= - \frac{\delta}{\lambda_\low} \sum_{n > m+1} \left({\textstyle\sum}_j \pi_{n,j}\right) \left( u_i(m) - u_i(n-1) - \lambda ( u_i(m+1) - u_i(n))\right)\\
  &\quad + \frac{\delta \lambda}{\lambda_\low} \sum_{n \geq m+1} \pi_{n,0} u_i(n)\\
  S_{i, 0}(\tilde{\pi}) - S_{i,0}(\pi)
  &= - \delta \sum_{n \geq m+1} \pi_{n,0} u_i(n).
\end{align*}
Now, for any $a \in [0,1)$, we have for all $n > m+1$,
\begin{align*}
  u_i(m) - u_i(n-1) - a ( u_i(m+1) - u_i(n))
  &>  u_i(m) - u_i(n-1) - ( u_i(m+1) - u_i(n))\\
  &=  u_i(m) - u_i(m+1) - ( u_i(n-1) - u_i(n))  \geq 0,
\end{align*}
where the first inequality follows from the fact that $u_i(n)$ is
strictly decreasing, and the second inequality follows from the fact
that $u_i(k) - u_i(k+1)$ is non-increasing. Furthermore, since
$\pi_{m+1,2} > 0$, we must have $\sum_j \pi_{m+2,j} > 0$. Coupled with
the fact that $\pi_{k,0} = 0$ for all $k \geq 0$, we obtain for all
small enough $\delta > 0$ and for $i \in \{\high, \low\}$, 
\begin{align*}
  S_{\low,2}(\tilde{\pi}) > S_{\low,2}(\pi), & &  S_{\low,1}(\tilde{\pi}) < S_{\low,1}(\pi), & &   S_{i,0}(\tilde{\pi}) = S_{i,0}(\pi).
\end{align*}
Since $\pi$ is obedient with $S_{H, 1}(\pi) > 0$, we conclude that
$\tilde{\pi}$ is obedient as well for small enough $\delta >
0$. Finally, we have
\begin{align*}
  W_\low(\tilde{\pi})
  &= \lambda_\low S_{\low,2}(\tilde{\pi}) > \lambda_\low S_{\low,2}(\pi) = W_\low(\pi)\\
  W_{\high}(\tilde{\pi})
  &= \lambda_\high (S_{\high,1}(\tilde{\pi}) + S_{\high,2}(\tilde{\pi}))\\
  &= \lambda_\high \left( S_{\high,1}(\pi) + S_{\high,2}(\pi) \right)\\
  &\quad +  \delta \sum_{n > m+1} \left({\textstyle\sum}_j \pi_{n,j}\right) ( u_\high(m+1) - u_\high(n))  + \delta  \sum_{n \geq m+1} \pi_{n,0} u_\high(n)\\
  &> W_{\high}(\pi),
\end{align*}
where the final inequality follows from the fact that $u_\high$ is
strictly decreasing and $\sum_j \pi_{m+2,j} > 0$. Thus, we obtain that
$\pi$ is Pareto dominated by $\tilde{\pi}$.}

\kiedit{Thus, we obtain that any $\pi$ with $\pi_{k,0} = 0$, $S_{\high,1}(\pi) > 0$ and for which there exists an $m \geq 0$ such
that $\pi_{m, 1} > 0$ and $\pi_{m+1, 2} > 0$ cannot be Pareto
efficient.}  \Halmos\endproof

\section{\textcolor{black}{Further Numerical Analysis for Section~\ref{subsec:finite}}}
\label{sec:finite:structure}

\kiedit{In this section, we numerically examine the structure of the optimal signaling mechanism in the fully persuadable population setting introduced in Section~\ref{subsec:finite}. We start by presenting two examples which show that $\sm(\theta)$ may not have the structure of a threshold mechanism as defined in Section~\ref{subsec:finite}.}

\kiedit{{\bf Examples:} Suppose $u_\low(k) = 1 - c(k+1)$ and $u_\high(k) = 1-c(k+1) - \ell_\high$ with $c=0.15$, and $\ell_\high=-0.7$. Further, let $\lambda_\high = 0.7$ and $\lambda_\low= 1 - \lambda_\high = 0.3$. Recall that $\sigma(n,s) \in [0,1]$ denotes the probability of sending signal $s \in \{0,1,2\}$ when the queue length is $n$. By solving the linear program introduced in Section~\ref{subsec:finite}, we obtain that:
\begin{enumerate}
    \item The mechanism $\sm(0.7)$
    is given by: 
    \begin{align*}
        \sigma(n,s) = \begin{cases}
        \ind (s = 2) & \text{for $0 \leq n < 4$;}\\
        \ind (s = 1) & \text{for $4 \leq n < 10$;}\\
        {0.444} \times \ind (s = 1) + {0.556} \times \ind(s = 0) & \text{for $ n = 10 $;}\\
        {0.190} \times \ind(s = 1) + {0.810} \times \ind(s = 0) & \text{for $ n = 11 $;}\\
    \ind (s = 0) & \text{otherwise,}
  \end{cases}
    \end{align*}
    \item The mechanism {$\sm(0.8)$}
    is given by: 
    \begin{align*}
        \sigma(n,s) = \begin{cases}
        \ind (s = 2) & \text{for $0 \leq n < 4$;}\\
        \ind (s = 1) & \text{for $4 \leq n < 9$;}\\
        {0.774} \times \ind (s = 1) + {0.226} \times \ind(s = 0) & \text{for $ n = 9 $;}\\
        \ind(s = 1)  & \text{for $ n = 10 $;}\\
        {0.199} \times \ind(s = 1) + {0.801} \times \ind(s = 0) & \text{for $ n = 11 $;}\\
    \ind (s = 0) & \text{otherwise,}
  \end{cases}
    \end{align*}
\end{enumerate}}
\kiedit{The above examples show that while the signaling mechanism still follows a ``monotone'' structure by sending signal $2$ (i.e., $\join$ for both types) for small queue length, and then signal $1$ (i.e., $\leave$ for $\low$-type and $\join$ for $\high$-type) for medium queue length and then signal $0$ (i.e., $\leave$ for both types) for sufficiently large queue lengths, the queue length at which the mechanism randomizes between the two signals does not necessarily follow the structure of the $\mathsf{Th}(x,y)$ defined at the beginning of Section~\ref{subsec:finite}.}

\kiedit{Even though the above examples show that  the optimal signaling can be ``slightly'' different from $\mathsf{Th}(x,y)$, 
our numerical analysis confirms that there will be little loss in limiting ourselves to the class of threshold signaling mechanism. As a representative example, for model primitives: $\lambda_\low = 1-\lambda_\high$ with $\lambda_\high \in [0,1]$, $\ell_\high = -0.7$, $u_\low(k) = 1 - c(k+1)$, and $u_\high(k) = 1-c(k+1)-\ell_\high$ with $c=0.15$,
we compute, for each $(\theta,\lambda_\high)$, the best threshold signaling mechanism (found through exhaustive search on a grid of two thresholds with $1/16$ increments) which we denote by $\tsm$. In Figure~\ref{fig:tsm:gap}, we plot the heat map of $\frac{W(\sm, \theta) - W(\tsm, \theta)}{W(\sm, \theta) - W(\fim, \theta)}$. (Note that we use $W(\sm, \theta) - W(\fim, \theta)$ as the normalization factor to ensure that the ratio is in $[0,1]$.) 
We observe that the normalized gap is zero  for most values of $(\theta,\lambda_\high)$; in the regime of $(\theta,\lambda_\high)$ where the gap is nonzero---which includes the examples presented above--- it is very small and notably far from $1$. Thus our numerical analysis suggests that if for practical reasons, using a threshold mechanism is more desirable, there exists a threshold mechanism which performs nearly as well as the optimal one, and better than the full-information mechanism.}

\kiedit{Our further numerical analysis---which we omit for the sake of brevity---shows that 
in the linear utility case, such deviations from a threshold mechanism only occurs when $|\ell_\high|$ is small. For example, for  model primitives used in Figure~\ref{fig:finite:1} where $\ell_\high \in \{-1, -5, -10\}$, for any $\theta \in \{1/12, 2/12, \ldots, 11/12\}$ the optimal signaling mechanism has a threshold structure.} 



\begin{figure}[t]
\centering
\includegraphics[height=0.6\linewidth]{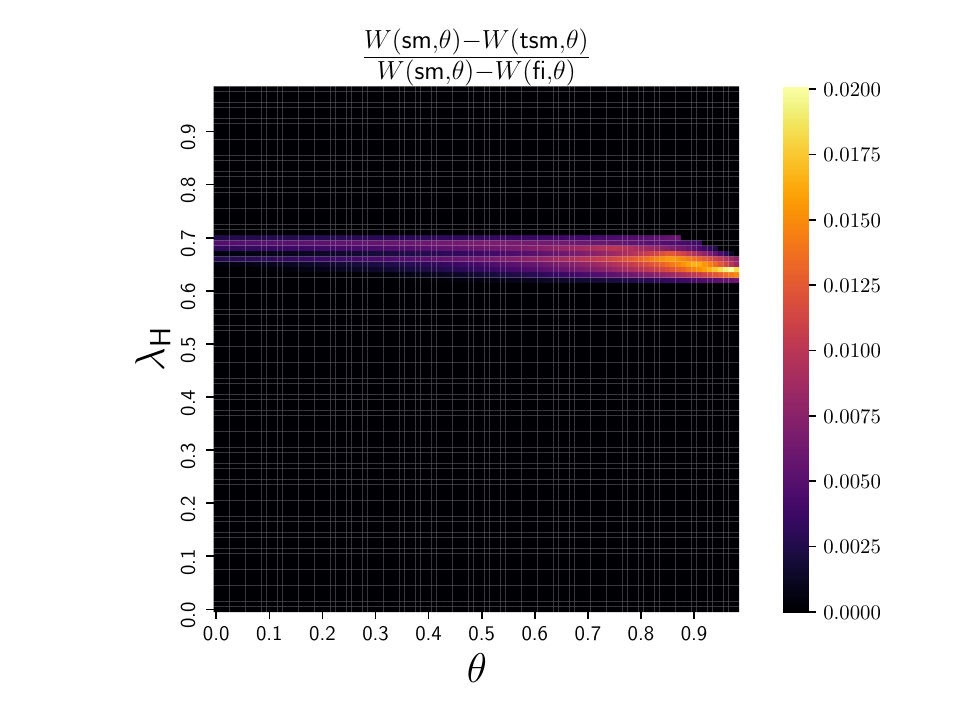}
\caption{Heat map of the normalized welfare gap between optimal signaling mechanism and the best threshold mechanism (found through exhaustive search on a grid of $1/16$ increments).   Model primitives: $\lambda_\low = 1-\lambda_\high$ with $\lambda_\high \in [0,1]$, $\ell_\high = -0.7$, $u_\low(k) = 1 - c(k+1)$ and $u_\high(k) = 1-c(k+1) - \ell_\high$ with $c=0.15$.}\label{fig:tsm:gap}
\end{figure}



\section{\textcolor{black}{Further Numerical Analysis for Section~\ref{subsec:priority}}}
\label{sec:priority:app}

\vmredit{In this section, we expand our numerical analysis for the model introduced 
in Section~\ref{subsec:priority}, where the two types have different service rates. First, using the linear program developed in Section~\ref{subsec:priority} we verify the power of information design for a wide range of gap between the two service rates. Next, we illustrate the effectiveness of information design in a FCFS system, i.e., without a priority scheme.}

\vmredit{In the left panel of Figure~\ref{fig:priority:extra}, we compare the welfare outcome of the Pareto-efficient signaling mechanism ($\sm$) and the 
Pareto-efficient admission policy ($\ar$) when we fix the service rate of $\high$-type to be $1$, but vary that of $\low$-type from $0.8$ to $1.25$. (We recall that under the preemptive priority scheme, the welfare of $\high$-type is unaffected by the signaling mechanism or the admission policy.) In particular, we plot the heat map of $W_{\low}(\ar) -  W_{\low}(\sm)$ on the plane $(\mu_{\low}, \lambda_{\high}) \in [0.8, 1.25] \times [0,1]$. We observe that for $\lambda_\high$ sufficiently large, $W_{\low}(\ar) =  W_{\low}(\sm)$ for any $\mu_{\low} \in [0.8,1.25]$, implying that information design is as powerful as the first-best even when the service rate for the $\low$-type users, $\mu_{\low}$, is considerably below or above its counterpart $\mu_{\high}$ for the $\high$-type users. To illustrate that information design remains effective even when $\lambda_{\high}$ is small, in the right panel of  Figure~\ref{fig:priority:extra}, we plot the welfare of the $\low$-type users under the Pareto-efficient signaling mechanism ($\sm$) and the 
Pareto-efficient admission policy ($\ar$) and the two benchmarks of full-information and no-information mechanisms when $\mu_{\low} \in [0.8, 1.25]$ and $\lambda_{\high} = 0.3$. We observe that for any $\mu_{\low}$, the welfare under $\sm$ remains close to that under $\ar$ and dominates that under the two benchmarks. }



\begin{figure}[t]
	\begin{subfigure}{0.48\linewidth}
    	\centering
    	\includegraphics[height=0.9\linewidth]{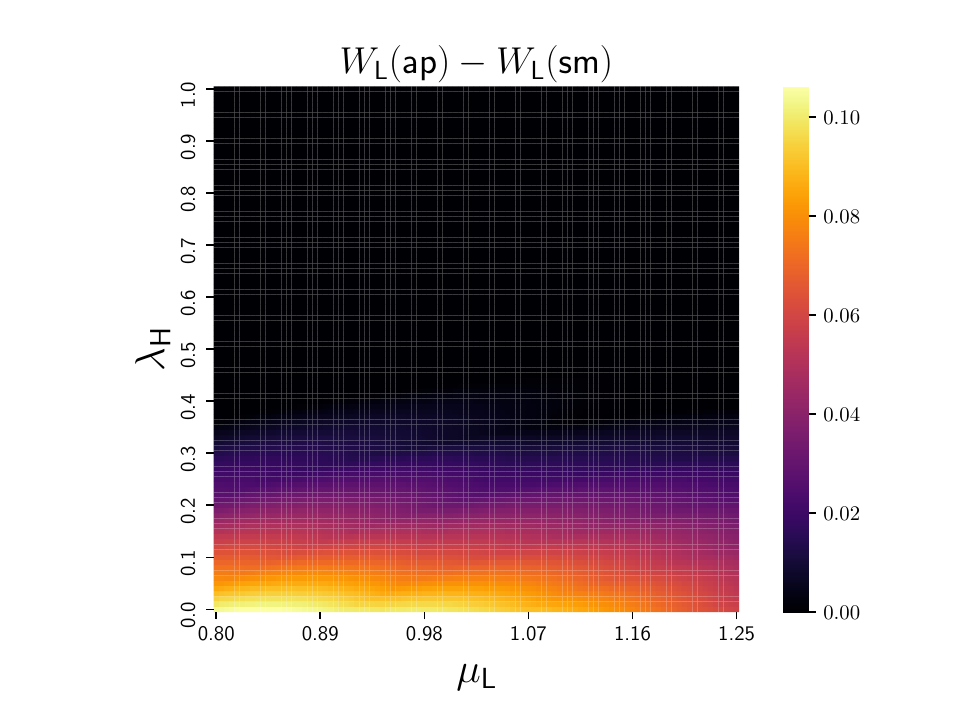}
    \end{subfigure}%
	\begin{subfigure}{0.05\linewidth}
	\end{subfigure}
	\begin{subfigure}{0.45\linewidth}
	\includegraphics[width = \textwidth]{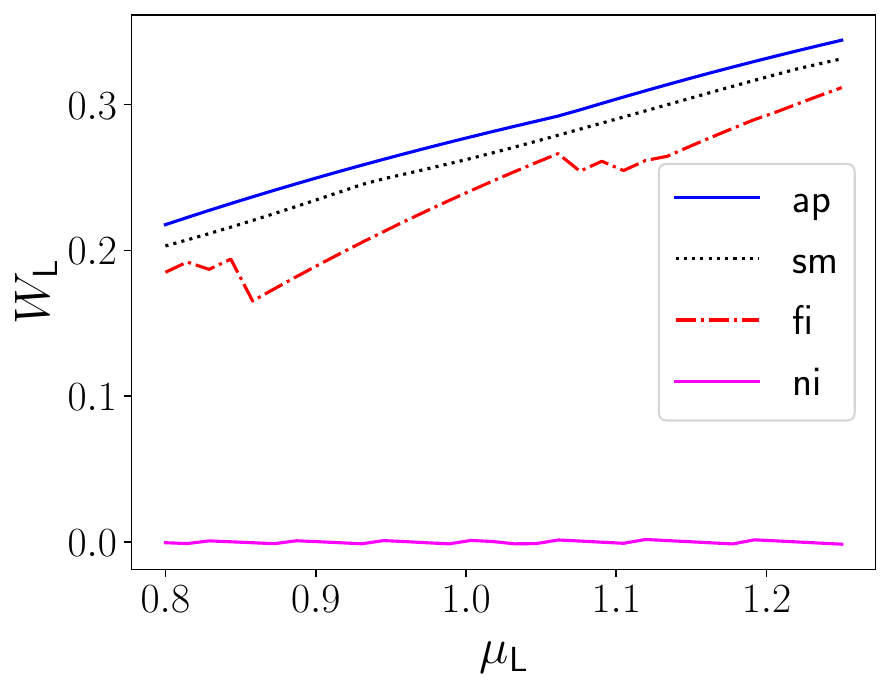}
	\end{subfigure}
	\caption{
\vmredit{Left: Heat map of the welfare gap between the optimal admission policy and optimal public signaling mechanism, i.e., $W_{\low}(\ar) -  W_{\low}(\sm)$ when varying $(\mu_{\low}, \lambda_{\high}) \in [0.8, 1.25] \times [0,1]$.  Other model primitives are the same as in Figure~\ref{fig:priority-queue}. 
Right: Welfare outcomes for $\low$-type under the optimal admission policy, optimal public signaling mechanism, full information, and no information when $\lambda_{\high} = 0.3$ and $\mu_{\low} \in [0.8,1.25]$. Other model primitives are the same as in the left panel. }\vmcomment{Jerry: can you please adjust sizes to make this look nicer?}}\label{fig:priority:extra}
\end{figure}

\vmredit{Next, we consider a setting where the two types differ in their service rates but not in their service priority, i.e., we revisit the FCFS queuing discipline when $\mu_{\low} \neq \mu_{\high}$.  First, we remark that analyzing this setting under the FCFS service discipline is prohibitively challenging because of an explosion in the state space -- it is no longer sufficient to keep track of the number of users in the queue (or even the number of users of different types). Instead, one must track the exact {\em sequence} of the types of users in the queue, as different type sequences (e.g., $\high\high\low$ vs $\high\low\high$) imply different transitions  in the underlying Markovian process. Because of this state space explosion, even numerically computing the Pareto efficient mechanisms under the FCFS discipline is challenging.}

\vmredit{Nevertheless, to study the impact of information design, we restrict our attention to the class of  threshold signaling mechanisms. As discussed before, this class of mechanisms are practically appealing due to their ease of implementation. To that end, we compute the Pareto-efficient threshold signaling mechanisms and compare its welfare outcomes with those of the two benchmarks of full- and no-information mechanisms as well as the welfare outcomes of the Pareto-efficient admission polices within the class of threshold policies.} \vmredit{In Figure~\ref{fig:priority:FCFS}, we present our numerical results for the aforementioned setting and mechanisms. In particular, we 
consider a system with $\lambda_{\low} = \lambda_{\high} = 0.5$, $\mu_{\high} = 1$, and $\mu_{\low} \in \{1, 1.1, 1.2\}$. (The utility functions are the same as the ones described in Section~\ref{subsec:priority}.)}\footnote{\vmredit{We note that due to lack of analytical tractability, we compute the Pareto-efficient threshold signaling mechanism and admission policies using discrete event simulations and exhaustive search over thresholds (on the expected wait time) with granularity of 0.1.The apparent non-convexity of the Pareto-frontier for admission policies  is due to unavoidable simulation noise.}}
\vmredit{We observe that even when restricted to the class of threshold signaling mechanisms, information design results in Pareto-improvement compared to the two benchmarks of providing full or no information for all considered service rates.}

\vmcomment{Question: do we need to talk more about the mechanics of computing FCFS in a two type system (in addition to footnote 19)? i.e., how we set up the discrete event simulation, check for obedience constraints, etc?} \jacomment{I think footnote 19 is good as is.}

\jacomment{I think the current plot is alright. I actually think the plots would be easier to interpret if we draw the lines connecting the Pareto frontier of ap (just like all the other plots) even when it might be slightly non-convex due to simulation errors.}





\begin{figure}
	\begin{subfigure}{0.33\linewidth}
    	\centering
    	\includegraphics[width=\linewidth]{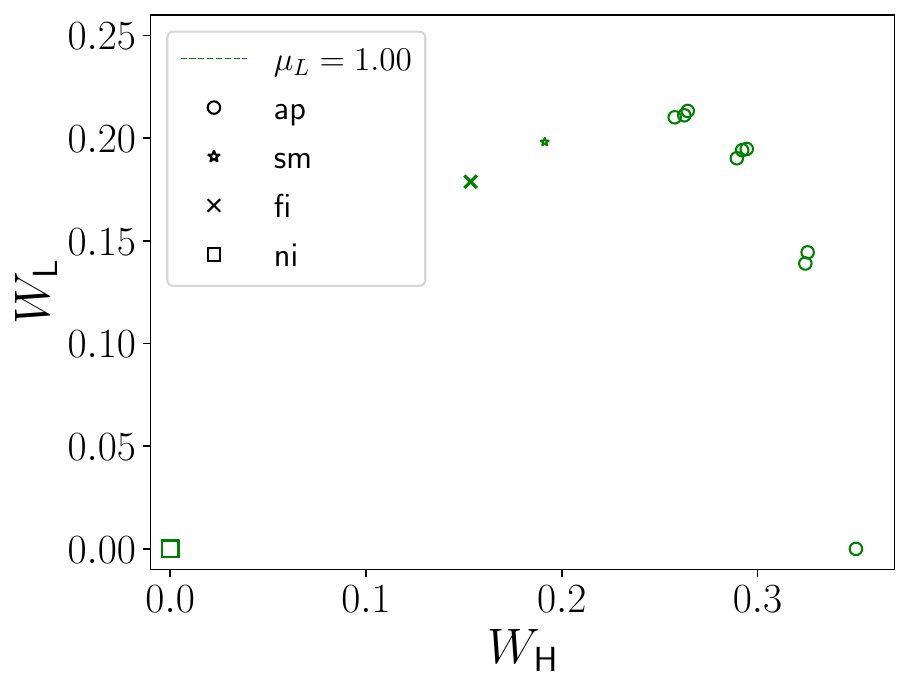}
    \end{subfigure}%
	\begin{subfigure}{0.33\linewidth}
    	\centering
    	\includegraphics[width=\linewidth]{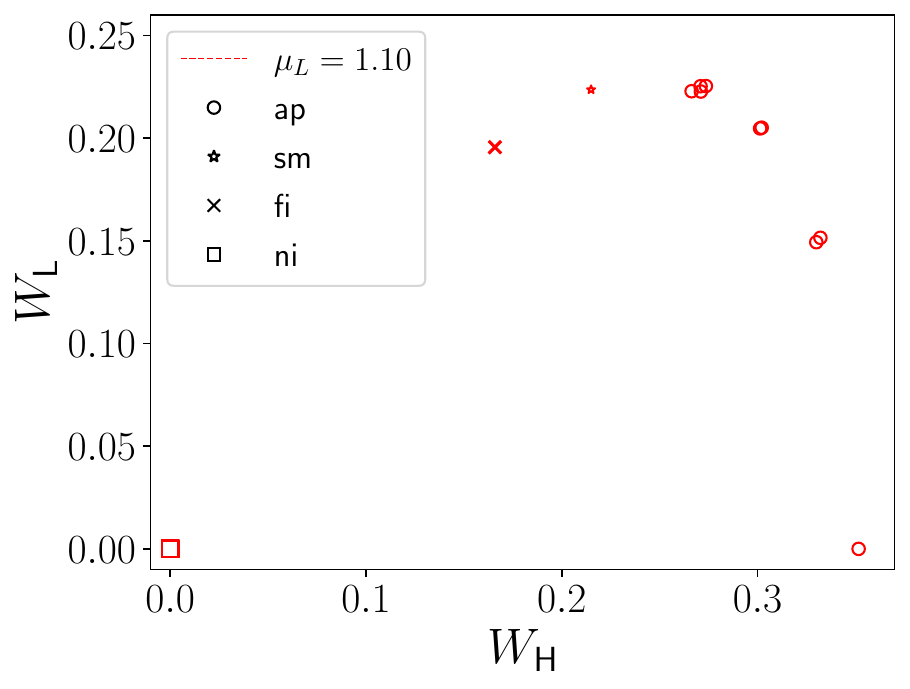}
    \end{subfigure}%
	\begin{subfigure}{0.33\linewidth}
    	\centering
    	\includegraphics[width=\linewidth]{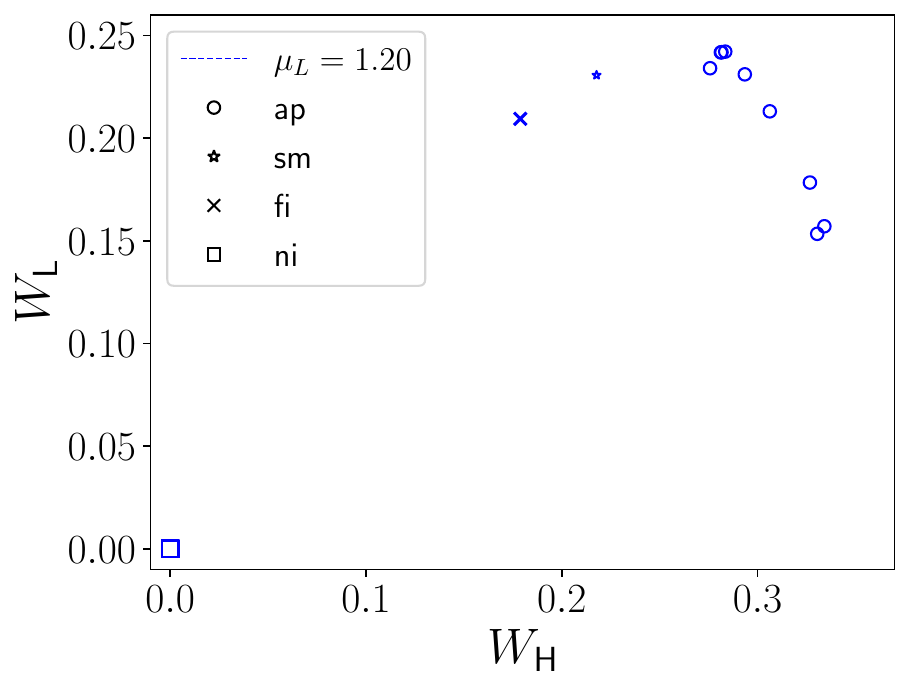}
    \end{subfigure}%
	\caption{\vmredit{Welfare of Pareto-efficient threshold signaling mechanisms, admission policies, full-information, and no-information mechanisms for a FCFS system with 
	$\mu_\low = 1$ (left), $\mu_\low = 1.1$ (center), and $\mu_\low = 1.2$ (right), $\lambda_\high =  \lambda_\low = 0.5$; Other model primitives are the same as in Figure~\ref{fig:priority-queue}.}}\label{fig:priority:FCFS}
\end{figure}

\section{Exogenous Abandonment}
\label{subsec:abandon}

In many situations, applicants of a social service may withdraw their request because they no longer need the service. For example, an individual seeking affordable housing may relocate to another city or move in with a partner. 
To include the possibility of such exogenous abandonment, 
in this section, we consider the same setting as introduced in Section \ref{sec:model}, with one key modification: each arriving user has an independent deadline $\tau$ after which she no longer needs the service. 
More specifically, if she has not already received service by time $\tau$ after her arrival, her need for service disappears and she abandons the queue.
We assume deadlines are i.i.d. and exponentially distributed with rate $\gamma$. 

In the presence of exogenous abandonment, we let $u_i(k)$ denote the expected utility of a type-$i$ user for joining when $k$ users are already ahead in queue. Note that some of these users ahead in queue may abandon before completing their service, and the waiting time for a user is lower than than in the no-abandonment case. Furthermore, a user may obtain some utility subsequent to the abandonment. We assume all these aspects are incorporated into the utility function.

Given these modifications, we can follow the same steps as described in Section \ref{sec:model} and  (i) establish a correspondence between signaling mechanisms and a set of all distributions satisfying obedience constraints, and (ii) characterize the Pareto frontier of the signaling mechanisms and that of the admission policies by formulating and solving linear optimization problems over feasible steady-state distributions. 
In particular, to obtain the Pareto frontier of signaling mechanisms, we solve the following linear program for each $\theta \in [0,1]$: 
\begin{align*}
    \max_{\pi} & \qquad \theta W_\low(\pi) + (1-\theta) W_\high(\pi) \\  
    \text{subject to, } &
     J(\pi) \defeq \sum_{n=0}^\infty \left((1+\gamma n)\pi_{n+1} - \lambda_\high \pi_n\right) u_\low(n) \geq 0, \tag{\textsf{JOIN}} \\
      & L(\pi) \defeq \sum_{n=0}^\infty \left(\lambda \pi_n - (1+\gamma n) \pi_{n+1}\right) u_\low(n)  \leq 0 \tag{\textsf{LEAVE}} \\
     &\lambda_\high \pi_n \leq (1+\gamma n)\pi_{n+1}  \leq \left(\lambda_{\low} +  \lambda_{\high}\right) \pi_{n} \quad \text{for all $n \geq 0$} \tag{\textsf{BALANCE}} \\
     &\sum_{n=0}^{\infty} \pi_{n} = 1,\quad  \pi_{n} \geq 0 \quad \text{for all $n\geq 0$,} 
\end{align*}
where $W_\high(\pi)$ is as defined in \eqref{eq:Wel2}, and $W_\low(\pi)$ is given by $W_\low(\pi) = J(\pi)$ as defined above. The main difference is in the detailed-balance conditions (\textsf{BALANCE}), which capture the fact that the effective arrival rate into the queue is between $\lambda_\high$ and $\lambda = \lambda_\low + \lambda_\high$, and the effective departure rate equals $1 + \gamma n$ when the queue-length equals $n$. Furthermore, the definitions of $J(\pi)$ and $L(\pi)$ reflect the fact that the joining rate of $\low$-type users into the queue is proportional to $(1+\gamma n)\pi_{n+1} - \lambda_\high \pi_n$ when queue-length is $n$, and the rate of leaving of such users is given by $\lambda \pi_n - (1+\gamma n) \pi_{n+1}$. Finally, note that the Pareto frontier of admission policies can be obtained as before by not imposing the two obedience constraints $(\textsf{JOIN})$ and $(\textsf{LEAVE})$ in the preceding program.

For our numerical analysis of this model, we continue to focus on the setting of linear utilities with the same value for service and waiting costs across the two types. It follows from a straightforward analysis that, with $n$ users already in queue, the probability that a joining user receives service (i.e., does not abandon before being served) is given by $\frac{1}{1 + (n+1)\gamma}$, and the expected time until service completion or abandonment is given by $\frac{(n+1)}{1 + (n+1)\gamma}$. Taken together, the utility function of type-$i$ users is given by $u_i(n) = \frac{1}{1 + (n+1)\gamma}\cdot \left(1 -c(n+1)\right) + \frac{(n+1)\gamma}{1 + (n+1)\gamma} \cdot a_i$, where $a_i$ denotes the utility obtained by a type-$i$ user on abandonment. Note that when $\gamma = 0$ (i.e., with no abandonment), this utility function reduces to the one considered in Section~\ref{sec:linear-costs}. 

In Figure~\ref{fig:welfare-comparison:abandon}, we plot the welfare of
Pareto-efficient signaling mechanisms (stars) and admission policies (circles)
 for different values of $\lambda_\low \in \{0.13, 0.20, 0.30\}$, with $\gamma = 0.02$, $c = 0.15$, and $a_i = 0$ for $i \in \{\low, \high\}$. Similar to the setting in Figure \ref{fig:welfare-comparison}, 
we fix $\lambda = 1$. For each value of $\lambda_\low$, we also plot the
full-information mechanism ($\fim$, cross) and the no-information mechanism
($\nim$, square).
We observe that for all three values of $\lambda_\low$, the full-information and no-information mechanisms are Pareto dominated by a signaling mechanism, illustrating the power of information design over these simple information sharing benchmarks. Further, the  Pareto frontier of signaling mechanisms still overlaps with that of admission policies. Taken together, this numerical example shows that our qualitative insights continue to hold in the presence of exogenous abandonment. Finally, compared with Figure~\ref{fig:welfare-comparison}, we observe that the welfare of both types improves as the service is less congested due to abandonment.

\begin{figure}
	\centering
		\includegraphics[width =0.7\textwidth]{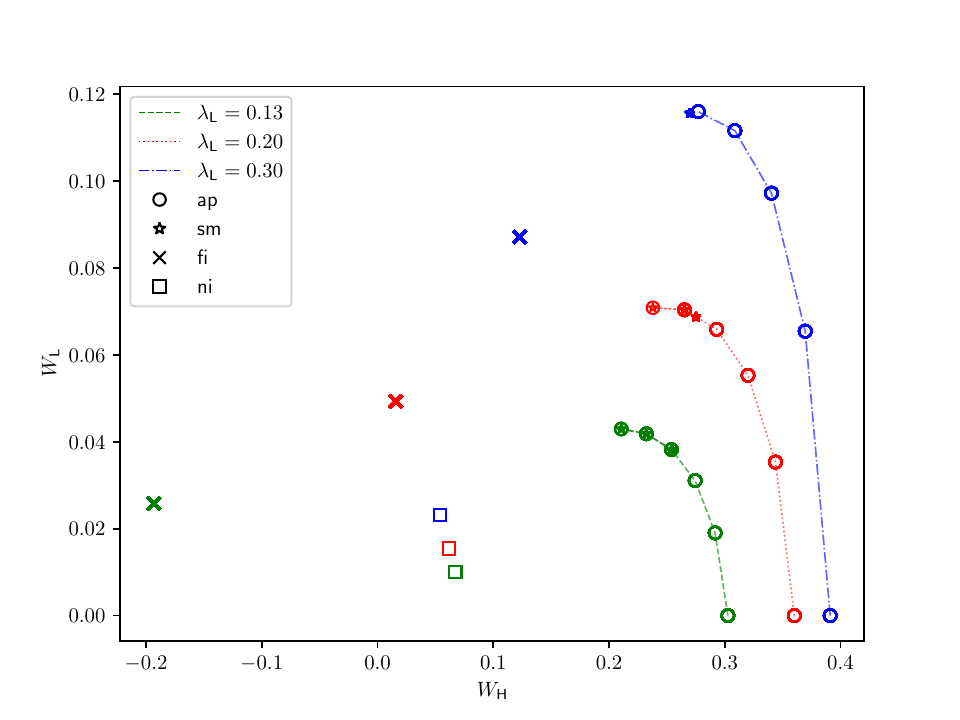}
\caption{
Welfare of Pareto-efficient signaling mechanisms and admission policies for $\lambda_\low \in \{0.13, 0.20, 0.30\}$, $\lambda_\high = 1 - \lambda_\low$, $\gamma=0.02$, $c= 0.15$, $a_\low = a_\high = 0$. 
Here, green (dashes) represents $\lambda_\low = 0.13$, red (dots) represents $\lambda_\low = 0.20$, and blue (dashdots) represents $\lambda_\low = 0.30$. Further, circles ($\circ$) represent efficient admission policies ($\ar$), stars ($\star$) represent efficient signaling mechanisms ($\sm$), cross ($\times$) represents the full-information mechanism ($\fim$), and square ($\square$) represents the no-information mechanism ($\nim$).}\label{fig:welfare-comparison:abandon}
\end{figure}


\section{General User Heterogeneity}
\label{subsec:greater-heterogeneity}

In our baseline model, we capture the extreme of user heterogeneity by considering two user types, one of which has no viable outside option and must join the service. However, in practice, it is reasonable to expect a range of user types with varying levels of need for service and access to outside options. For instance, even among patients with less severe conditions who may be persuaded to avail the alternatives to an emergency room visit, the value of such alternatives might vary substantially based on the patients' symptoms. To incorporate such considerations, in this section, we extend our model to allow for multiple user types that differ in their outside options and value for service. We analyze this model numerically and show that our qualitative insights regarding the effectiveness of information design for welfare improvement continue to hold.

Suppose we have $I$ user types, where a user of type $i \in [I]$ arrives at rate $\lambda_i$, gets utility $u_i(n)$ upon joining the queue with $n$ users ahead of her, and has an outside option of $\ell_i \in \reals \cup \{-\infty\}$. Here, $\ell_i = -\infty$ captures the case where type-$i$ users have no viable outside option. 
(We assume no abandonment in this section.) Our baseline model corresponds to the case $I=2$ with $\ell_1 = 0$ and $\ell_2 = -\infty$. 

In practice, a social service provider may not always be able to observe the type of a user. Moreover, ethical concerns may limit a service provider from making information provision depend on the users' outside options.  Such limitations may make private signaling infeasible, and due to such practical considerations, we focus on {\em public signaling mechanisms}. Note that in our baseline model,  public and private signaling are the same because high-need users have no outside option and always join irrespective of the belief. 

For public signals, using the revelation principle, one can show that it suffices to consider signaling mechanisms where signals correspond to subsets $S \subseteq [I]$, and which are obedient in the sense that when the signal is $S \subseteq [I]$ only users with type $i \in S$ find it optimal to join the queue, whereas users with type $j \notin S$ find it optimal to leave. Focusing on such signaling mechanisms, similar arguments as in Section~\ref{sec:model} allow us to formulate a linear program to compute the Pareto-efficient (public) signaling mechanisms. To see this, for a public signaling mechanism, let $x_{n,S}$ denote the joint probability (in steady-state) that the queue-length upon arrival of a user is $n$ and the user receives the signal $S \subseteq [I]$, and note that  $ \sum_{S \subseteq [I]} x_{n, S}$ denotes the probability that the queue-length is $n$ in steady-state. The detailed-balance condition can then be written as $\sum_{S \subseteq [I]} \lambda_S x_{n,S} = \sum_{S \subseteq [I]} x_{n+1, S}$, where $\lambda_S = \sum_{i \in S}\lambda_i$ denotes the total arrival rate of the users with type $i \in S$. The welfare of type-$i$ users can then be written as a function of $x = \{x_{n, S} : n \geq 0, S \subseteq [I]\}$ as follows: 
\begin{align*}
    W_i(x) = \lambda_i \left( \sum_{n \geq 0} \sum_{S \ni i}  x_{n, S} u_i(n) +  \sum_{n \geq 0} \sum_{S \not\ni i}  x_{n, S} \ell_i\right).
\end{align*}
Further, upon receiving a signal $S \subseteq [I]$, since a user with type $i \in S$ finds it optimal to join the queue, this implies $\sum_{n \geq 0} x_{n, S} ( u_i(n) - \ell_i) \geq 0$ for $i \in S$. Similarly, since a user with type $i \notin S$ finds it optimal to leave, we have $\sum_{n \geq 0} x_{n, S} ( u_i(n) - \ell_i) \leq 0$ for $i \notin S$. Putting it all together, it follows that the Pareto-efficient public signaling mechanisms correspond to the optimal solutions of the following linear program for different choices of non-negative weights $\theta = (\theta_i : i \in [I])$:
\begin{align*}
    \max_x &\quad \sum_{i \in [I]} \theta_i W_i(x)\\
    \text{subject to, } & \sum_{n\geq 0} x_{n, S} (u_i(n) - \ell_i) \geq 0, \quad \text{for $i \in S$ and $S \subseteq [I]$,}\\
    & \sum_{n\geq 0} x_{n, S} (u_i(n) - \ell_i) \leq 0, \quad \text{for $i \not\in S$ and $S \subseteq [I]$,}\\
    & \sum_{S \subseteq [I]} \lambda_S x_{n,S} = \sum_{S \subseteq [I]} x_{n+1,S}, \quad \text{for all $n \geq 0$,}\\
    & \sum_{n\geq 0} \sum_{S \subseteq [I]} x_{n, S} = 1, \text{ and } x_{n, S} \geq 0 \text{ ~for all $n \geq 0$, and $S \subseteq [I]$.}
\end{align*}

Using this linear program, we numerically investigate the effectiveness of information design 
by comparing it to the full-information mechanism and the Pareto-efficient admission polices.\footnote{{We compute the Pareto-efficient admission polices for any given weight $\theta = (\theta_i : i \in [I])$ by dropping the obedience constraints from the preceding linear program.}} In particular, we consider an example with three users types, all with the same linear utility function $u(n) = 1 - c(n+1)$ with $c = 0.15$ but different outside options given by $(\ell_1, \ell_2, \ell_3) = (0, -0.25, -\infty)$.  In particular, the first two types have viable outside options, while the third type has no viable outside option. In keeping with the terminology of our baseline model for easier comparison, we refer to the third-type as the $\high$ type, and the first two types as $\low$ types. The arrival rates  are given by $(\lambda_1, \lambda_2, \lambda_3) = (\lambda_\low, \lambda_\low, \lambda_\high)$ with $\lambda_\low = (1-\lambda_\high)/2$.

In Figure~\ref{fig:welfare-theta-comparison:three}, we plot the welfare $W(\pi, \theta)$ as a function of the arrival rate $\lambda_\high$ for the full-information mechanism, the optimal (public) signaling mechanism $\sm(\theta)$ and the optimal admission policy $\ar(\theta)$ for $\theta= (\theta_1, \theta_2, \theta_3) \in \left\{ (0, 0, 1), \left(\tfrac{1}{4}, \tfrac{1}{4}, \tfrac{1}{2}\right),  \left( \tfrac{1}{2}, \tfrac{1}{2},0\right)\right\}$. Similar to our observations in Figure~\ref{fig:welfare-theta-comparison}, we observe that information design results in welfare improvement over the full-information mechanism when the user population is fairly balanced (given our parametrization of the arrival rates, this corresponds to $\lambda_\high$ being not too large).  Further, for large enough $\lambda_\high$, the optimal signaling mechanism $\sm(\theta)$ achieves the same welfare as that of the optimal admission policy $\ar(\theta)$. This point is further illustrated in Figure~\ref{fig:heatmap:three} (left panel), where, for the weight parametrization $\theta = (\frac{\theta_\low}{2}, \frac{\theta_\low}{2}, 1 - \theta_\low)$, we display the region in the $(\theta_\low, \lambda_\high)$ plane  where the optimal signaling mechanism achieves the same welfare as the optimal admission policy. Finally, Figure~\ref{fig:heatmap:three} (right panel) shows that even in regions where the two policies do not achieve the same welfare, the welfare gap is fairly small.



\begin{figure}
\centering
	\begin{subfigure}{0.32\linewidth}
	\includegraphics[width = \textwidth]{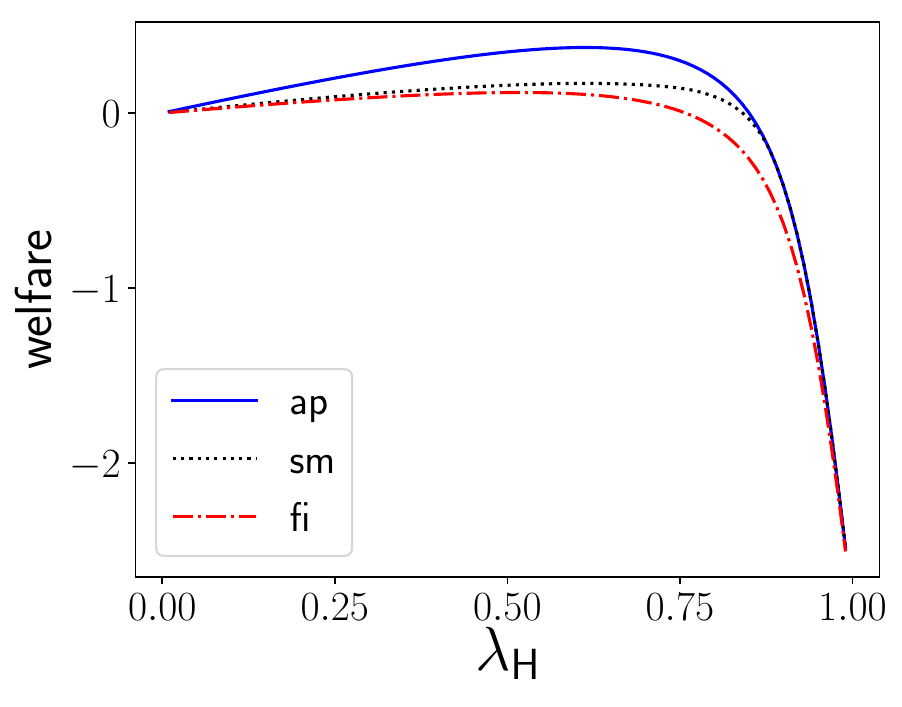}
	\caption{$\theta=(0,0,1)$}
	\label{fig:welfare-theta-comparison-1-three}
	\end{subfigure}%
	\begin{subfigure}{0.32\linewidth}
	\includegraphics[width = \textwidth]{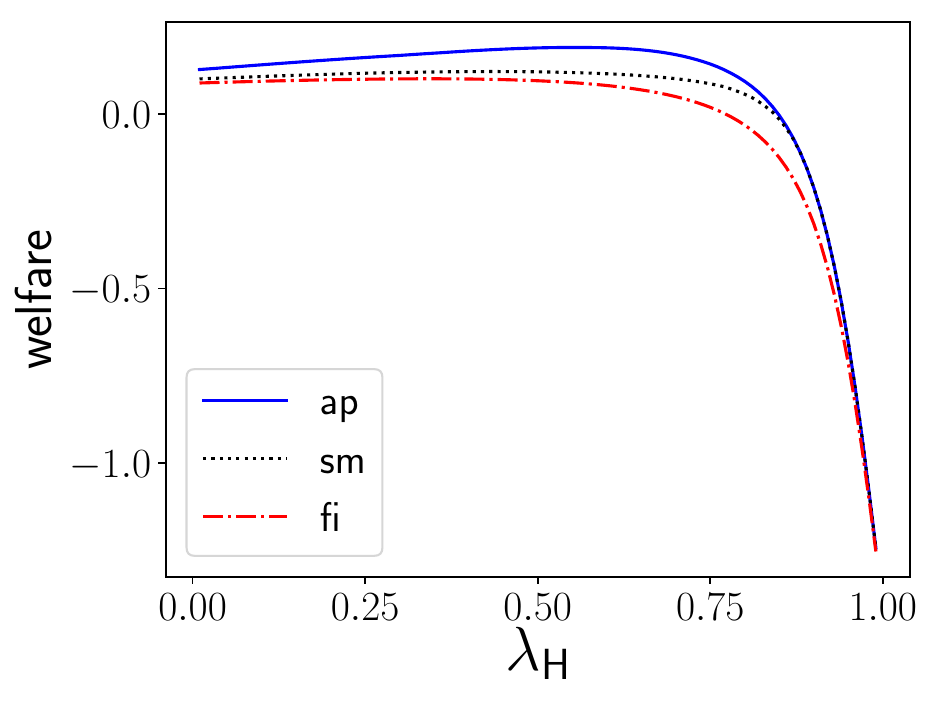}
	\caption{$\theta=(\frac{1}{4},\frac{1}{4},\frac{1}{2})$}
	\label{fig:welfare-theta-comparison-2-three}
    \end{subfigure}%
    \begin{subfigure}{0.32\linewidth}
	\includegraphics[width = \textwidth]{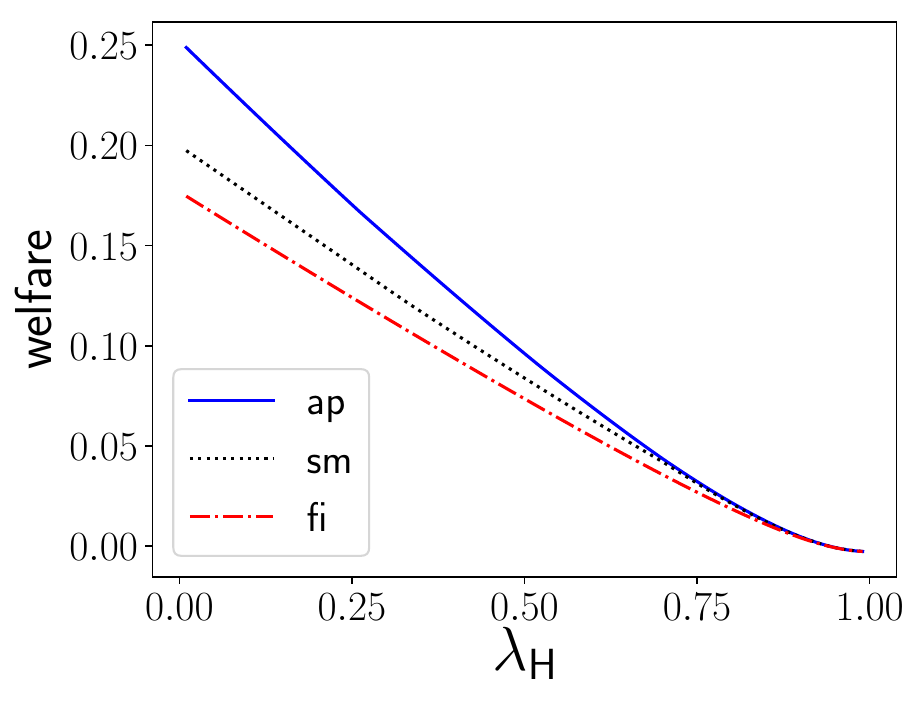}
	\caption{$\theta=(\frac{1}{2},\frac{1}{2},0)$}
	\label{fig:welfare-theta-comparison-3-three}
    \end{subfigure}%
\caption{Welfare of the Pareto-efficient (public) signaling mechanism $\sm(\theta)$, the Pareto-efficient admission policy $\ar(\theta)$, and the full-information mechanism $\fim$ for 
three types $I =\{1,2,3\}$. Here, the arrival rates are given by $(\lambda_1, \lambda_2, \lambda_3) = (\lambda_\low, \lambda_\low, \lambda_\high)$ with $\lambda_\low = (1-\lambda_\high)/2$. The outside options equal $(\ell_1, \ell_2, \ell_3) = (0, -0.25, -\infty)$, and $u_i(k) = 1 - c(k+1)$ with $c=0.15$.}\label{fig:welfare-theta-comparison:three}
\end{figure}

\begin{figure}
	\begin{subfigure}{0.48\linewidth}
    	\centering
    	\includegraphics[height=0.9\linewidth]{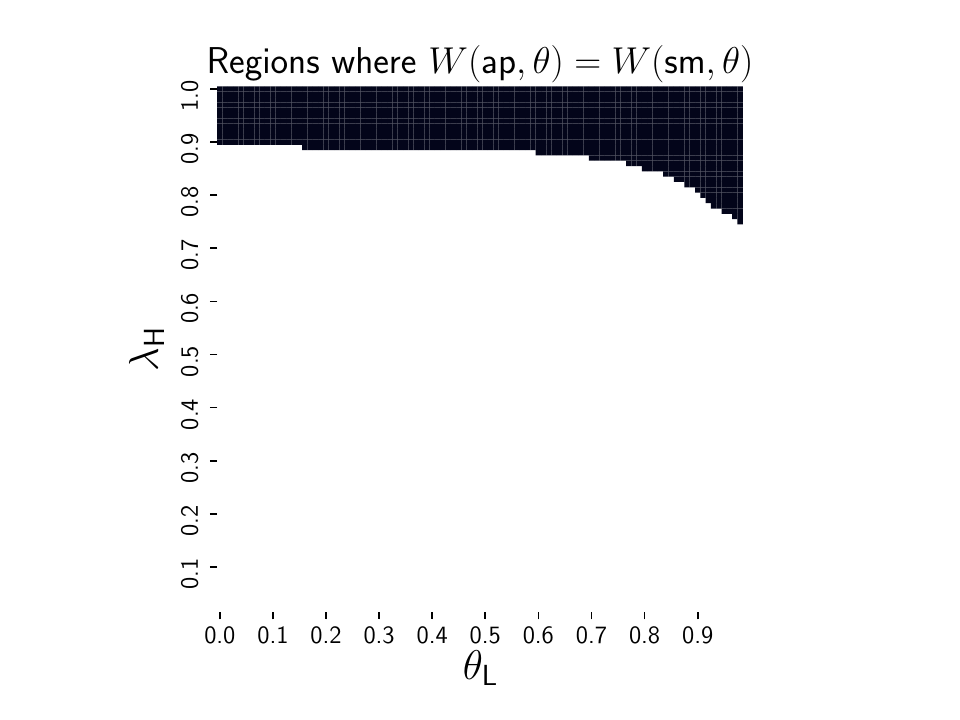}
    \end{subfigure}%
	\begin{subfigure}{0.48\linewidth}
		\centering
		\includegraphics[height=0.9\linewidth]{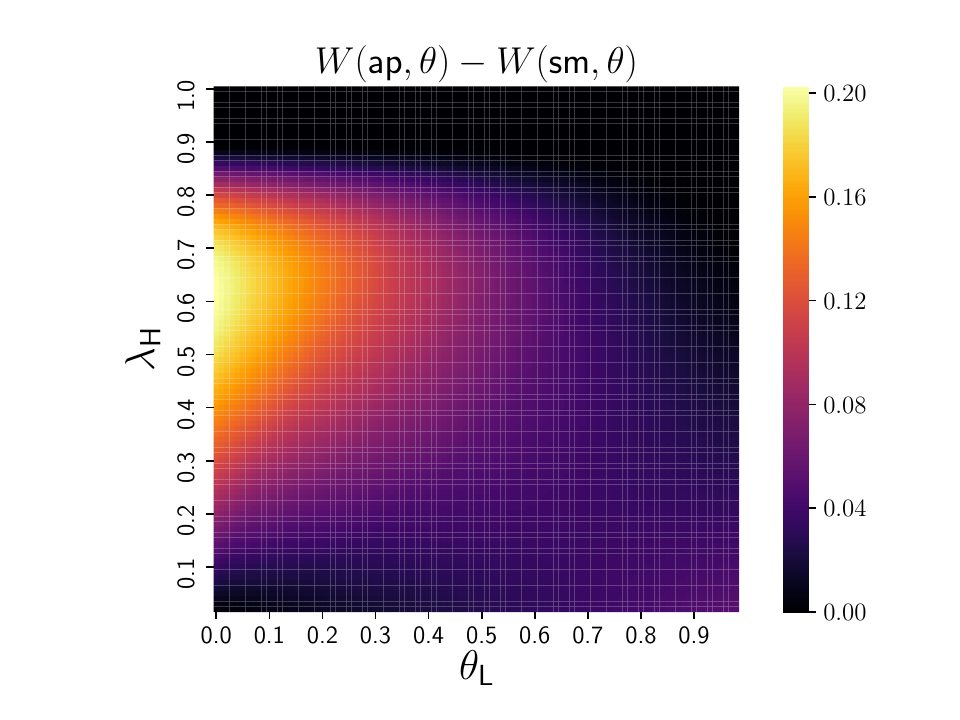}
	\end{subfigure}
	\caption{
Left: Regions of the $(\theta, \lambda_{\high})$ plane for which $\sm(\theta) = \ar(\theta)$, i.e., the signaling mechanism $\sm(\theta)$ is Pareto-efficient within $\Pidb$. 
Right: Heat map of the welfare gap between the optimal admission policy and optimal public signaling mechanism, i.e., $W(\ar, \theta) - W(\sm, \theta)$. 
Model primitives: The arrival rates are given by $(\lambda_1, \lambda_2, \lambda_3) = (\lambda_\low, \lambda_\low, \lambda_\high)$ with $\lambda_\low = (1-\lambda_\high)/2$. The outside options equal $(\ell_1, \ell_2, \ell_3) = (0, -0.25, -\infty)$, $u_i(k) = 1 - c(k+1)$ with $c=0.15$, and the welfare weights are given by $\theta = (\theta_\low/2, \theta_\low/2, 1-\theta_\low)$.
}\label{fig:heatmap:three}
\end{figure}

\end{APPENDIX}




\end{document}